\PassOptionsToPackage{dvipsnames}{xcolor}
\documentclass[amsmath,amssymb,aps,prx,twocolumn,superscriptaddress,showpacs,notitlepage,floatfix]{revtex4-2}

\usepackage{amsmath,amssymb,amsthm,amsfonts,amstext,mathrsfs,dsfont,physics}

\usepackage{txfonts}
\usepackage{times}
\usepackage{booktabs}
\usepackage{multirow}
\usepackage{lineno}

\usepackage{graphicx}
\usepackage{epstopdf}
\usepackage{wrapfig}
\usepackage{adjustbox}
\usepackage{hyperref}
\usepackage{tikz}
\usetikzlibrary{svg.path}
\definecolor{orcidlogocol}{HTML}{A6CE39}
\tikzset{
  orcidlogo/.pic={
    \fill[orcidlogocol] svg{M256,128c0,70.7-57.3,128-128,128C57.3,256,0,198.7,0,128C0,57.3,57.3,0,128,0C198.7,0,256,57.3,256,128z};
    \fill[white] svg{M86.3,186.2H70.9V79.1h15.4v48.4V186.2z}
                 svg{M108.9,79.1h41.6c39.6,0,57,28.3,57,53.6c0,27.5-21.5,53.6-56.8,53.6h-41.8V79.1z M124.3,172.4h24.5c34.9,0,42.9-26.5,42.9-39.7c0-21.5-13.7-39.7-43.7-39.7h-23.7V172.4z}
                 svg{M88.7,56.8c0,5.5-4.5,10.1-10.1,10.1c-5.6,0-10.1-4.6-10.1-10.1c0-5.6,4.5-10.1,10.1-10.1C84.2,46.7,88.7,51.3,88.7,56.8z};
  }
}
\makeatletter
\newlength{\orcid@xh}
\DeclareRobustCommand\orcidlink[1]{%
  \texorpdfstring{%
    \setlength{\orcid@xh}{\fontcharht\font`X}%
    \href{https://orcid.org/#1}{\mbox{%
      \begin{tikzpicture}[yscale=-0.00390625*\orcid@xh,xscale=0.00390625*\orcid@xh,transform shape]
        \pic{orcidlogo};
      \end{tikzpicture}}}%
  }{}}
\makeatother

\usepackage{xcolor}
\usepackage{soul} 
\sethlcolor{yellow}

\usepackage{algpseudocode}
\usepackage{algorithm}
\usepackage{algorithmicx}

\usepackage{bm}
\usepackage{dcolumn}
\usepackage{psfrag}
\usepackage{latexsym}
\usepackage{xr}

\algrenewcommand\algorithmiccomment[1]{\hfill$\triangleright$ #1}

\DeclareMathOperator{\argmax}{arg\,max}
\begin{document}
\title{Real-time noise adaptive quantum error correction by multi-agent learning}

\author{Manuel Guatto \orcidlink{0009-0008-7043-1451}}
\email{m.guatto@fz-juelich.de}
\thanks{Corresponding author}
\affiliation{Forschungszentrum J\"ulich GmbH, Peter Gr\"unberg Institute, Quantum Control (PGI-8), 52425 J\"ulich, Germany}
\affiliation{Institute for Theoretical Physics, University of Cologne, D-50937 Cologne, Germany}

\author{Francesco Preti\orcidlink{0000-0002-0343-9049}}
\affiliation{Forschungszentrum J\"ulich GmbH, Peter Gr\"unberg Institute, Quantum Control (PGI-8), 52425 J\"ulich, Germany}

\author{Michael Schilling\orcidlink{0009-0006-2875-5909}}
\affiliation{Forschungszentrum J\"ulich GmbH, Peter Gr\"unberg Institute, Quantum Control (PGI-8), 52425 J\"ulich, Germany}
\affiliation{Institute for Theoretical Physics, University of Cologne, D-50937 Cologne, Germany}

\author{Tommaso Calarco \orcidlink{0000-0001-5364-7316}}
\affiliation{Forschungszentrum J\"ulich GmbH, Peter Gr\"unberg Institute, Quantum Control (PGI-8), 52425 J\"ulich, Germany}
\affiliation{Institute for Theoretical Physics, University of Cologne, D-50937 Cologne, Germany}
\affiliation{Dipartimento di Fisica e Astronomia, Universit\`a di Bologna, 40127 Bologna, Italy}

\author{F. A. C\'ardenas-L\'opez \orcidlink{0000-0002-2916-2826}}
\affiliation{Forschungszentrum J\"ulich GmbH, Peter Gr\"unberg Institute, Quantum Control (PGI-8), 52425 J\"ulich, Germany}

\author{Felix Motzoi \orcidlink{0000-0003-4756-5976}}
\email{f.motzoi@fz-juelich.de}
\thanks{Corresponding author}
\affiliation{Forschungszentrum J\"ulich GmbH, Peter Gr\"unberg Institute, Quantum Control (PGI-8), 52425 J\"ulich, Germany}
\affiliation{Institute for Theoretical Physics, University of Cologne, D-50937 Cologne, Germany}

\date{\today}

\begin{abstract}
Quantum error correction is essential for scalable quantum computing, yet existing approaches rely on static assumptions about noise that break down in realistic hardware where error channels drift over time. We introduce a unified framework that separates QEC into two learning timescales: offline code discovery and online adaptation. Offline, Multi-Agent Reinforcement Learning (MARL) autonomously discovers complete QEC cycles as explicit quantum circuits, with separate agents responsible for encoding, syndrome extraction, and error recovery, and without prescribing a code family or circuit ansatz. Online, a lightweight adaptive layer, termed Bandit Retraining for Adaptive
Variational Error Correction (BRAVE), realigns the discovered code with the
drifting error channel through a transversal single-unit rotation, while a
bandit controller decides from the syndrome record when an update is worth
its cost. Updates are applied during operation, without decoding the logical
state or retraining the MARL stack, and follow drifts on timescales of order
$0.1$\,s and longer. This yields a “discover once, adapt online” strategy that combines the flexibility of learned codes with real-time adaptation to non-stationary noise. At sufficiently high sampling rates relative to noise drift, our method reduces logical infidelity by up to 3.1-fold for qubit codes and 3.4-fold for qutrit codes compared to the strongest static alternative at each error rate, including a perfect code using two more physical units, while substantially extending robustness to noise fluctuations. These results establish a paradigm in which QEC is no longer static but dynamically optimized for realistic quantum hardware.

\end{abstract}
\maketitle

\section{Introduction}\label{sec1}
Quantum error correction (QEC) is essential for preserving quantum information in the presence of decoherence and environmental noise, remaining as the primary obstacles to scalable quantum computation. Standard QEC frameworks encode logical information into enlarged Hilbert spaces to protect against errors~\cite{NielsenChuang2010,Devitt_2013}. Over the past decades, a variety of codes have been developed to correct single- and multi-qubit errors~\cite{Shor1995,Steane1996,Knill2001}, notably leading to the stabilizer formalism~\cite{gottesman1997stabilizer}, which provides a unified framework for constructing and analyzing QEC codes~\cite{Kitaev2003,Fowler2012,Acharya2024,Brady_2024,Breuckmann2021}. These approaches, however, are typically designed under the assumption that the underlying noise is stationary and accurately characterized.

In realistic quantum hardware, this assumption is frequently violated.  Fluctuations in control electronics, calibration drift, crosstalk, and environmental couplings generate time-dependent noise processes whose dominant error channels evolve during operation.  As a result, QEC protocols optimized for a fixed noise model can become progressively suboptimal. This challenge is exacerbated by the complexity of QEC circuits and the difficulty of identifying the relevant error mechanisms \textit{a priori}, particularly in architectures with partially unknown or evolving noise spectra.

Recent machine-learning-based approaches have demonstrated the potential of data-driven optimization in quantum technologies, including  quantum control~\cite{Preti_SOMA, Guatto_Rob, Calzavara2022Optimizing, Eickbusch2022Fast, Dalgaard2020-fo, Porotti2022deepreinforcement, Nam_Nguyen2024-bz}, quantum compilation~\cite{Prati_Compilation, Preti2024hybriddiscrete, Fürrutter2024, wang2024quantumcompilingreinforcementlearning, PhysRevLett.125.170501}, and entanglement purification~\cite{Preti2022, Preti_Bernard_2024}.  Within QEC, machine learning has been used to optimize individual components such as encoding~\cite{olle2024simultaneousdiscoveryquantumerror, olle2025scalingautomateddiscoveryquantum, Nautrup2019optimizingquantum, meyer2025learningencodingsmaximizingstate, Cao_2022}, decoding~\cite{puviani2023boostinggottesmankitaevpreskillquantumerror, Gicev2023scalablefast, matekole2022decodingsurfacecodesdeep, Lange_2025, Sweke2021-aj}, and circuit synthesis~\cite{zen2024quantumcircuitdiscoveryfaulttolerant}. However, existing approaches generally assume fixed noise models and optimize isolated parts of the QEC stack. This limits their capacity to address time-dependent or unknown error processes, and reduces the ability to co-design the QEC stack. At the same time, higher-dimensional systems offer a route to more efficient encodings~\cite{Gottesman99}, motivating  scalable approaches capable of discovering and optimizing QEC codes in increasingly complex Hilbert spaces beyond qubits. Closer to the present work, Sivak \textit{et al.}~\cite{sivak2026rlqec} use reinforcement learning to steer the physical control parameters of a superconducting processor during QEC, with syndrome-derived signals as feedback, keeping the gate infidelities low as the control parameters drift. The code itself, however, remains fixed throughout. Here we address the complementary question of adapting the code: calibration keeps the physical error rates low, while matching the code to the error channel that remains requires changing its basis as that channel drifts.

In this work, we introduce a framework for real-time adaptive quantum error correction that combines code discovery with online noise adaptation. The code-discovery stage is not initialized with a prescribed QEC ansatz and is not restricted to a specific code family, although in the present benchmarks we validate the approach within the stabilizer formalism and for specific target noise families.

Our approach separates QEC into two complementary stages. In the first stage, a multi-agent reinforcement learning (MARL) framework autonomously discovers complete QEC cycles, with explicit circuit realizations for encoding, syndrome extraction, and recovery. Unlike most previous approaches that optimize these isolated subroutines, the MARL agents co-design the full QEC stack from first principles. In the second stage, the discovered code is augmented with an adaptive variational layer, termed Bandit Retraining for Adaptive Variational Error Correction (BRAVE), which tracks the drifting error channel by updating a low-dimensional set of parameters. The adaptive layer admits variational optimization, although the main benchmarks use the closed-form solution available for the noise models considered. Here, \emph{real-time} refers to adaptation during ongoing QEC on the timescale of the noise drift, rather than within an individual QEC round (Sec.~\ref{sec:discussion_BRAVE}). This online adaptation does not require retraining the full MARL architecture, enabling lightweight real-time recalibration during operation. 

The main methodological contributions are therefore a multi-agent framework for the automated construction of complete QEC cycles and a modular strategy for adapting the resulting protocols to drifting noise. In the offline stage, specialized reinforcement-learning agents construct the encoding, syndrome-extraction, and recovery circuits, with each agent addressing a distinct task within the same protocol. In the online stage, the learned circuit structure is retained while a parameterized unitary layer adjusts the code frame. This stage separates the estimation of the noise drift, the selection of the corresponding frame, and the decision of when to update it. For the channels benchmarked here, the frame is obtained analytically, while the controller learns an update policy that accounts for the cost of reconfiguration. Together, these stages implement a “discover once, adapt online” strategy: the protocol adapts using syndrome information without retraining the circuit-design agents.

 We first demonstrate that the MARL framework recovers canonical qubit codes and extends naturally to qutrit architectures, but this component primarily establishes the quality of the learned base code. Decomposing the discovery of a complete QEC cycle into three specialized optimization tasks generalizes previous automated design approaches, such as Ref.~\cite{F_sel_2018}: the modular structure reduces the effective learning complexity, allows task-specific reward design, and admits hierarchical constructions such as code concatenation. It is also formulated for general $d$-level systems, so the discovery procedure applies beyond qubit-based architectures. We then investigate the central question of this work: whether a learned QEC protocol  can remain robust under continuously drifting noise. We show that the adaptive BRAVE layer maintains high logical fidelity under non-stationary noise, typical in superconducting qubits, lowering the logical infidelity by up to $3.1\times$ for qubits and $3.4\times$ for qutrits relative to the strongest static code available at each error rate, including a perfect code that uses two more physical units. More broadly, our results establish a platform-agnostic framework in which QEC is no longer treated as a static design problem, but as an online adaptive closed-loop protocol for realistic quantum hardware.

The remainder of the paper is organized as follows. In Sec.~\ref{sec:methods}, we introduce the framework and describe its individual components. In Sec.~\ref{sec:non_sta_noise}, we present the physical noise model used as a case study, focusing on non-stationary noise processes relevant to superconducting hardware. Section~\ref{sec:brave_method} describes BRAVE: the adaptive layer, the update controller, and
its syndrome-only implementation. Sections~\ref{sec:marl_results} and~\ref{sec:brave_results} contain the main results. Section~\ref{sec:discussion} discusses the scope of the two stages and possible extensions, including scalability, hardware-aware discovery and the timescale of the adaptation, and Sec.~\ref{sec:conclusions} summarizes our conclusions.

 \begin{figure*}[!t]
    \centering
    \includegraphics[width = \textwidth]{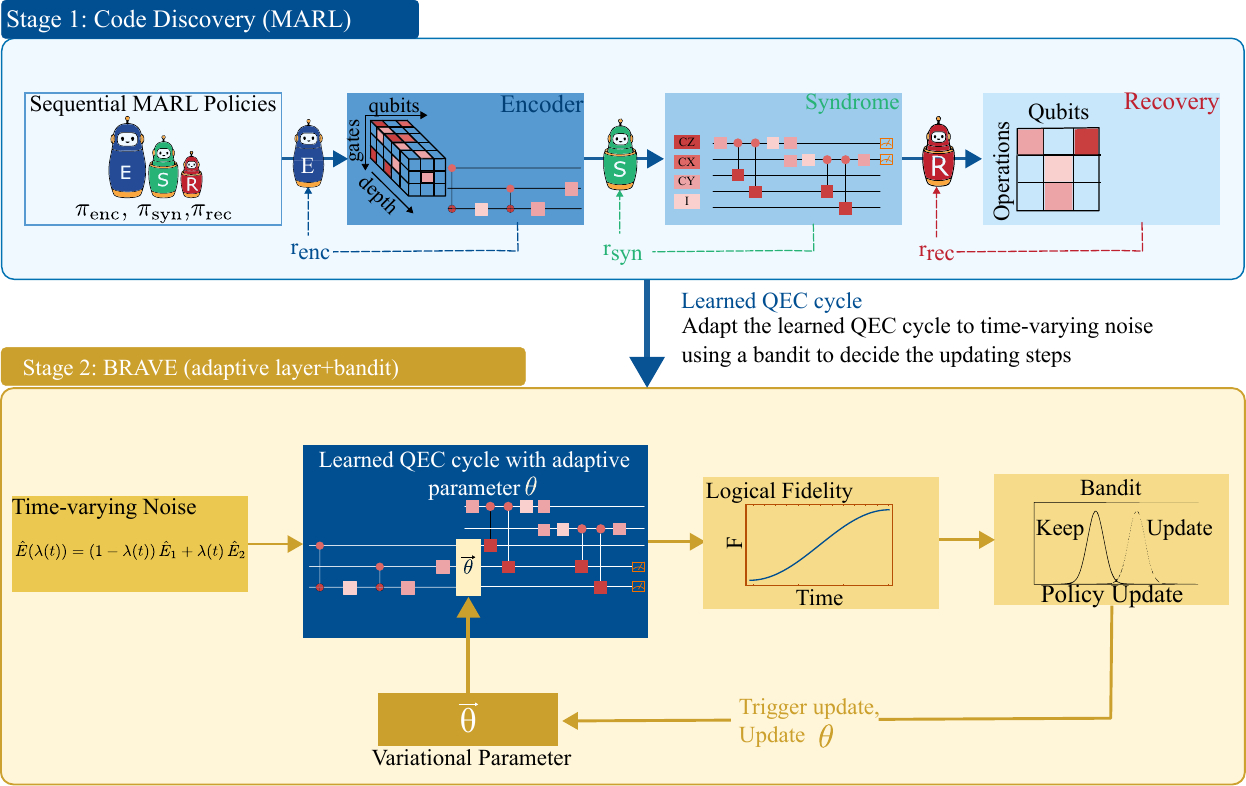}
    \caption{
\textbf{Two-stage framework for adaptive quantum error correction.}
\textbf{(Stage 1) Sequential multi-agent reinforcement learning (MARL) for QEC code discovery.}
Three agents act in sequence to construct a complete QEC cycle. The encoder agent ($\pi_{\mathrm{enc}}$, blue) generates the encoding circuit, the syndrome agent ($\pi_{\mathrm{syn}}$, green) builds the syndrome extraction procedure, and the recovery agent ($\pi_{\mathrm{rec}}$, red) selects the correction operation conditioned on the measured syndrome. Each agent interacts with its corresponding environment (colored blocks) and receives a stage-specific reward ($r_{\mathrm{enc}}$, $r_{\mathrm{syn}}$, $r_{\mathrm{rec}}$), which is used to update the respective policy. This sequential decision process yields an optimized QEC circuit under a stationary noise model.
\textbf{(Stage 2) Adaptive QEC via bandit-based retraining (BRAVE).}
The learned QEC cycle is augmented with a variational layer parameterized by $\boldsymbol{\theta}$ and deployed under time-varying noise. A bandit controller decides at every control step whether to retain the current parameters or to realign the layer. Its feedback is the logical infidelity of the code under the layer in use, shown in the diagram as the logical fidelity $F$: in the benchmark with known drift it is evaluated from the noise model, while in the syndrome-only implementation it is replaced by its posterior expectation inferred from the syndrome record, so that the logical state is never measured. This feedback loop adapts the QEC cycle to non-stationary noise during operation.
}
    \label{fig:RL_general_scheme}
\end{figure*}

\begin{figure*}[t]
    \centering
    \includegraphics[width = \linewidth]{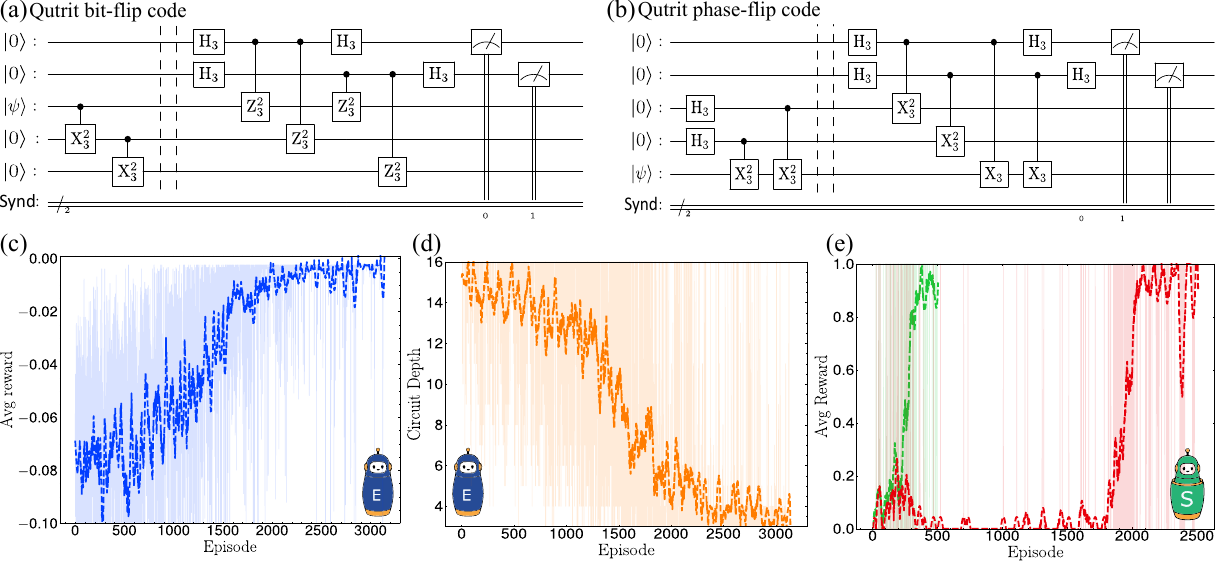}
    \caption{\textbf{RL-based discovery} of \textbf{(a)} the qutrit bit flip and \textbf{(b)} the qutrit phase flip  code. The circuits in the upper part represent the best solution delivered by the agents. In both cases, the encoder is shown before the dashed lines, while the syndrome measurement circuit follows after. In the lower part we see some relevant learning curves: \textbf{(c)} shows the reward maximization for the encoder based on the Knill-Laflamme conditions, \textbf{(d)} the minimization of the circuit depth for the encoder and \textbf{(e)} the reward maximization for the syndrome measurement circuits using the M\&M method: the green line represents the first syndrome, $S_0$ and the red line the second syndrome, $S_1$.}
    \label{fig: Qt_Flip_Phase}
\end{figure*}

\section{Adaptive Reinforcement Learning for QEC}\label{sec:methods}
\subsection{Framework Overview}
We begin by outlining the two-stage pipeline that combines multi-agent reinforcement learning (MARL) for QEC code discovery with  Bandit Retraining for Adaptive Variational Error Correction (BRAVE) for adaptation to non-stationary noise, as illustrated in Fig.~\ref{fig:RL_general_scheme}.\\

 \textit{Stage I: Code discovery}: In the first stage, three reinforcement-learning agents sequentially construct a complete QEC cycle consisting of encoding, syndrome extraction, and recovery under a stationary noise model. Each agent in the pipeline acts on a circuit-construction environment by placing elementary gates [Eqs.~(\ref{eq:1quditgate}) and (\ref{eq:2qudit_gates})], receiving a task-specific reward, and passing the circuit to a subsequent stage. In contrast to conventional approaches, the QEC architecture is not restricted to a predefined code family but is discovered directly through the coordinated actions of specialized policies $\pi_\text{enc},\pi_\text{syn}$ and $\pi_\text{rec}$.\\

  The encoder agent $E$ learns a circuit that maps the logical information into an enlarged Hilbert space (spanned by $n$ additional computational units) 
  enforcing the Knill--Laflamme conditions for correctability. The resulting circuit defines the code space and serves as the input for the subsequent syndrome-extraction stage.

  Given the learned encoding, the syndrome agent $S$ constructs a quantum circuit that entangles the logical state with a set of auxiliary computational units such that their measurement provides information concerning which error subspace is inhabited. 
  Thus, the agent identifies operators whose measurement outcomes distinguish the code subspace from the different error subspaces, without revealing or disturbing the encoded logical information.

  Finally, the recovery agent
  learns a correction circuit conditioned on the syndrome information generated by the circuit from $S$. Its objective is to move corrupted codewords back to the code space via appropriate correction operation and maximize the logical fidelity after error correction.

  Together, the three agents define a complete learned QEC cycle,
\[
(R \circ S \circ E)[\rho],
\] which serves as the base architecture for the adaptive BRAVE stage. \\

\textit{Stage II: Online Adaptation (BRAVE).} After the MARL agents have learned the full QEC cycle (blue circuit in stage 2 of Fig.~\ref{fig:RL_general_scheme}), we consider operation under time-dependent noise channel. In this regime, a code optimized for the initial noise model may become suboptimal, because the error subspaces for which it was designed may no longer provide the best isolation of the dominant errors.

To address this, we augment the MARL-discovered QEC cycle with a variational unitary transformation $U(\boldsymbol{\theta})$. 
This adaptive component is governed by a bandit-based controller that tracks the misalignment between the code and the drifting error axis during execution and decides whether to keep the current layer or realign it. This mechanism enables online adaptation of the learned QEC cycle without retraining the full MARL-discovered architecture.

 The approach naturally induces a separation of timescales: a slow offline stage, in which MARL discovers the global QEC architecture, and a fast online stage, in which BRAVE recalibrates a small set of variational parameters in response to noise drift. This resulting hierarchical structure adapts the code to non-stationary noise while maintaining a fixed circuit backbone, yielding a closed-loop adaptive QEC protocol.

  We demonstrate the approach on qubit and qutrit systems using a transmon-inspired flux-noise model, where fluctuations of the external magnetic flux induce time-dependent variations in the dominant error channel (Sec.~\ref{sec:non_sta_noise}).

\subsection{Automated discovery of QEC codes}
\label{sec:MARL_sec}
Quantum error correcting codes protect quantum information against noise by encoding logical states into an extended Hilbert space $\ket{i_{\rm log}} \in \mathcal{H}_d^{\otimes n}$, such that noise affects only specific degrees of freedom. Any stabilizer-based QEC framework requires three components -- encoding, syndrome measurement, and recovery -- each assigned to a dedicated agent that specializes in its role while collectively optimizing the
overall QEC performance (see Appendix~\ref{sec:stabilizer_appendix}
for a detailed treatment of the stabilizer formalism).

We adopt a sequential training strategy in which each agent is
trained independently, treating the policies of all previously
trained agents as fixed~\cite{marl-book}. Formally, agent
$i \in \mathcal{N}$ learns a policy $\pi_i: \mathcal{O}_i \to
\mathcal{A}_i$ that is a best response to the frozen policies
$\boldsymbol{\pi}_{-i}$ of all preceding agents:
\begin{equation}
V^i_{\pi_i^*, \boldsymbol{\pi}_{-i}} \geq
V^i_{\pi_i, \boldsymbol{\pi}_{-i}} \quad
\text{for all } \pi_i \in \Pi_i,
\end{equation}
where $V^i$ denotes the expected cumulative reward for agent $i$.
The three agents are trained in sequence: first the encoder, then
the syndrome measurement agent conditioned on the learned encoding,
and finally the recovery agent conditioned on both the encoding and
the syndrome strings extracted from the stabilizer measurements.
This yields a composed joint policy $\pi=\pi_\text{rec}(\pi_\text{syn}(\pi_\text{enc}))$.
Further details on the architecture and training procedure are provided in Appendix~\ref{sec:RL},~\ref{sec:General_overview}.

\textbf{Encoder agent.}
A valid encoding must satisfy the Knill-Laflamme (KL) conditions~\cite{Knill_2000},
\begin{equation}
\bra{i}\hat{E}^\dagger_k \hat{E}_l \ket{j} = C_{kl}\delta_{ij},
\label{eq:K-L_conditions}
\end{equation}
where $\{|i\rangle\}$ spans the code space $\mathcal{H}_C$ and
$\hat{E}_k$ are Kraus operators of the noise channel
$\mathcal{N}[\rho] = \sum_k \hat{E}_k \rho \hat{E}_k^\dagger$.
These conditions ensure that correctable errors remain distinguishable without disturbing the logical information. The natural
search space for valid encodings is the Clifford group: since
stabilizer codes are defined by their stabilizer generators,
which are Pauli operators, and Clifford circuits are precisely
those that map Pauli operators to Pauli operators under conjugation,
any valid stabilizer encoding can be expressed as a Clifford circuit.
The encoder agent therefore builds a Clifford circuit by sequentially
placing gates from $\mathcal{C} = \{\text{CNOT}, S, H\}$, a
universal gate set for the Clifford group, represented as a
three-index tensor $(i,j,k)$ encoding gate type, qubit index, and
circuit depth. Rather than searching over abstract encodings, the agent directly explores the space of physical circuits, and is rewarded according to the degree to which the resulting logical states satisfy the KL conditions. In this way, they serve not just a correctness criterion for the learned code, but also as the direct optimization target. The encoder reward combines three KL contributions corresponding to logical-state orthogonality, error detectability, and error distinguishability:
\begin{align}
r_t^{(\mathrm{enc})} = -\gamma_t \sum_{m=1}^{3} KL_m +
r_{\mathrm{success}},
\label{eq:enc_rew}
\end{align}
where the discount factor $\gamma_t$ penalizes unnecessary circuit
depth, encouraging the agent to find minimal encodings. A curriculum
learning strategy progressively exposes the agent to larger error
sets, avoiding local minima and guiding convergence to minimal-depth
encoders (Appendix~\ref{sec:encoder_method}).

\textbf{Syndrome agent.} Once a valid encoding exists, errors must
be identified without measuring the logical state itself. A
detectable error $\hat{E}_k$ anticommutes with at least one
stabilizer generator $\hat{S}_\ell$, flipping its eigenvalue from
$+1$ to $-1$. The pattern of flipped eigenvalues (syndrome
string) identifies the error up to stabilizer equivalence,
without disturbing the logical information. Errors that commute with
all stabilizers but act nontrivially on $\mathcal{H}_C$ constitute
undetectable logical errors and define the fundamental limit of any
syndrome-based approach. The standard circuit primitive for
extracting a stabilizer eigenvalue is a controlled-Pauli measurement
on an ancilla-data qubit pair: the ancilla is prepared, entangled
with the data register via a controlled-Pauli gate, and then
measured. The gate set $\mathcal{A}_s = \{H\text{-}CX\text{-}H,\,
H\text{-}CZ\text{-}H,\, H\text{-}CY\text{-}H,\, I\}$ implements
exactly this primitive for each Pauli operator, making it the
natural action space for constructing stabilizer generators. The
syndrome measurement agent constructs generators $\{S_j\}$ by
placing gates from $\mathcal{A}_s$ on ancilla-data qubit pairs,
represented as a 2D tensor $(j,k)$. Each candidate stabilizer is accepted with reward $1$ only if it acts as the identity on the code space, distinguishes at least one correctable error, is independent of previously found stabilizers, and commutes with all of them ($C_i$ conditions):
\begin{align}
r_t^{(\mathrm{syn})} = \begin{cases}
1, & C_1, C_2, C_3, C_4 \text{ all satisfied}, \\
0, & \text{otherwise.}
\end{cases}
\end{align}
The full measurement circuit is assembled by composing $n-k$
sub-policies $\{\pi_j\}$ via a Mix\&Match~\cite{Mix_Match_Curr}
scheme that activates each sub-policy over its corresponding ancilla
(Appendix~\ref{sec:syndrome_method}).

\textbf{Recovery agent.} Given the syndrome string, the final step
is to apply a correction that returns the corrupted state to the
code space. Crucially, $\mathcal{R} \neq \mathcal{N}^{-1}$:
recovery is a conditional operation that exploits the classical
syndrome information, not an inversion of the noise channel. Since
the syndrome identifies the error only up to stabilizer equivalence,
the recovery operation need only return the state to $\mathcal{H}_C$, not undo the noise exactly. This structure means that for each syndrome string, the agent must learn which correction operator restores the logical state with highest probability. Given the syndrome string $\mathbf{s}(E_i)$, the recovery agent selects a correction operator from $\mathcal{A}_r = \{\hat{X}, \hat{Z}\}$ and the qubit $q_i$ on
which to apply it. The policy is implemented as a syndrome-conditioned weighted mixture,
\begin{equation}
\pi_{\mathrm{mm}}(a|s,\mathbf{s}) =
\sum_{i=1}^K w_i(\mathbf{s})\,\pi_i(a|s),
\end{equation}
and the agent is trained to maximize the fidelity between the corrected and target states. The reward is given by the state fidelity
\begin{equation}
r_t^{(\mathrm{rec})} = F_t =
\mathrm{tr}\!\left[
\sqrt{\sqrt{\rho}\,
\rho_{\mathrm{target}}\,
\sqrt{\rho}}
\right]^2,
\end{equation}
where $F_t$ denotes the Uhlmann fidelity between the recovered state $\rho$ and the target logical state (Appendix~\ref{sec:recovery_method}).

The action space of all the three agents already carries the structure on which hardware constraints act: at each step the agent selects a gate type together with the units it acts on, so restricting the available gate types to the native set of a platform, or forbidding the pairs that are not physically connected, amounts to masking actions rather than changing the formulation. We do not impose such restrictions in the present benchmarks, where the aim in this work is to validate the pipeline without hardware constraints, however we discuss a possible implementation and some future directions in Sec.~\ref{sec:MARL_discussion}.

\section{Non-stationary noise}
\label{sec:non_sta_noise}
\begin{figure*}[!t]

    \includegraphics[width=0.8\linewidth]{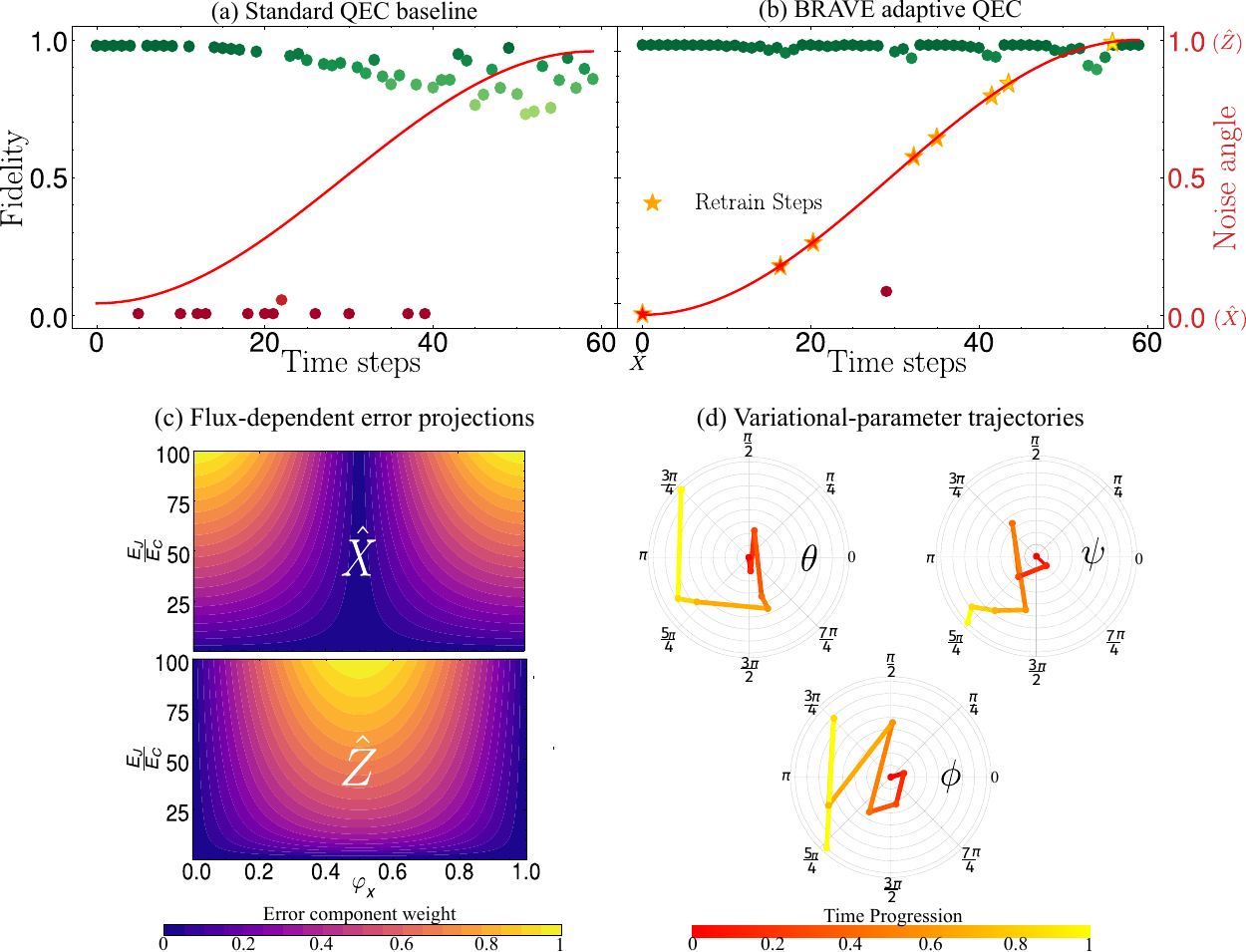}
    \caption{\textbf{BRAVE adaptation of code to  fluctuating noise channel. }
\textbf{(a)} Fidelity trajectory over time under the fixed standard QEC baseline in the presence of non-stationary noise with error probability $p = 0.2$. As the noise changes, fidelity continuously degrades.
\textbf{(b)} Fidelity trajectory using the adaptive BRAVE approach. Stars indicate retraining steps and are colored using the same temporal gradient as in panel~(d), so that their color encodes the corresponding time step along the variational-parameter trajectory. When fidelity begins to drop, the agent responds by retraining the variational layer, leading to a recovery in fidelity.
\textbf{(c)} $\hat{X}$ and $\hat{Z}$ projection of the interacting Hamiltonian of the Transmon circuit as a function of the external flux $\varphi_{x}(t)$ for different $E_J/E_C$ ratios. This illustrates that for different values of the control parameter, the noise profile changes drastically.
\textbf{(d)} 2D trajectory of the variational parameters $\boldsymbol{\theta}$ defining the correction action. As the noise shifts from an $\hat{X}$-type to a $\hat{Z}$-type error, the points move from the center of the circle outward toward its boundary, indicating parameter adaptation. Examining the adaptation at $t = 0$ to $ t = 1.0 $ in the figure, we observe that at $ t = 0 $, when the noise is $ \hat{X} $, the parameters are initialized at $ \vec{0} $. This implies that the unitary transformations act as the identity, and the stabilizers function in the standard way to detect bit-flip errors. In contrast, at $ t = 1 $, when the noise corresponds to $ \hat{Z} $, the parameters take the values $ \boldsymbol{\theta} = \left( \frac{3\pi}{4}, \frac{5\pi}{4}, \frac{5\pi}{4} \right) $. This configuration is equivalent to the effect of the Hadamard gate in the standard phase-flip code, rotating both the error and the stabilizer projectors s.t. the error is detectable and correctable}
\label{fig:Fig_2_1}
\end{figure*}

In this section, we will show that the noise model consisting in a linear combination of both $\hat{X}$ and $\hat{Z}$ errors can be mapped in a superconducting circuit architecture based on a tunable-frequency transmon circuit~\cite{PhysRevA.76.042319} subjected to noise on the external magnetic flux. Roughly speaking this artificial atom consists on a shunted capacitance coupled to a superconducting quantum interference device (SQUID) threaded by an external magnetic flux $\phi_{x}$. This circuit is described by the following Hamiltonian
\begin{eqnarray}
    \hat{H} = 4E_{c} \hat{n}^2 - E_{J1}\cos(\hat{\varphi}) - E_{J2}\cos(\hat{\varphi} -\pi\varphi_x).
\end{eqnarray}
Here $E_{C}=e^2/2C$ is the charge energy with $e$ being the electron charge, $\hat{n}$ is the charge operator that quantifies the number of Cooper-pairs on the device. $E_{Jk}$ is the Josephson energy of the $k$th junction, moreover $\hat{\varphi}$ is the phase operator and $\varphi_x=\phi_{x}/\varphi_{0}$ is the external phase with $\varphi_{0}=\hbar/2e$ is the quantum flux. We should note that we can separate the Hamiltonian contributions that depends or not on $\varphi_x$. i.e., $\hat{H}_0 = 4E_{c} \hat{n}^2 - E_{J1}\cos(\hat{\varphi})$ and $\hat{H}_n = - E_{J2}[\cos(\pi{\varphi_x})\cos(\hat{\varphi}) + \sin(\pi\varphi_x)\sin(\hat{\varphi})]$. Thus, we can represent the noise term in the eigenbasis of the free part $\hat{H}_0=\sum_{\ell}\epsilon_{\ell}\ketbra{\ell}{\ell}$, obtaining $\cos(\hat{\varphi})\propto \sum_{\ell}\ketbra{\ell}{\ell}\approx \hat{Z}$ and $\sin(\hat{\varphi})\propto \sum_{\ell}\ketbra{\ell}{\ell+1}\approx \hat{X}$ so that if $\varphi_x \in (0, 1/2)$ we have a smooth transition between both types of errors as depicted in Fig.~\ref{fig:Fig_2_1}(c). In summary, the error model considered in our adaptive reinforcement learning scheme is naturally present in the context of superconducting qubits, which further strengthens the experimental relevance of our algorithmic approach. While our framework is demonstrated specifically for superconducting qubits, this noise model is general to dipolar coupling between qubits and two-level systems. In such a case, typically near resonance results in energy exchange (X-like errors) while off-resonant coupling gives Z error, with a smooth transition when the frequency difference shifts between the two \cite{RevModPhys.93.025005}.

\section{BRAVE}
\label{sec:brave_method}
When the dominant error channel drifts in time, the error subspaces targeted by the originally learned stabilizer code may no longer coincide with the most relevant physical errors. As a result, a fixed QEC cycle can become suboptimal even if it was well adapted to the initial noise model. Starting from the QEC cycle discovered in the first stage by MARL, we introduce BRAVE which augments the learned encoder, syndrome-extraction, and recovery operations with a variational calibration layer that adapts the effective error basis under time-dependent noise, such as superconducting qubits error under the fluctuation of an external magnetic flux. This approach leverages a variational ansatz in which the QEC cycle is augmented by a tunable unitary $\hat{U}_{\boldsymbol{\theta}}$, parametrized through the generators of the Lie algebra $SU(d)$. The unitary contains \(d^2-1\) degrees of freedom~\cite{Tilma_2002} and acts as a calibration layer that adjusts the effective error basis seen by the code whenever an update is applied. In the qubit case, these generators reduce to the Pauli matrices, while for qutrit systems they correspond to the Gell-Mann matrices. The variational layers modify the encoder $\mathcal{E}'$, inducing corresponding changes to the stabilizers and recovery operators as
\begin{equation}
    S_i' = \hat{U}_{\boldsymbol{\theta}} S_i \hat{U}_{\boldsymbol{\theta}}^\dagger \quad
    \text{and} \quad E_{\mathbf{s}}' = \hat{U}_{\boldsymbol{\theta}} E_{\mathbf{s}} \hat{U}_{\boldsymbol{\theta}}^\dagger \: .
\end{equation}
This ensures that all parts of the QEC cycle remain consistent and co-adapted under parameter changes.

The layer acts identically on all physical units, $\hat{U}_{\boldsymbol{\theta}} = \hat{u}(\boldsymbol{\theta})^{\otimes n}$, which delimits the class of drift it can compensate. It absorbs exactly those drifts that act as a change of basis of a fixed error set, identical on all units: the error operators rotate, while their number and their algebraic structure are unchanged, as for the flux-noise model of Sec.~\ref{sec:non_sta_noise}, where the dominant error rotates from $\hat{X}$ to $\hat{Z}$. A change of the Kraus weights at fixed error set does not affect correctability, since the Knill--Laflamme conditions do not depend on the probabilities, and changes only the logical error rate through the probability of uncorrectable multi-unit events; a change of the Kraus rank is a different matter, since a new error operator appears whose correctability is set by the distance of the base code rather than by the layer, and if the enlarged error set exceeds the capability of the code no choice of $\boldsymbol{\theta}$ restores the conditions. Drifts acting differently on different units would require independent layers on each unit, correlated errors are not compensated by a transversal single-unit rotation, and leakage lies outside the framework altogether, since population leaving the computational subspace cannot be restored by any unitary acting within it. For non-unitary processes such as amplitude damping no code of this size satisfies the conditions exactly, and the natural setting is approximate quantum error correction, with the layer optimized against a fidelity objective rather than against the violation of the Knill--Laflamme conditions.

This distinction between a code that becomes suboptimal and one that fails is worth making precise, since the drift considered here does the latter. Writing the drifted error in the frame of the code as $\cos\delta\,\hat{X}' + \sin\delta\,\hat{Z}'$, where $\delta$ is the misalignment between the code and the instantaneous error axis, the component along the stabilizer axis commutes with all stabilizers and acts on the code space as a logical operator: it is undetectable, and it contributes a logical error at first order in the physical error probability $\simeq 2p\sin^2\delta$ for the three-qubit code, instead of at second order as for a correctable error set. A change of the error probabilities at fixed error type, by contrast, leaves the conditions satisfied and makes the code merely suboptimal.

Because the layer is transversal, applying it does not require decoding the logical information. A state encoded in the frame $\boldsymbol{\theta}$ is carried into the frame $\boldsymbol{\theta}'$ by the transversal rotation $\big(\hat{u}(\boldsymbol{\theta}')\hat{u}(\boldsymbol{\theta})^{\dagger}\big)^{\otimes n}$ applied to the physical units, with the stabilizers and recovery operators following by conjugation, so that the code space is mapped onto the new code space and the logical information is preserved exactly, as demonstrated by the continuous-protection benchmark of Appendix~\ref{sec:meth_var} (Fig.~\ref{fig:continuous}), where the layer is updated repeatedly on a single encoded state that is never decoded. The ideal update maps the encoded state between the corresponding code spaces without decoding. In our simulations, the update cost includes only the physical error of this additional layer of single-unit gates: the QEC round that carries the layer suffers an additional logical infidelity $\Delta_{\mathrm{gate}}$, with $\Delta_{\mathrm{gate}} \simeq 4e_{\mathrm{gate}}/3$ for the three-qubit code, where $e_{\mathrm{gate}}$ is the gate error (Appendix~\ref{app:update_cost}). Since the layer is then held fixed for the $M$ QEC rounds of a control step, this amounts to a cost per round $\lambda = \Delta_{\mathrm{gate}}/M$, Eq.~\eqref{eq:lambda} below, and $\lambda$ always denotes this per-round quantity. In an experiment, reprogramming the basis-change gates of the stabilizer measurement and the recovery pulses would add to it, so the value used in our simulations is its minimal one.

The layer determines \emph{what} the code is realigned to. A bandit controller
decides \emph{when}. Time is divided into control steps, $f_s$ per unit time,
each containing $M$ QEC rounds during which the layer is held fixed. At the
start of step $k$ the controller takes a binary action
$a_k\in\{\textit{keep},\,\textit{update}\}$: \textit{keep} leaves
$\boldsymbol{\theta}$ unchanged, while \textit{update} realigns the layer with
the current estimate of the error axis. The update layer is applied in one of
the $M$ rounds of the step, so its cost per round is
\begin{equation}
  \lambda=\frac{\Delta_{\rm gate}}{M},\qquad
  \Delta_{\rm gate}\simeq
  \begin{cases}
    \tfrac{4}{3}\,e_{\rm gate} & \text{three-qubit code},\\[2pt]
    1.7\,e_{\rm gate} & \text{three-qutrit code},
  \end{cases}
  \label{eq:lambda}
\end{equation}
where $\Delta_{\rm gate}$ is the additional logical infidelity of a round that
carries one transversal layer of single-unit gates with depolarizing error
$e_{\rm gate}$ ( more details on the derivation are present in Appendix~\ref{app:update_cost}).

The controller is a contextual bandit. It acts on a context $\mathbf{x}_k$ that
summarizes the state of the code at step $k$ through a logistic policy,
\begin{equation}
  \pi(\textit{update}\mid\mathbf{x}_k)=\sigma\!\left(\mathbf{w}\cdot\mathbf{x}_k+b\right),
  \qquad \sigma(z)=\left(1+e^{-z}\right)^{-1},
  \label{eq:policy}
\end{equation}
and receives after each step the reward
\begin{equation}
  r_k=-\frac{L_k+\lambda\,a_k}{p},
  \label{eq:reward}
\end{equation}
where $L_k$ is the logical infidelity per round of the layer in use during the
step and $a_k=1$ for an update. The parameters $(\mathbf{w},b)$ are trained by
the REINFORCE rule on simulated trajectories and then frozen; all results are
obtained on independent trajectories. The context has three components: the
gain expected from realigning, a measure of how reliable that gain is, and the
number of steps elapsed since the last update. Its precise form depends on the
information available to the controller, and we consider two settings; the full
procedure is given in Algorithm~\ref{alg:brave}.

\emph{Exact drift} (Fig.~\ref{fig:brave_main}). The drift parameter $\alpha(t)$ is
known, so the misalignment $\delta_k$ of the deployed layer and the infidelity
$L_k$ are known exactly, and an update sets the layer to the closed-form
aligned one. This setting isolates the timing decision from the estimation of
the drift; the context and the training are specified in
Appendix~\ref{sec:controller_exact}.

\emph{Syndrome only} (Fig.~\ref{fig:comparison}). The drift is not known and is
inferred from the syndrome record, as described below; the gain and $L_k$ are
then posterior expectations (Appendix~\ref{sec:syndrome_impl}).

An update requires the layer that realigns the code with the current error
axis, which in general is found by numerically minimizing the violation of the
Knill--Laflamme conditions over $\boldsymbol{\theta}$. For the drift models used in this work the optimum of this optimization is known analytically, $\hat{u} = R_y(-\varphi)$ with $\varphi = \pi\alpha/2$ for qubits and its qutrit counterpart, and restores the Knill--Laflamme conditions exactly for the instantaneous channel (Appendix~\ref{sec:meth_var}); the numerical optimizer reproduces it to machine precision and is retained as a general-purpose tool for channels without a closed form, while the results of Figs.~\ref{fig:brave_main} and~\ref{fig:comparison} use the closed form directly. It acts on the channel model rather than on the quantum system, so it consumes no quantum samples, and warm-started from the previous solution it converges in about $10^2$ evaluations of the objective. The cost of an update is therefore dominated not by the optimization but by the estimation of the drift, which we now describe.

\emph{Syndrome-only implementation.} The exact-drift setting evaluates the infidelity from the known noise model, which is possible in simulation but not in an experiment, where neither the drift nor the logical state is known. We therefore reformulate the online stage in terms of syndrome-native observables, which are produced by the hardware in every QEC round. The syndrome record is processed in batches of $M$ rounds. For the three-qubit code each qubit is flagged with probability $q = p\cos^2\delta$, with $\delta$ the misalignment introduced above, so that a round shows a non-trivial syndrome with probability
\begin{equation}
    P_{\mathrm{det}}(\delta) = 3q(1-q),
    \label{eq:pdet}
\end{equation}
and the number of detection events in a batch is binomial with this parameter and is a sufficient statistic for $\delta$. The drift is modelled by its functional form, a $\sin^2$ transition with unknown start time and duration, and these two parameters are inferred from the detection counts by a Bayesian particle filter updated after every batch, at no experimental cost beyond the syndromes already measured; the layer is then set in closed form from the posterior. The decision to update uses the posterior expected infidelity in place of the measured fidelity, so that the controller trades the expected gain of realigning the code against the cost $\lambda$ introduced above, and the codeword fidelity is used only to evaluate performance. The controller itself is the contextual bandit of Eq.~\eqref{eq:policy}, which decides from the predicted misalignment of the current layer, the spread of the posterior, and the time elapsed since the last update, and can therefore distinguish a layer that has just been set from one that has become stale.

Two consequences of Eq.~\ref{eq:pdet} determine what this estimation costs. The detection rate depends on the misalignment only at second order, so resolving a misalignment $\delta$ requires of order $1/(12\,p\,\delta^4)$ rounds: the number of syndrome rounds needed per update grows rapidly as the physical error probability decreases, and it is this requirement, rather than the classical optimization, that sets the drift timescales the method can follow. The benchmark of Sec.~\ref{sec:brave_results} rests on four assumptions, which we state explicitly: ideal syndrome extraction, independent rounds, a known physical error probability $p$, and a known functional form of the drift. The first two define the code-capacity setting used throughout this work. The last two are provided by device characterization. We provide a more detailed explanation in Appendix~\ref{sec:syndrome_impl}.

\emph{Context-free controller and regret.} A simpler controller is obtained by dropping the context: a two-armed gradient bandit whose action probabilities depend on no state, which uses the sampled codeword fidelity as its reward and resets its preferences after every update (Algorithm~\ref{alg:brave_cf}). We use it for three purposes only: the illustrative trajectories of Fig.~\ref{fig:Fig_2_1}, the regret analysis, for which it is analytically tractable, and the comparison with a deterministic fidelity trigger. The results of Figs.~\ref{fig:brave_main} and~\ref{fig:comparison} are obtained with the contextual controller of Algorithm~\ref{alg:brave}.

To characterize the efficiency of the retraining mechanism, we consider the
context-free controller of Algorithm~2 as an adaptive scheduler. Each retraining
carries a cost, and the fidelity on which the controller acts is a single noisy
sample, so a rule that retrains whenever the fidelity drops would retrain often
and partly in response to statistical fluctuations. The bandit instead learns from
the observed fidelities whether a drop is informative enough to justify a new
optimization of $\boldsymbol{\theta}$, rather than acting as a fixed deterministic
trigger; its retraining cost relative to such a trigger is analyzed in
Appendix~\ref{sec:bandit_eval}.

We quantify this decision-making process through the notion of regret, defined as the cumulative difference between the expected reward of the optimal keep/retrain policy and the rewards obtained by the learned bandit policy. The regret study is carried out for the simpler context-free controller of Algorithm~\ref{alg:brave_cf}, whose policy dynamics can be written as a differential equation. The regret analysis in Appendix~\ref{sec:regret_bandits} provides an analytically tractable characterization of the dominant scaling trends under non-stationary reward functions induced by time-dependent noise. In particular, it clarifies how the efficiency of the retraining policy depends on the learning rate $\eta$ and the noise variation frequency $\nu$.
Qualitatively, the regret analysis identifies the same operational regime highlighted by the syndrome-only implementation discussion above: BRAVE performs best when the characteristic noise-drift timescale remains sufficiently separated from the retraining dynamics, allowing the controller to track the evolving error basis while maintaining low adaptation overhead.
Appendix~\ref{sec:bandit_eval} further compares the same context-free controller with a deterministic threshold-based retraining rule, focusing on the trade-off between logical fidelity and the number of retraining events isolating the bandit performance.

\section{MARL Discovers Canonical Qubit Codes and Novel Qutrit Code Realizations}
\label{sec:marl_results}
Our triple-agent MARL framework is designed to autonomously
construct the full QEC pipeline -- encoding, syndrome measurement,
and recovery -- without prior knowledge of the target code. We
first validate the framework on qubit systems, where known canonical
codes serve as a benchmark, and then extend it to qutrit systems,
where systematic RL-based discovery has not been previously explored.
The extension to $d$-dimensional qudits is made possible by a
natural generalization of the Clifford gate set, which we introduce
before presenting the qutrit results. Table~\ref{tab:codes_table}
summarizes representative qubit and qutrit codes discovered by our
agents, while additional circuit-level details are collected in the
appendices.

\subsection{Rediscovering Qubit Codes}
On standard qubit benchmarks, the MARL pipeline rediscovers the
three-qubit repetition codes, the four-qubit detection code, the
five-qubit perfect code, and Shor's nine-qubit code. The corresponding circuits are reported in Appendix~\ref{sec:qubit_codes}.
These results validate the framework: the agents recover
well-established codes from scratch, confirming that the sequential
training strategy and the reward structure introduced in
Sec.~\ref{sec:MARL_sec} are sufficient to navigate the space of valid
stabilizer encodings without explicit supervision.

\subsection{Qutrit Code Realizations via Generalized Clifford Gates}
Extending the framework to qudits requires generalizing the gate
set available to the encoder agent. For a $d$-dimensional qudit,
the generalized Pauli operators are
\begin{equation}
\hat{X} = \sum_{n=0}^{d-1}\ketbra{n}{n+1}, \quad
\hat{Z} = \sum_{n=0}^{d-1}\omega^n\ketbra{n}{n},
\label{eq:1quditgate}
\end{equation}
with $\omega = e^{2\pi i/d}$, arithmetic modulo $d$,
$\hat{X}^d = \hat{Z}^d = \hat{I}$, and
$\hat{X}\hat{Z} = \omega\hat{Z}\hat{X}$, recovering the standard
Pauli matrices for $d=2$. For an $n$-qudit system the Pauli group
extends as $\mathscr{P}_n = \otimes^n \mathscr{P}_1$. The
corresponding two-qudit Clifford gates are
\begin{equation}
\begin{split}
\mathrm{CNOT}_d &= \sum_{j,k=0}^{d-1} \ket{j}\bra{j} \otimes
\ket{k}\bra{k+j}, \\
\mathrm{H}_d &= \frac{1}{\sqrt{d}}\sum_{j,k=0}^{d-1}
\omega^{jk} \ketbra{j}{k}, \quad
\mathrm{S}_q = \sum_{j=0}^{d-1}\ket{j}\bra{jq},
\end{split}
\label{eq:2qudit_gates}
\end{equation}
with products modulo $d$. Table~\ref{tab:gates_qutrits_stab} in
Appendix~\ref{app:qutrit_algebra} shows their action on stabilizers
for $d=3$. The gates used for the generation of the qutrit circuits
are a subset of this generalized Clifford group, and the MARL agents
operate over exactly the same action space structure as in the qubit
case, with the gate set replaced by its $d=3$ counterpart.

Applying the learning framework to qutrits, the agents produce a range of
code realizations that, to our knowledge, have not been previously reported through systematic automated discovery. The optimized circuits for the qutrit bit-flip and phase-flip codes are shown in
Fig.~\ref{fig: Qt_Flip_Phase}~(a) and~(b).
Fig.~\ref{fig: Qt_Flip_Phase}~(c) shows the reward maximization
performed by the PPO algorithm for the encoder circuit of the
phase-flip code, where the metric is the negative sum of the
Knill-Laflamme conditions (Appendix~\ref{sec:encoder_method}).
Fig.~\ref{fig: Qt_Flip_Phase}~(d) shows the corresponding
minimization of circuit depth achieved through the depth penalty in
the reward signal. Fig.~\ref{fig: Qt_Flip_Phase}~(e) shows the
discovery of the syndrome measurement circuit using the Mix\&Match
approach, where the green curve represents the first stabilizer
$\hat{S}_0$ and the red curve represents $\hat{S}_1$
(Appendix Table~\ref{tab:syndrome_eras_qutrit}). The full set of
discovered codes, including circuits, code descriptions, and their
use, is reported in Appendix~\ref{sec:qubit_codes} for qubit codes
and Appendix~\ref{sec:qutrit_codes} for qutrit codes.

\begin{table}[!b]
\centering
\begin{tabular}{|cccc|cccc|}
\hline
\multicolumn{4}{|c|}{Qubit} &
  \multicolumn{4}{c|}{Qutrit} \\ \hline
\multicolumn{1}{|c|}{$n$} &
  \multicolumn{1}{c|}{$k$} &
  \multicolumn{1}{c|}{$D$} &
  Error channel &
  \multicolumn{1}{c|}{$n$} &
  \multicolumn{1}{c|}{$k$} &
  \multicolumn{1}{c|}{$D$} &
  Error channel \\ \hline
\multicolumn{1}{|c|}{4} &
  \multicolumn{1}{c|}{2} &
  \multicolumn{1}{c|}{2} &
  (Detecting) $\hat{X}, \hat{Z}, \hat{Y}$ &
  \multicolumn{1}{c|}{3} &
  \multicolumn{1}{c|}{1} &
  \multicolumn{1}{c|}{2} &
  (Detecting) $\hat{X}, \hat{Z}, \hat{Y}, \hat{X}^2, \hat{Z}^2,
  \hat{Y}^2$ \\ \hline
\multicolumn{1}{|c|}{3} &
  \multicolumn{1}{c|}{1} &
  \multicolumn{1}{c|}{3} &
  $\hat{X}$ &
  \multicolumn{1}{c|}{3} &
  \multicolumn{1}{c|}{1} &
  \multicolumn{1}{c|}{3} &
  $\hat{X}, \hat{X}^2$ \\ \hline
\multicolumn{1}{|c|}{3} &
  \multicolumn{1}{c|}{1} &
  \multicolumn{1}{c|}{3} &
  $\hat{Z}$ &
  \multicolumn{1}{c|}{3} &
  \multicolumn{1}{c|}{1} &
  \multicolumn{1}{c|}{3} &
  $\hat{Z}, \hat{Z}^2$ \\ \hline
\multicolumn{1}{|c|}{5} &
  \multicolumn{1}{c|}{1} &
  \multicolumn{1}{c|}{3} &
  $\hat{X}, \hat{Z}, \hat{Y}$ &
  \multicolumn{1}{c|}{9} &
  \multicolumn{1}{c|}{1} &
  \multicolumn{1}{c|}{3} &
  $\hat{X}, \hat{Z}, \hat{Y}, \hat{X}^2, \hat{Z}^2, \hat{Y}^2$
  \\ \hline
\multicolumn{1}{|c|}{9} &
  \multicolumn{1}{c|}{1} &
  \multicolumn{1}{c|}{3} &
  $\hat{X}, \hat{Z}, \hat{Y}$ &
  \multicolumn{1}{c|}{} &
  \multicolumn{1}{c|}{} &
  \multicolumn{1}{c|}{} &
   \\ \hline
\end{tabular}
\caption{Comparison of quantum error-correcting (QEC) codes
discovered via reinforcement learning for qubit and qutrit systems.
The table lists the number of physical units, number of logical
qubits, code distance, and types of detectable error operators.
For qutrits, due to the geometry of SU(3), both single and double
powers of Pauli-like operators are considered.}
\label{tab:codes_table}
\end{table}

\begin{figure*}[!t]
    \centering
    \includegraphics[width=\linewidth]{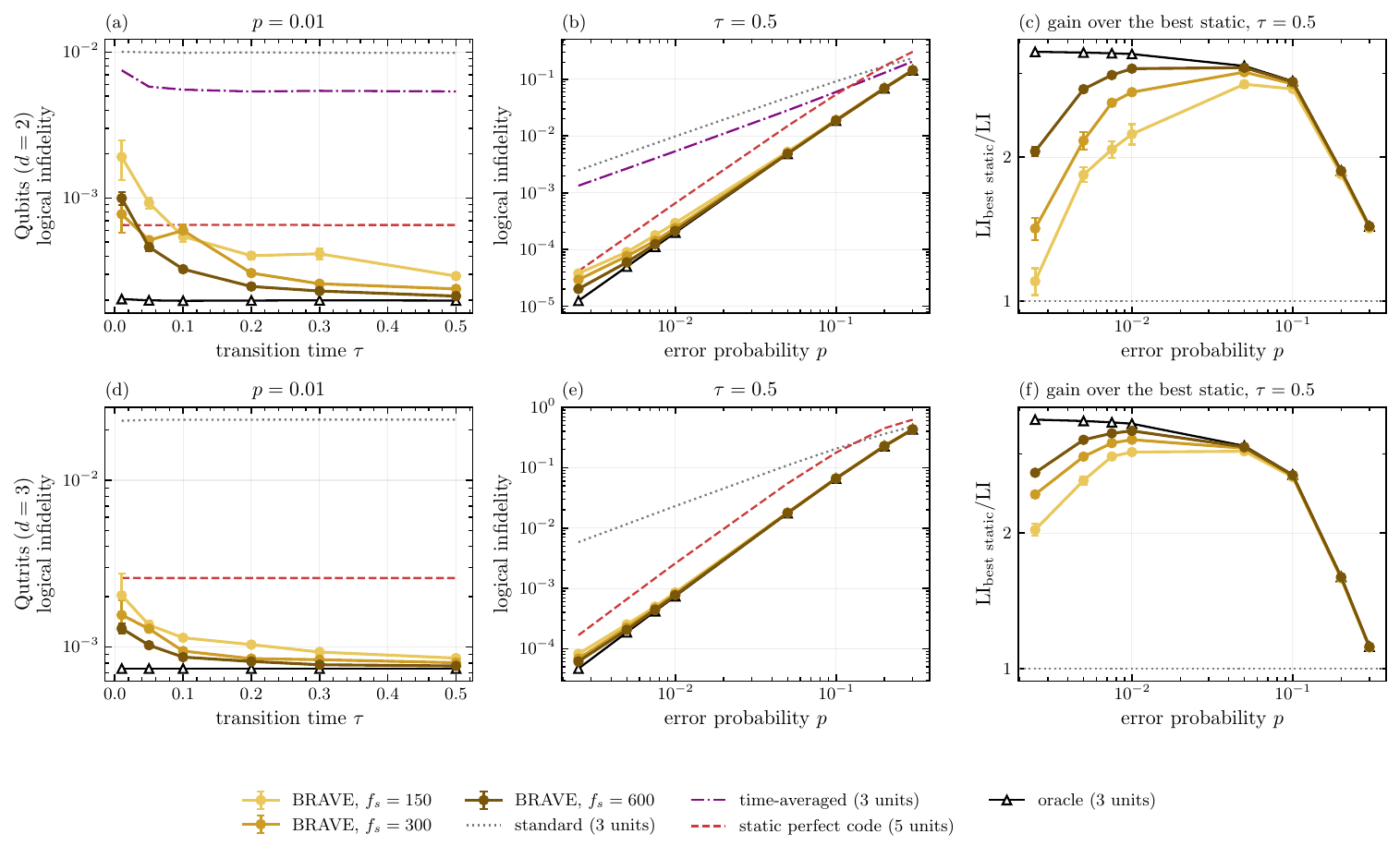}
    \caption{\textbf{BRAVE against static and ideal baselines, for qubits and qutrits.}
    Logical infidelity per QEC round of the three-unit code augmented by the BRAVE variational layer (gold, one curve per sampling rate $f_s$), compared with the same code used as a fixed code (standard), with the code optimized for the time-averaged channel (qubits only), with the static perfect code that corrects any single-unit error using two additional units, and with an oracle that realigns the layer exactly at every step at no cost. Top row: qubits. Bottom row: qutrits. BRAVE uses the contextual controller with exact knowledge of the drift (Appendix~\ref{sec:controller_exact}), with $M = 100$ QEC rounds per control step and a gate error $e_{\mathrm{gate}} = 10^{-3}$ per update.
    \textbf{(a,d)} Dependence on the noise transition time $\tau$ at $p = 10^{-2}$. The advantage of adaptation grows with $\tau$ and with the sampling rate, since both increase the number of updates available per transition, and the curves converge onto the oracle. 
    \textbf{(b,e)} Dependence on the physical error probability $p$ at $\tau = 0.5$.
    \textbf{(c,f)} Gain over the best static code, $\mathrm{LI}_{\mathrm{best\ static}}/\mathrm{LI}$, as a function of $p$ at $\tau = 0.5$, together with the oracle ceiling, which is the best attainable by any adaptive strategy built on the same three-unit code. Values above unity indicate an advantage in using BRAVE adaptation. 
    Standard QEC uses the same MARL-discovered encoding, syndrome-extraction, and recovery circuits, but without the BRAVE variational layer and without online retraining; the time-averaged code uses the best fixed layer for the channel averaged over the drift (Appendix~\ref{sec:time_avg}) and is a qubit-only baseline. The perfect code is the five-qubit code and its five-qutrit analogue. Error bars are standard errors of the mean over 50 Monte Carlo trajectories. Results for all transition times are reported in Appendix~\ref{sec:meth_var}.}
    \label{fig:brave_main}
\end{figure*}

\begin{figure*}[!t]
    \centering
    \includegraphics[width=\linewidth]{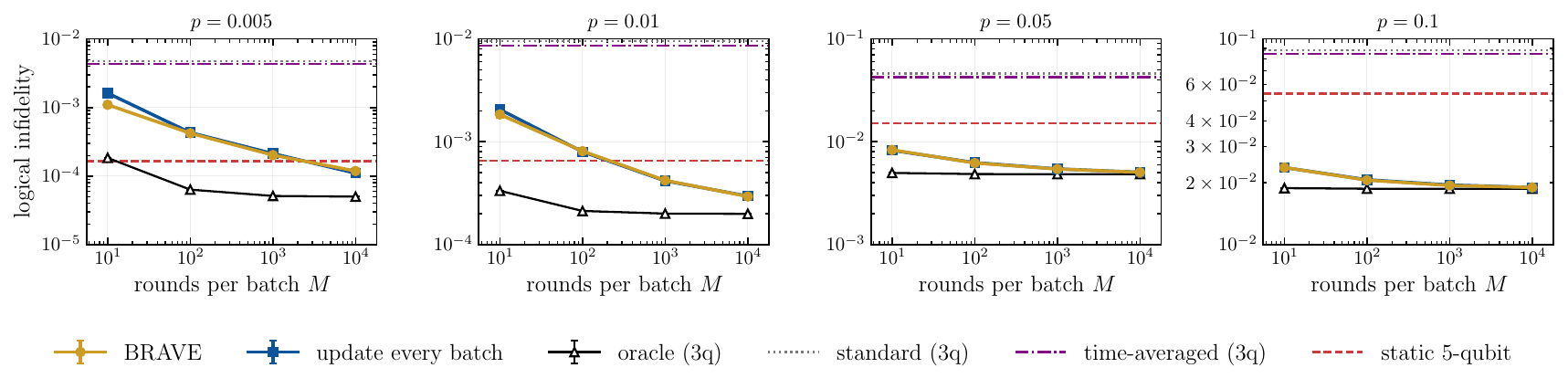}
    \caption{\textbf{Adaptation from syndrome data only.}
    Logical infidelity per QEC round as a function of the number of syndrome rounds per batch $M$, at four physical error probabilities, for $\tau = 0.3$, $f_s = 600$ and a gate error $e_{\mathrm{gate}} = 10^{-3}$ per update. The logical fidelity is used only to evaluate performance and never as feedback. The oracle realigns the layer exactly at every batch and pays the update cost
    $\lambda = 4e_{\mathrm{gate}}/(3M)$ each time.The infidelity decreases with $M$ as the estimate of the misalignment sharpens, and the number of rounds needed to outperform the five-qubit code grows as the error probability decreases, from $M \gtrsim 10$ at $p = 0.1$ and $p = 0.05$ to $M \gtrsim 10^{3}$ at $p = 10^{-2}$ and $M \approx 10^{4}$ at $p = 5\times10^{-3}$, where BRAVE reaches $1.2\times10^{-4}$ against $1.7\times10^{-4}$. This follows from the second-order sensitivity of the detection rate to the misalignment, Eq.~\ref{eq:pdet}. At $M = 10^{4}$ BRAVE lies within $4\%$ of the oracle for $p \geq 0.05$, namely $5.0\times10^{-3}$ against $4.8\times10^{-3}$ at $p = 0.05$ and $1.89\times10^{-2}$ against $1.87\times10^{-2}$ at $p = 0.1$.
    The comparison with the direct scheme isolates the role of the controller: the two reach the same logical infidelity, within $2\%$ over most of the range, while BRAVE updates the layer in only $2$--$50\%$ of the batches, and where data are scarce relative to the update rate it is also more accurate, reaching $1.10\times10^{-3}$ against $1.62\times10^{-3}$ at $p = 5\times10^{-3}$ and $M = 10$ while updating in $6\%$ of the batches.
    Error bars are standard errors of the mean over 50 trajectories. The benchmark assumes ideal syndrome extraction, independent rounds, a known physical error probability $p$ and a known functional form of the drift (Sec.~\ref{sec:brave_method}).}
    \label{fig:comparison}
\end{figure*}

\section{BRAVE Adapts QEC Codes to Time-Dependent Noise}
\label{sec:brave_results}

Fixed QEC codes are designed for stationary noise channels:
once the noise profile drifts, the orthogonality conditions
imposed by the KL conditions can no longer be satisfied, and
the code fails. The BRAVE framework addresses this by
augmenting the MARL-discovered codes with a variational
layer that dynamically adapts to changes in the noise
environment, driven by a bandit controller that decides
when the layer is realigned. Throughout this
section, we use as baseline the corresponding \emph{standard
QEC} scheme --- the same encoder, syndrome measurement, and
recovery circuits discovered by MARL, but used as a fixed
code without the BRAVE variational layer and without online
retraining as the noise drifts in time.

Since a code that has become mismatched is a weak reference, we compare BRAVE with three further baselines. The \emph{time-averaged} code uses the best fixed layer for the channel averaged over the drift, which for the qubit model is $\hat{u} = R_y(-\pi/4)$ with a Hadamard-type correction (Appendix~\ref{sec:time_avg}). It is used as a baseline for qubits only, and for qutrits the static references are the unrotated code and the five-qutrit code. The \emph{static perfect code}, the five-qubit code and its five-qutrit analogue, corrects any single-unit error using two additional units and therefore never suffers the first-order failure described in Sec.~\ref{sec:brave_method}. The \emph{oracle} realigns the layer exactly at every step and bounds what any
adaptive strategy built on the same three-unit code can achieve, separating the
loss due to imperfect tracking from the loss due to the code itself. In the
exact-drift benchmark of Fig.~\ref{fig:brave_main} its updates are free; in the
syndrome-only benchmark of Fig.~\ref{fig:comparison} it realigns at every batch
and is charged the update cost $\lambda$ each time, so that its infidelity is
$\simeq 2p^2+\lambda$ and decreases with $M$.

The variational aspect of this method stems from our
definition of the corrective unitaries. For qubit and qutrit
systems, these generators correspond to the Pauli matrices
and the Gell-Mann matrices respectively. In this framework,
we can decide whether the corrective unitaries are applied
uniformly to all units or individually to each of them. The
latter option offers more fine-grained control but increases
the complexity of the optimization task. In practice,
realizing such unitaries may require deep circuits, which can
introduce additional errors, especially correlated ones that
can potentially impair the effectiveness of QEC. Throughout
this work we therefore follow the former approach, where the
recalibrated angles are obtained, when no closed form is
available, through the Powell
algorithm~\cite{Powell1964}, which has found extensive
use in variational optimization and
control~\cite{Caneva2011Chopped, Sivak2022ModelFree,
9259985}. We should note that this variational approach may
address different types of errors, and pre-knowledge about
the dominant source of error can reduce the computational
resources required by excluding irrelevant SU($d$)
generators. In fact, the selection rules imposed by specific
quantum platforms limit the different types of transitions
and errors appearing on the system.

In the following examples, we consider a realistic noise
model relevant to superconducting transmon
qubits~\cite{PhysRevA.76.042319, RevModPhys.93.025005},
where fluctuations in the external magnetic flux give rise
to a combination of bit-flip and phase-flip errors, as discussed in Sec.~\ref{sec:non_sta_noise}. This hybrid noise
channel naturally emerges from the system Hamiltonian and
allows us to explore the impact of continuously tunable
error combinations on the performance of QEC strategies. We
should note that this framework can also be extended to
other quantum platforms that experience similar noise
profiles. In particular, for the trapped-ion quantum
platform~\cite{Trapped, fossfeig2024progresstrappedionquantumsimulation},
spurious electric fields in the trap can induce decoherence
and heating~\cite{PhysRevA.61.063418, PhysRevA.93.043415,
Sato_2019}. Likewise, for Rydberg
atoms~\cite{RevModPhys.82.2313, Adams_2020}, their large
electric dipole moments lead to both relaxation and
decoherence mechanisms~\cite{RydgerError, PhysRevApplied.22.064021}.

In the two-level description of the system, we approximate
the flux noise with a time-varying noise channel represented
by the following time-dependent Kraus operators using a
unitary interpolation~\cite{Schilling2024}:
\begin{eqnarray}\label{eq:qubit_alpha_channel}
\hat{E}_1(t) = \sqrt{1-p}\,\hat{I}, \quad
\hat{E}_2(t) = \sqrt{p}\,(\hat{Z}\hat{X}^{\dagger})^{\alpha(t)}
\hat{X},
\end{eqnarray}
where $p$ is the error probability and $\alpha(t)$ is a
continuous parameter that governs the transition between the
channel operators $\hat{X} \rightleftarrows \hat{Z}$ --- see
Fig.~\ref{fig:Fig_2_1}~(c) for the detailed
dependence of the noise channel on the external magnetic
flux. For a three-level system this generalizes to
\begin{eqnarray}\nonumber\label{eq:qutrit_alpha_channel}
\hat{E}_1(t) &=& \sqrt{1-p_1-p_2}\,\hat{I}, \quad
\hat{E}_2(t) = \sqrt{p_1}\,(\hat{Z}\hat{X}^{\dagger})^{\alpha(t)}
\hat{X}, \\
\hat{E}_3(t) &=& \sqrt{p_2}\,(\hat{Z}^2\hat{X})^{\alpha(t)}
\hat{X}^2,
\end{eqnarray}
where the additional Kraus operator $\hat{E}_3$ is required
to satisfy the trace-preserving condition, since
$\hat{X}^2 \neq \hat{I}$ for qutrits. Again, $p_1$ and
$p_2$ are the probabilities for different errors to occur
after the measurement phase and $\alpha(t)$ changes with
time. For testing the range of validity of our multi-armed
bandit approach, we consider the bit- and phase-flip QEC
codes already found by our MARL approach  (see
Fig.~\ref{fig: Qt_Flip_Phase}) and on top of them add
the variational layer adapting to the noise profile. This affects
 the system for different probabilities in the
range $(0.0025, 0.3)$, which are related to the total
coherence time $T = T_1 + 2T_2$ via $T = -1/\log(1-p)$,
which can be calculated from the Bloch-Redfield
equations~\cite{Krantz2019Quantum} and gives coherence
times in the range $T = 3$--$400\ \mu$s~\cite{PhysRevA.76.042319,
Hight1}, respectively. We consider
$\alpha(t) = \sin^2(\pi t/2\tau)$ for different transition
times $\tau$, over which the dominant error rotates from
$\hat{X}$ to $\hat{Z}$. The sampling rate $f_s$, the number of control steps per unit time defined in Sec.~\ref{sec:brave_method}, sets how finely the drift is resolved: one transition spans $f_s\tau$ control steps, so that the noise changes smoothly between steps at high $f_s$ and abruptly at low $f_s$.

To illustrate the behavior of the adaptive method, we refer
to the trajectories shown in Fig.~\ref{fig:Fig_2_1}, which are
obtained with the context-free controller of
Algorithm~\ref{alg:brave_cf}, where the sampled fidelity is the
reward. In
subplot~(a), the logical fidelity is plotted as a function of time. Here, the fixed-code standard QEC baseline fails to
detect and correct the expected errors, particularly due to
a drift toward dominant $\hat{Z}$-type errors. 
Standard QEC codes are designed to keep the orthogonality
conditions only for stationary noise channels; therefore,
if these channels evolve in time, the orthogonality
requirements can no longer be satisfied, leading to a code
that does not fulfil the KL conditions for correctable
errors.
In contrast, in subplot~(b), the trajectories obtained using our
BRAVE approach show a more robust
performance, adapting to the noise. Although a slight drop in fidelity occurs over
time, this drop acts as a trigger for the bandit algorithm
to update the variational parameters $\boldsymbol{\theta}$,
enabling adaptation to the evolving noise profile. This
adaptive behaviour results in an observed increase in
recovery fidelity. Further insights into the bandit
algorithm's effect on the variational parameters can be
gained by examining their trajectories in subplot~(d) in
Fig.~\ref{fig:Fig_2_1}. Initially, the parameters are
clustered around the pole of the Bloch sphere
($\boldsymbol{\theta} = 0$), indicating a dominance of
$\hat{X}$-type errors. As time progresses, the trajectory
shifts toward the boundary of the sphere, reflecting the
system's adaptation to the increasing prevalence of
$\hat{Z}$-type errors and modifying the basis through an effective Hadamard gate.

Building on the initial performance analysis, we further
analyze the robustness of the BRAVE approach under a
broader range of conditions. Specifically, we investigate
its sensitivity to varying error probabilities $p$, the
rate of change of the noise profile $\tau$, and the number
$f_s$ of control steps per unit time. We
perform 50 simulations for each configuration. The full set
of parameters that we utilize is listed in
Table~\ref{tab:var_sim_param}.

In these benchmarks the logical infidelity is evaluated on freshly encoded logical states (for qubits, one Haar-random state per control step, shared by its $M$ rounds; for qutrits, the exact average over logical states), and the reported infidelity is the average over these states and over the syndrome outcomes at the instantaneous noise. The sweep therefore measures the logical error per round of the adaptive protocol as the noise drifts, rather than the accumulated infidelity of a single state. The complementary case, in which one logical state is encoded once and
protected continuously while the layer is updated on it, is shown in Appendix~\ref{sec:meth_var} (Fig.~\ref{fig:continuous}).

\begin{table}[!b]
\centering
\scriptsize
\begin{tabular}{|c|cccccccc|}
\hline
Parameters & \multicolumn{8}{c|}{Values} \\ \hline
$p$ &
  \multicolumn{1}{c|}{0.0025} &
  \multicolumn{1}{c|}{0.005} &
  \multicolumn{1}{c|}{0.0075} &
  \multicolumn{1}{c|}{0.01} &
  \multicolumn{1}{c|}{0.05} &
  \multicolumn{1}{c|}{0.1} &
  \multicolumn{1}{c|}{0.2} &
  0.3 \\ \hline
$\tau$ &
  \multicolumn{1}{c|}{0.01} &
  \multicolumn{1}{c|}{0.05} &
  \multicolumn{1}{c|}{0.1} &
  \multicolumn{1}{c|}{0.2} &
  \multicolumn{1}{c|}{0.3} &
  \multicolumn{1}{c|}{0.5} &
  \multicolumn{2}{c|}{} \\ \hline
$f_s$ &
  \multicolumn{1}{c|}{150} &
  \multicolumn{1}{c|}{300} &
  \multicolumn{1}{c|}{600} &
  \multicolumn{5}{c|}{} \\ \hline
\end{tabular}
\caption{Summary of the simulation parameters used to
evaluate the performance of the proposed adaptive framework
(BRAVE). We conducted 50 simulations for each configuration
across varying error probabilities $p$, transition times
$\tau$, over which the dominant error rotates from $\hat{X}$ to
$\hat{Z}$, and sampling
rates $f_s$, which are the number of control steps per unit time. Both qubit and qutrit systems were tested for all three sampling rates}.
\label{tab:var_sim_param}
\end{table}

We begin by analyzing the qubit case under varying sampling rates. A
key challenge that can affect our bandit-based approach is the lack of timely
information; under such conditions, the agent may fail to decide whether to
retrain the variational layer, resulting in a loss of fidelity. As shown in
Fig.~\ref{fig:brave_main}~(a), all our adaptive methods achieve lower Logical
Infidelities (LI) compared to the fixed-code standard QEC baseline across all
values of $\tau$, and increasing the sampling rate reduces the error rate
further: at $f_s = 600$ and $p = 10^{-2}$ the infidelity falls from
$1.0\times10^{-3}$ at $\tau = 0.01$ to $2.1\times10^{-4}$ at $\tau = 0.5$, while
the standard code remains at $9.9\times10^{-3}$ throughout. At the lowest
sampling rates the curves for $f_s = 150$ and $f_s = 300$ cross within their
error bars for $\tau \lesssim 0.1$, where the number of batches per transition
is smallest. The most favorable scenario arises when the noise frequency is slow and the sampling rate is high, where the improvement over the unadapted code reaches a factor of $46$ at $p = 10^{-2}$ and of $123$ at $p = 2.5\times10^{-3}$.

We then perform the same analysis for the qutrit case, in
Fig.~\ref{fig:brave_main}~(d). The same trend emerges, with the adaptive method
reaching the oracle to within $10\%$ from $\tau = 0.2$ at $f_s = 600$ and within
$16\%$ by $\tau = 0.5$ even at $f_s = 150$, and the advantage over the fixed
baseline again largest for slow transitions and high sampling rates. The
convergence is smoother than for qubits, the ratio to the oracle falling
monotonically from $1.7$ to $1.0$ across the range of $\tau$ at $f_s = 600$. The
reason is statistical: for qutrits the infidelity of each round is evaluated
exactly as an average over the logical states, whereas for qubits it is
estimated from one Haar-random state per step, so that the qutrit curves carry
a smaller sampling error. In both systems the update installs the exactly
aligned layer (Appendix~\ref{sec:meth_var}), so that the residual error comes
from tracking alone.

Panels~(b), (c), (e) and~(f) of Fig.~\ref{fig:brave_main} quantify the
robustness of the adaptive strategy across different physical error
probabilities, at $\tau = 0.5$, and against the stronger baselines introduced
above. For qubits, BRAVE at $f_s = 600$ lowers the logical infidelity by up to
$3.1\times$ relative to the best static code, which is the five-qubit code up to
$p = 0.1$ and the time-averaged code beyond it, while using two fewer physical
qubits; the gain exceeds $2.8\times$ over the whole intermediate range
$p \sim 7.5\times10^{-3}$--$10^{-1}$. For qutrits the advantage is slightly
larger, up to $3.4\times$ at $p = 10^{-2}$, measured against the five-qutrit
perfect code up to $p = 0.1$ and the unrotated code beyond it. At
$\tau = 0.5$ the adaptive code is preferable to every static alternative at
every error probability and sampling rate considered: the smallest gain is
$1.10\times$ for qubits, at $f_s = 150$ and $p = 2.5\times10^{-3}$, and
$1.11\times$ for qutrits, at $f_s = 150$ and $p = 0.3$. The residual
misalignment left by a finite update rate still costs a logical error of first
order in $p$, whereas the perfect code fails only at second order, and this is
what narrows the advantage at low noise. For faster drifts it reverses the
ordering there: over the full sweep the five-unit code is preferable to BRAVE in
29 of the 144 configurations for qubits, all with $p \leq 10^{-2}$ and
$\tau \leq 0.3$, and in 3 of 144 for qutrits, all with $p = 2.5\times10^{-3}$
and $\tau \leq 0.05$ (Appendix~\ref{sec:meth_var}).
Relative to the oracle that realigns the code exactly at every step, BRAVE at
$f_s = 600$ lies within $1\%$ for $p \geq 0.1$ and within $7\%$ at $p = 10^{-2}$
for qubits, and within $1\%$ for $p \geq 0.05$ and $4\%$ at $p = 10^{-2}$ for
qutrits, departing from it only in the low-noise regime, where it is short by
$1.6\times$ and $1.3\times$ respectively at $p = 2.5\times10^{-3}$, and it does
so while applying the update layer in about a quarter to a third of the
control steps. At the
highest error probabilities the gain falls back towards unity in both systems,
to $1.4\times$ for qubits and $1.1\times$ for qutrits at $p = 0.3$, since
multi-unit errors then dominate and no realignment of the code helps. The same
data for all transition times are reported in Appendix~\ref{sec:meth_var}.

The results above assume that the drift is known. Figure~\ref{fig:comparison} shows the same comparison when the controller observes only syndromes, with the drift inferred by the particle filter of Sec.~\ref{sec:brave_method} from batches of $M$ QEC rounds, and with the logical fidelity used only to evaluate performance. The infidelity decreases monotonically with $M$, as the estimate of the misalignment sharpens, and the number of rounds required to outperform the static codes depends strongly on the error probability: BRAVE overtakes the five-qubit code from $M \gtrsim 10$ at $p = 0.05$ and $p = 0.1$, from $M \gtrsim 10^{3}$ at $p = 10^{-2}$, and only at $M \approx 10^{4}$ at $p = 5\times10^{-3}$, where it reaches $1.2\times10^{-4}$ against $1.7\times10^{-4}$. This ordering follows from Eq.~\ref{eq:pdet}: the detection rate is sensitive to the misalignment only at second order, so the number of rounds needed to resolve it grows as $p$ decreases. At $M = 10^{4}$ the syndrome-only controller lies within $4\%$ of the oracle for $p \geq 0.05$ and within a factor of $1.5$ at $p = 10^{-2}$.

The same figure isolates what the controller contributes, by comparing it with the direct scheme that updates the layer after every batch. The two reach the same logical infidelity, within $2\%$ over most of the range, while the controller updates in only $2$--$50\%$ of the batches; and where data are scarce relative to the update rate it is also more accurate, reaching $1.10\times10^{-3}$ against $1.62\times10^{-3}$ at $p = 5\times10^{-3}$ and $M = 10$, a $32\%$ lower infidelity while updating in $6\%$ of the batches. The cost $\lambda$ used here includes only the gate error of the update layer. In an experiment, reconfiguring the code also requires reprogramming the basis-change gates of the stabilizer measurement and the recovery pulses, and may involve control latency or idle time; these overheads are device-specific and are not modelled. Additional reconfiguration overheads could further motivate selective updating, while their effect on tracking performance would require a hardware-specific assessment.

To disentangle the role of the bandit meta-controller from that of adaptive variational tuning alone,  Appendix~Sec.~\ref{sec:bandit_eval} compares the context-free controller of Algorithm~\ref{alg:brave_cf} with a deterministic threshold-based retraining policy built on the same variational layer. That control can maintain somewhat higher fidelity in some regimes, but only at the cost of a rapidly increasing number of retraining events as the noise strength grows. The role of the bandit is therefore not to maximize fidelity at all costs, but to provide a more favorable performance--cost tradeoff by limiting retraining overhead while preserving high logical
performance. The result of our analysis shows that our bandit algorithm is more efficient than the deterministic policy in the high-noise regime, where the deterministic trigger is no longer economical.

\section{Discussion and Future Directions}
\label{sec:discussion}
\subsection{MARL}
\label{sec:MARL_discussion}
\paragraph{Beyond stabilizer codes}
\label{sec:beyond_stabilizer}
The framework as presented searches the Clifford group and therefore discovers
stabilizer codes, but nothing in the formulation requires this: the agents are
rewarded by the Knill--Laflamme conditions and by the logical fidelity, neither
of which presupposes a stabilizer structure. The restriction enters only through
the action spaces, the Clifford gate set for the encoder and the
controlled-Pauli primitive for the syndrome agent. Enlarging the encoder's
action space beyond the Clifford group, and replacing the stabilizer-based
acceptance conditions of the syndrome agent with criteria that identify
correctable error subspaces without revealing logical information, would let the
same pipeline explore codes outside the stabilizer formalism, where
correctability can hold only approximately and no compact algebraic description
of the code space exists. This is the direction we are currently pursuing, and
it is also the natural setting for the non-unitary channels of
Sec.~\ref{sec:brave_method}, for which no small code satisfies the conditions exactly.

\paragraph{Towards larger stabilizer codes}
\label{sec:scalability}
Scalability of the discovery stage is limited by three distinct factors, and
separating them makes clear which extension addresses which, and at what cost.

The first is the cost of evaluating the rewards, which we compute on explicit
states at a cost exponential in the number of units. Within the stabilizer
formalism, the binary symplectic representation reduces every correctability
test to linear algebra over $\mathbb{F}_d$, and a Clifford gate acts on the
check matrix as a symplectic matrix, so the encoder can track its effect on the
stabilizers without constructing a state. This route is unavailable outside the
formalism: a code space that is not the joint eigenspace of an abelian Pauli
subgroup, or a non-Pauli error set, must still be evaluated explicitly.

The second is the size of the search space, which a cheaper reward does not
change, since the number of Clifford circuits grows super-exponentially with
$n$. Structuring the search is what addresses this, and restricting it to CSS
codes does so for both agents: the encoder can be reduced to Hadamards on a
fixed subset of ancillas followed by a CNOT network, or can search directly over
check matrices with synthesis performed afterwards, while the syndrome agent
constructs purely $X$- and $Z$-type generators independently, with
$H_XH_Z^{T} = 0$ guaranteeing commutation by construction. The cost is that this
gives up not only non-stabilizer codes but also the non-CSS stabilizer codes,
the five-qubit code among them.

The third is the informativeness of the reward. The encoder reward is already
graded, through the normalized count of violated conditions, the
depth-dependent weighting and the terminal bonus. The syndrome reward is the
restrictive one, since a candidate is accepted only when all validity
conditions hold; a simpler formulation based on syndrome uniqueness, described
in Appendix~\ref{sec:elementary_approaches}, became trapped in suboptimal
solutions, so relaxing the objective is not sufficient on its own. A more
promising route keeps the full conditions as the terminal objective and adds
potential-based shaping, $r'_t = r_t + \gamma\Phi(s_{t+1}) - \Phi(s_t)$, which
leaves the optimal policies unchanged; taking $\Phi$ to be the fraction of
validity conditions already met rewards a gate that makes a candidate stabilizer
commute with one more group element immediately rather than at the end of the
episode. For the recovery agent, averaging the fidelity over a representative
ensemble of logical states, or using the entanglement fidelity, would likewise
sharpen a reward whose attainable values are nearly discrete for restricted
error models.

None of these strategies is benchmarked here, and we do not claim they eliminate
local optima or by themselves resolve the scalability limitations.

\paragraph{Hardware-aware discovery}
\label{sec:hardware}
The framework is formulated at the circuit level, which makes it natural to
constrain the search by the properties of a given device, although we have not
done so in the benchmarks reported here. Connectivity limits and native gate
sets enter as masks on the action space, since each action already specifies a
gate type together with the units it acts on; both shrink the search space, so
the cost is borne by the quality of the reachable codes rather than by the
training procedure. Gate errors enter the reward instead. Assuming independent
failures, the probability that a circuit runs without error is
$\prod_i(1-p_i)$, so to first order in the $p_i$ the natural penalty is additive
over the gates,
\begin{equation}
    \mathcal{H} = -\eta\sum_{i} p_i,
    \label{eq:hazard}
\end{equation}
which can be added to the encoder reward of Eq.~\ref{eq:enc_rew}. This is more
informative than the depth penalty already present, which treats all gates
alike, even though on superconducting hardware a two-unit gate typically carries
an error one to two orders of magnitude larger than a single-unit one; the
weight $\eta$ must be chosen deliberately, since the correctability
contributions are of order unity while $\mathcal{H}$ is of order $10^{-2}$.

Hardware constraints change the syndrome extraction stage the most. Under
unconstrained access to gates and ancillas a valid extraction circuit follows
directly from the encoder; under connectivity and ancilla limits it no longer
does, since a weight-$w$ generator requires an ancilla coupled to $w$ data units
or a sequence of swaps, and a single ancilla fault can spread to several data
units and turn a correctable error into an uncorrectable one. None of this is
visible to the present reward. A hardware-aware version would keep the validity
conditions $C_1$--$C_4$ as the terminal objective and add the resources
consumed,
\begin{equation}
    r^{(\mathrm{syn})}_{t} = \prod_{i=1}^{4}c_i - \eta\sum_{g} p_g
    - \mu\sum_{\ell}\max\!\big(0,\, w_\ell - t\big),
    \label{eq:syn_hw_reward}
\end{equation}
where the second term is the hazard of Eq.~\ref{eq:hazard} summed over the
gates of the extraction circuit and the third penalizes fault propagation:
$w_\ell$ is the number of data units to which a single fault at position $\ell$
spreads, and $t = \lfloor (d-1)/2 \rfloor$, so that only schedules turning one
fault into more than $t$ errors are penalized. Evaluating $w_\ell$ requires no
simulation, since it follows from propagating a single Pauli fault through the
remaining Clifford gates. This reward is meaningful only together with a
circuit-level noise model, since under the code-capacity model used here the
cost of a deeper circuit does not appear in the logical error rate at all.

\paragraph{Variable code parameters}
\label{sec:variable_nk}
The numbers of physical and logical units, $n$ and $k$, are fixed before the
encoder is trained. This follows from how the reward is evaluated: the logical
states are built as
$\ket{i}_{\mathrm{log}} = \hat{U}_t\ket{i}\bigotimes_j\ket{0}_j$, so the
partition of the input registers into logical units and ancillas, which is
precisely $k$, must be fixed before a single reward can be computed, and the
encoder is trained before any stabilizer is available. The same construction
makes the reward exponential in $k$, since it enumerates $d^{k}$ codewords.

Two routes make $k$ variable. The first leaves the circuit search untouched and
promotes the partition to a single decision taken at the start of an episode, so
the reward remains well defined within an episode while the rate varies across
episodes; rewarding the rate on success, for instance through
$r_{\mathrm{success}} = k/n$, makes the trade-off explicit, and $k = n$ is
excluded by the correctability conditions alone. The second searches directly
over check matrices, where correctability is algebraic, no codeword enumeration
is required, and $k = n - \operatorname{rank}(H)$ emerges from the discovered
generators; this is the same reformulation discussed in
the paragraph \emph{Towards larger stabilizer codes}, with the same restriction to stabilizer codes. In
either case the reward becomes multi-objective, and reporting the Pareto front
in the $(k/n, d)$ plane is more informative than optimizing a fixed combination.
The number of physical units remains an outer-loop variable, set either by a
meta-agent or by training independent agents in parallel over a set of $(n,k)$
pairs. Low-density constructions require more than variable parameters, since
sparsity is a property of the generators: it must be imposed on the syndrome
stage by penalizing generator weight and unit degree, which acts in the same
direction as the hardware-aware terms treated in the previous section.

\subsection{BRAVE}
\emph{The timescale of the adaptation}
\label{sec:discussion_BRAVE}
The rate at which BRAVE can follow a drift is set by the accumulation of syndrome statistics, not by the classical computation. Following one transition takes $f_s\tau$ batches of $M$ rounds, that is $Mf_s\tau$ rounds in total: at $p = 10^{-2}$, where $M \gtrsim 10^{3}$ rounds per batch are required, and with the $f_s\tau = 180$ batches per transition used in Sec.~\ref{sec:brave_results}, this amounts to about $2\times10^{5}$ rounds, or $\sim0.2$\,s for a QEC cycle of $1\,\mu$s; at $p = 5\times10^{-3}$ the same estimate gives about $2$\,s, since fewer errors per round means fewer samples. For the main benchmarks, the closed-form update avoids repeated numerical optimization. In an experimental implementation, the combined latency of estimation, frame computation, and physical reconfiguration must fit within the available update interval.

This fixes the regime in which the method is useful. BRAVE addresses drifts evolving on timescales of order $0.1$\,s and longer, such as slow flux drifts and the fluctuations of two-level-system defects, while faster components remain part of the instantaneous channel and are corrected by the code itself; it is in this sense that the adaptation is real-time, proceeding during operation on the timescale of the drift envelope rather than within a single QEC round. The requirement also becomes milder as the physical error rate improves in one respect and harsher in another: better hardware means fewer detection events per round and therefore more rounds per update, but also a faster cycle and typically a slower drift, so the relevant comparison is always between $Mf_s\tau$ rounds and the time over which the error axis moves appreciably. A useful design rule is to choose $M \sim 1/(12\,p\,\delta^4)$ for the largest misalignment $\delta$ one is willing to tolerate, and then to verify that $M$ rounds are shorter than that time.

\emph{Relation to dynamical and Floquet codes}
\label{sec:floquet}
Dynamical and Floquet codes~\cite{Hastings2021dynamically} offer a related
perspective, since there the protected logical information is defined through a
sequence of measurements and the instantaneous stabilizer group evolves during
operation. The time dependence, however, serves a different purpose. In standard
Floquet constructions the measurement sequence follows a prescribed periodic
schedule, fixed in advance and independent of the noise the device actually
experiences; it is part of the definition of the code rather than a response to
the channel. BRAVE addresses the complementary question of when and how to
realign the code frame with an error basis that is itself drifting, with the
update triggered by inferred noise drift rather than by the position in a cycle.
The two mechanisms are therefore not alternatives, and combining them could be a
natural direction: a Floquet code whose measurement schedule were retuned in
response to the observed channel would inherit the fault tolerance of the
dynamical construction together with the adaptivity demonstrated here. The
difficulty is that an update cannot be applied to a single round in isolation,
since the logical operators of a Floquet code are defined by the measurement
history: changing the schedule changes which operators are instantaneously
protected, and the decoder's matching graph with them, so consistency across the
update would have to be established rather than assumed.

\section{Conclusions}
\label{sec:conclusions}
We have introduced a framework for quantum error correction that integrates automated code discovery with self-adaptive correction under time-varying noise. By combining autonomous discovery with dynamical adaptation, our approach overcomes a central limitation of conventional QEC, namely the reliance on static noise assumptions. The first tier employs a MARL architecture to autonomously discover complete QEC
cycles rewarded by task-specific objectives, without imposing a predefined
code family. The codes it can reach are set by its action spaces, here the
Clifford gate set and the controlled-Pauli measurement primitive. In the present implementation, this offline discovery stage is trained for a chosen family of target errors, namely stationary Pauli-like bit-flip/phase-flip channels and their qutrit generalizations. Indeed by extending the Clifford gate set to arbitrary $d$-dimensional systems, the framework naturally generalizes beyond qubits and enables systematic exploration of higher-dimensional encoding strategies.

We validated this approach by demonstrating that MARL recovers canonical qubit codes and extends them to three-level systems, where for the benchmark cases considered here it reconstructs known qutrit code families from scratch, including explicit circuits for encoding, syndrome measurement, and recovery, capable of correcting both single-type and hybrid noise processes. In this sense, the contribution is the first-principles discovery of the full operational cycle by the agent, rather than a claim of new inequivalent qutrit code spaces.

To address the challenge of non-stationary noise in realistic quantum hardware, we introduced BRAVE, an adaptive variational layer that dynamically reshapes the effective error basis and  restores approximate Knill--Laflamme conditions during operation. In this way, QEC is promoted from a static encoding strategy to a closed-loop control process, in which encoding and recovery operations are recalibrated in response to environmental drift. The use of a bandit-based meta-optimization further ensures that this adaptation is achieved with minimal retraining overhead.

Quantitatively, our framework demonstrates a substantial enhancement in
robustness, maintaining high logical fidelity in regimes where conventional QEC
becomes ineffective. Measured against the strongest static alternative at each
error probability, including the five-unit perfect code, which uses two more
physical units than the adaptive three-unit code, BRAVE lowers the logical
infidelity by up to $3.1\times$ for qubits and $3.4\times$ for qutrits, and for
the slowest drift considered, $\tau = 0.5$, it
is preferable to every static code considered at every error probability and
sampling rate of the sweep. For faster drifts at low error probability the
five-unit perfect code remains preferable. The advantage is mostly concentrated in the intermediate
regime $p \sim 10^{-2}$--$10^{-1}$, where a drifting error basis is the limiting
factor rather than the raw error rate; at low noise it narrows, since the
residual misalignment left by a finite update rate costs a logical error of
first order in $p$ whereas a perfect code fails only at second order, and at
$p \gtrsim 0.2$ it falls back towards unity because multi-unit errors dominate
and no realignment of the code helps. Once the noise is strong enough that
updates are frequent relative to the drift, BRAVE agrees with an oracle that
realigns the code exactly at every step to within $1\%$, while reconfiguring the
code only in a fraction of control steps. Crucially, the adaptation survives when the
drift is not known in advance: a controller that observes only the syndrome
record, with the misalignment inferred from detection counts at no experimental
cost beyond the measurements already performed, stays within $4\%$ of the oracle
for $p \geq 0.05$ at $M = 10^{4}$ rounds per batch, where it reconfigures the
code in about half of the batches; over the range of $M$ considered it does so
in $2$--$50\%$ of them.
 This last result assumes ideal syndrome extraction and independent rounds, as in the code-capacity setting used throughout, together with a known physical error probability and a known functional form of the drift, both of which a device characterization provides. The number of syndrome rounds it requires grows rapidly as $p$ decreases, since the detection rate responds to the misalignment only at second order. These results highlight the effectiveness of adaptive QEC
in realistic, time-dependent noise environments.

More broadly, our results establish a new paradigm in which quantum error correction is no longer a fixed construct tailored to a specific noise model, but a dynamically optimized process capable of tracking and compensating for slowly evolving error channels during operation. This perspective bridges concepts from quantum control and machine learning with QEC, suggesting a route toward self-calibrating error-correction protocols. This opens a pathway toward practical, hardware-efficient QEC schemes that reduce calibration requirements and improve resilience in near-term quantum devices. The discovery stage is at present limited to small codes, up to nine units,
because its rewards are evaluated on explicit states at a cost exponential in the
number of units, while the number of Clifford circuits grows super-exponentially.
Section~\ref{sec:MARL_discussion} identifies a route for each limitation: a check-matrix evaluation of
correctability within the stabilizer formalism, a restriction of the search to
CSS codes, and potential-based reward shaping for the syndrome agent. Future work will explore integration with fault-tolerant architectures, scalability to larger systems, and experimental implementation in real-time control loops. In such settings, adaptive QEC and machine learning may play a central role in stabilizing quantum computation under realistic operating conditions.

\begin{acknowledgments}
We thank Jos\'e Jesus, Dimitrios Georgiadis, Florian Marquardt, Matteo Puviani for useful discussions. This work was supported by the German Federal Ministry of Education and Research (BMBF) through the QSolid project (Grant No. 13N16149), the German Research Foundation (DFG) under Germany’s Excellence Strategy—Cluster of Excellence Matter and Light for Quantum Computing ML4Q2
 (EXC 2004/2 – 390534769), and the Jülich Supercomputing Center (JSC). Additional funding was provided by the Horizon Europe program via projects QCFD (101080085, HORIZON-CL4-2021-DIGITAL-EMERGING-02-10) and OpenSuperQPlus100 (101113946, HORIZON-CL4-2022-QUANTUM-01-SGA). This research utilized resources of the Oak Ridge Leadership Computing Facility at Oak Ridge National Laboratory, supported by the U.S. Department of Energy, Office of Science (Contract No. DE-AC05-00OR22725). Simulations were performed in \textsc{Python} using \textsc{Qiskit} \cite{qiskit2024}, \textsc{PyTorch} \cite{paszke2019pytorchimperativestylehighperformance}, and \textsc{Stable-Baselines3} \cite{stable-baselines3}.

\section{Author contributions}
M.G., F.A.C.L., and F.M. conceived the research and planned the work at all stages.
M.G., F.P. and M.S. performed the numerical simulations and the theoretical analysis.
F.M., T.C. supervised the work.
All the authors wrote and revised the manuscript.

\section{Data and Code Availability}
All the resources needed to reproduce the plots in this paper are available on GitHub~\cite{Guatto2025RealTimeAdaptiveQEC}.

\end{acknowledgments}

\clearpage
\appendix

\renewcommand{\thefigure}{A\arabic{figure}}
\setcounter{figure}{0} 

\section{Additional discovered QEC codes}

\subsection{Qubit codes}
\label{sec:qubit_codes}

\textbf{Three-qubits codes: }In this section, we analyze the results for three-qubit quantum error-correcting codes, considering both Pauli $\hat{X}$ (bit-flip) and $\hat{Z}$ (phase-flip) errors. According to the quantum Hamming bound~\cite{aly2007notequantumhammingbound}, it is possible to construct a three-qubit code capable of correcting single-qubit Pauli $\hat{X}$ or $\hat{Z}$ errors. We focus on a specific error model known as the Pauli error channel, where each noise process is considered to be independent. In this case, the action of each noise channel is given by :
\begin{equation}
    \mathcal{N}_{j}(\rho) = (1-p_{j})\rho + p \hat{O}_{j} \rho \hat{O}_{j},
\end{equation}
where $ p_{j} $ is the occurrence probability of an $\hat{O}_{j}$ Pauli error. For a three-qubit code, we aim to correct $ \hat{X} $ or $ \hat{Z} $ errors on any of the qubits in the codeword. The overall error channel acting on the three-qubit state $ \rho $ is represented as the tensor product of individual error channels acting on each qubit. Therefore, the final error channels for the three-qubit code are given by:

\begin{equation}
    \mathcal{N}_\text{total} = \bigotimes_{j=0}^2 \mathcal{N}_{j}(\rho) \:,
\end{equation}

\textbf{Bit-Flip and Phase Codes: }To correct either $\hat{X}$ or $\hat{Z}$ errors, we constructed an encoding circuit using the first agent of our RL chain by direct satisfying the Knill-Laflamme conditions 
during the training. The resulting encoder maps the qubit states onto the following codewords:
\begin{table}[!b]
\centering
\begin{tabular}{|c|c|c||c|c|c|}
\hline
\multicolumn{3}{|c||}{$\mathcal{E}_\text{X final}$} & \multicolumn{3}{c|}{$\mathcal{N}_\text{Z final}$} \\ \hline
Error                      & $S_1$ & $S_2$ & Error                      & $S_1$ & $S_2$ \\ \hline
$\hat{I}\ket{\psi_\text{log}}$   & 0     & 0     & $\hat{I}\ket{\psi_\text{log}}$   & 0     & 0     \\ \hline
$\hat{X}_1\ket{\psi_\text{log}}$ & 0     & 1     & $\hat{Z}_1\ket{\psi_\text{log}}$ & 0     & 1     \\ \hline
$\hat{X}_2\ket{\psi_\text{log}}$ & 1     & 0     & $\hat{Z}_2\ket{\psi_\text{log}}$ & 1     & 0     \\ \hline
$\hat{X}_3\ket{\psi_\text{log}}$ & 1     & 1     & $\hat{Z}_3\ket{\psi_\text{log}}$ & 1     & 1     \\ \hline
\end{tabular}

\caption{Syndrome tables for the $\mathcal{E}_\text{X final}$ and $\mathcal{N}_\text{Z final}$ channels.}
\label{tab:repcode}
\end{table}
\begin{eqnarray*}
    &&\ket{0_L} = \ket{000}, \quad \ket{1_L} = \ket{111},\\
    &&\ket{0_L} = \ket{+++}, \quad \ket{1_L} = \ket{---}. 
\end{eqnarray*}
This corresponds to a three-qubit repetition code for each Pauli channel. After constructing the encoding circuit, we employed a second RL agent, called the syndrome measurement agent, to determine the corresponding syndrome measurement circuit. Since we are encoding $ n = 3 $ qubits into $ k = 1 $ logical qubit, we expect to measure $ n - k = 2 $ stabilizers. Each stabilizer consists of a tensor product of $n$ Pauli operators, meaning that $ (n - k) n = 6 $ Pauli operators should be measured. Thus, the expected maximum length of the syndrome measurement circuit is equal to six. From the syndrome measurement, we obtain the stabilizers

\begin{eqnarray*}
    &&S_1 = \hat{Z}_3 \hat{Z}_2, \quad S_2 = \hat{Z}_3 \hat{Z}_1,\\
    &&S_1 = \hat{X}_3 \hat{X}_2, \quad S_2 = \hat{X}_3 \hat{X}_1.
\end{eqnarray*}

\begin{figure}[!t]
    \centering
    \includegraphics[width=\columnwidth]{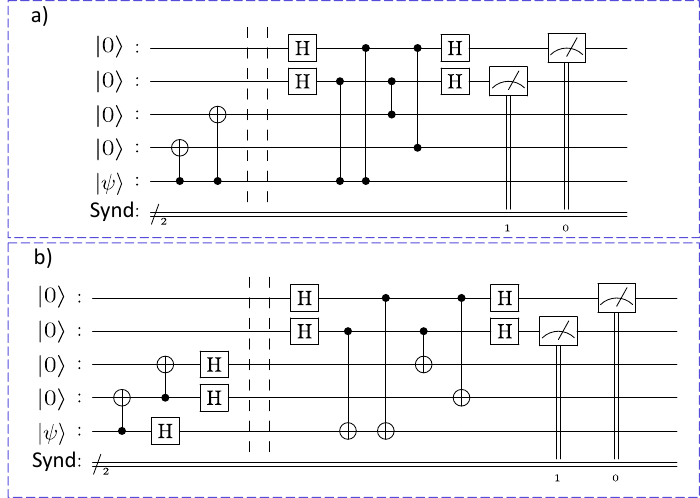}
    \caption{a) Bit flip code discovered by our agent. (b) Phase flip code discovered by our agent}
    \label{fig:circ_3xz}
\end{figure}

We summarize the outcome of the syndrome measurement in Table~\ref{tab:repcode} and Fig.~\ref{fig:circ_3xz} provides the explicit circuit structure for the repetition code. Finally, we run the recovery agent for all syndromes confirmed that the recovery operators match the errors as shown in the table. This verifies the correctness of the recovery procedure for the Pauli-$Z$ channel as well.\par
 
\textbf{Four-qubits code: }Now, let us increase the complexity of the circuit discovery by considering a harder QECC case, the four-qubit code where we encode two logical qubits but we can only detect errors. Here, every qubit is susceptible to being affected by all the Pauli errors described by the channel
\begin{equation*}
    \mathcal{N}(\rho) = \sum_{k=0}p_{k}\hat{O}_{k}^{\dag}\rho\hat{O}_{k}.
    \label{eq: Depo}
\end{equation*}
Where $\hat{O}_{k}$ is a different Pauli matrix including the identity and $p_k$ is their occurrence probability. Note that this four-qubit code aims to correct single-qubit Pauli erasure error --assuming that all other qubits remain unaffected-- or detect if an error corrupted the logical qubits. 
\begin{figure*}[ht]
    \centering
    \includegraphics[width=\textwidth]{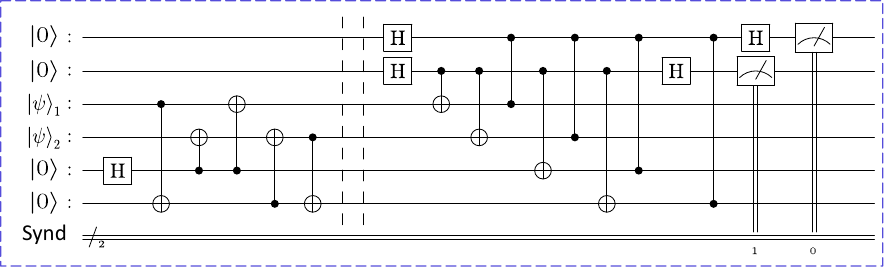}
    \caption{This figure illustrates the encoder for a 4-qubit code, followed by the syndrome measurement circuit shown after the two dashed lines. This code is designed to detect a single Pauli error or correct a single erasure error, encoding 2 logical qubits into 4 physical qubits.}
    \label{fig:circ_4q}
\end{figure*}

Fig.~\ref{fig:circ_4q} shows the quantum circuit that implements all of the QECC parts. Before the dashed lines we have the encoder that consist in a Hadamard gate followed by five CNOTs. We note that this encoding is different w.r.t. the classical way of building the code however our circuit still uses the least number of gates. Afterwards, we have the syndrome measurement which corresponds to the known syndrome measurement circuit. From this latter part we can argue that our RL agent found the following stabilizers:
\begin{eqnarray*}
    S_1 = \hat{X}_1 \hat{X}_2 \hat{X}_3, \quad S_2 = \hat{Z}_1 \hat{Z}_2 \hat{Z}_3.
\end{eqnarray*}
The measurement outcome of these syndromes are summarized in Table~\ref{tab:syndrome_4q}, showing that they are able to detect if one logical qubit has been affected by the noise. The RL agent uses such measurement information to build the recovery action -- in the case of an erasure on a single qubit -- or to post-select the qubits without errors. 
\begin{table}[!b]
\centering
\begin{tabular}{|c|c|c||c|c|c|}
\hline
\textbf{Error} & $S_1$ & $S_2$ & \textbf{Error} & $S_1$ & $S_2$ \\ \hline
$\hat{I}\ket{\psi_\text{log}}$ & 0 & 0 & $\hat{Z}_0\ket{\psi_\text{log}}$ & 0 & 1 \\ \hline
$\hat{X}_0\ket{\psi_\text{log}}$ & 1 & 0 & $\hat{Z}_1\ket{\psi_\text{log}}$ & 0 & 1 \\ \hline
$\hat{X}_1\ket{\psi_\text{log}}$ & 1 & 0 & $\hat{Z}_2\ket{\psi_\text{log}}$ & 0 & 1 \\ \hline
$\hat{X}_2\ket{\psi_\text{log}}$ & 1 & 0 & $\hat{Z}_3\ket{\psi_\text{log}}$ & 0 & 1 \\ \hline
$\hat{X}_3\ket{\psi_\text{log}}$ & 1 & 0 & & & \\ \hline
\end{tabular}

\caption{Syndrome table discovered by the syndrome measurement agent for the four-qubit code with the Pauli error channel.}
\label{tab:syndrome_4q}
\end{table}

\begin{figure*}[!t]
    \centering
    \includegraphics[width=\textwidth]{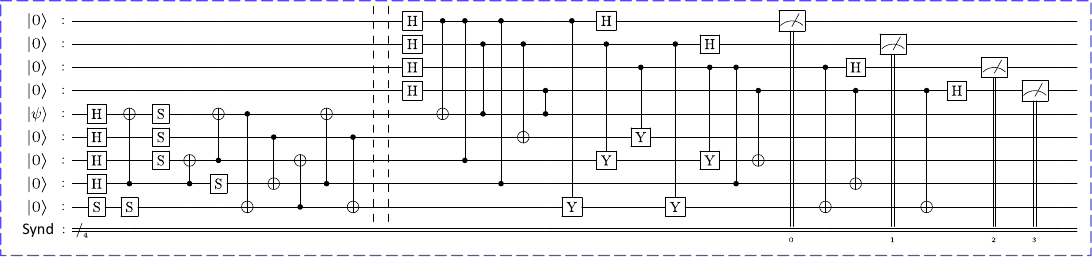}
    \caption{
The figure displays the encoder for the $[[5,1,3]]_2$ code, capable of correcting a single Pauli error on any qubit, shown before the two dashed lines. Following the dashed lines is the syndrome measurement circuit.}
    \label{fig:circ_5q}
\end{figure*}
\textbf{Five-qubits code: }The MARL framework was also employed to discover a five-qubit quantum error-correcting code capable of handling a general Pauli error channel. Similar to the four-qubit case, we use the  MARL for the encoder, the syndrome measurement, and the recovery, and thus they are collaboratively constructed during the entire error-correcting scheme.

The encoder generates codewords ensuring protection against all single-qubit Pauli errors, demonstrating that the MARL approach offers robust error-correcting codes for a modest numbers of qubits. Moreover, the syndrome measurement agent generates a circuit whose observables correspond to $n - k = 4$ stabilizer given by
\[
\begin{aligned}
    S_1 &= \hat{X}_1 \hat{Z}_3 \hat{Z}_4 \hat{Y}_5, & S_2 &= \hat{Z}_1 \hat{X}_2 \hat{Y}_3 \hat{Y}_5, \\
    S_3 &= \hat{Y}_2 \hat{Y}_3 \hat{Z}_4 \hat{X}_5, & S_4 &= \hat{Z}_1 \hat{X}_3 \hat{X}_4 \hat{X}_5.
\end{aligned}
\]
The measurement outcome of these syndromes are summarized on Table.~\ref{tab:syndrome_5q_horizontal}. Finally, the recovery agent utilizes the information from the measured syndromes to learn the appropriate recovery operations for each error type. The agent was able to map each unique syndrome to the correct Pauli operation needed to reverse the effect of the error, thus restoring the logical qubit to its original state. The circuit representation of the encoder and the syndrome measurement process is the one depicted in the Fig.~\ref{fig:circ_5q}.
 
\begin{figure*}
\centering \includegraphics[width=\textwidth]
{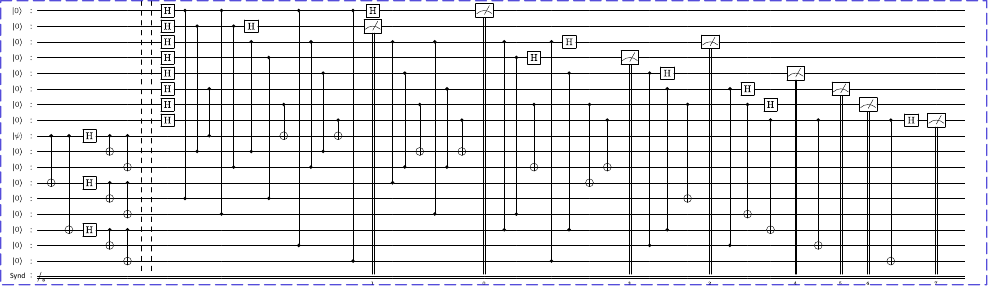} 
\caption{The encoding circuit for the nine-qubit code, constructed by the encoder agent. The circuit encodes one logical qubit into nine physical qubits through two levels of concatenation, first protecting against bit-flip errors and then against phase-flip errors.}
\label{fig:circ_9q} 
\end{figure*}
\begin{table}[!b]
\centering
\begin{tabular}{|c|c|c|c|c||c|c|c|c|c|}
\hline
\textbf{$\hat{X}_j$ Error} & $S_1$ & $S_2$ & $S_3$ & $S_4$ & \textbf{$\hat{Z}_j$ Error} & $S_1$ & $S_2$ & $S_3$ & $S_4$ \\ \hline
$\hat{I}\ket{\psi_\text{log}}$  & 0 & 0 & 0 & 0 & $\hat{Z}_0\ket{\psi_\text{log}}$ & 1 & 1 & 1 & 1 \\ \hline
$\hat{X}_0\ket{\psi_\text{log}}$ & 1 & 1 & 0 & 1 & $\hat{Z}_1\ket{\psi_\text{log}}$ & 0 & 0 & 0 & 1 \\ \hline
$\hat{X}_1\ket{\psi_\text{log}}$ & 1 & 0 & 1 & 0 & $\hat{Z}_2\ket{\psi_\text{log}}$ & 1 & 0 & 1 & 1 \\ \hline
$\hat{X}_2\ket{\psi_\text{log}}$ & 1 & 1 & 1 & 0 & $\hat{Z}_3\ket{\psi_\text{log}}$ & 0 & 1 & 1 & 0 \\ \hline
$\hat{X}_3\ket{\psi_\text{log}}$ & 0 & 0 & 1 & 0 & $\hat{Z}_4\ket{\psi_\text{log}}$ & 1 & 0 & 0 & 0 \\ \hline
$\hat{X}_4\ket{\psi_\text{log}}$ & 0 & 1 & 0 & 1 & & & & & \\ \hline
\end{tabular}

\caption{Syndrome table for the five-qubit code: comparison of $X_j$ and $Z_j$ single-qubit errors.}
\label{tab:syndrome_5q_horizontal}
\end{table}

\textbf{Nine-qubits code: }This code is the most general ones able to correct any Pauli error acting independently on each qubit. The MARL agent builds the quantum circuit that encodes each of these errors onto orthogonal subspaces depicted in the first part of Fig.~\ref{fig:circ_9q}. The syndrome measurement agent finds the following set of observables corresponding to the stabilizers 

\begin{eqnarray*}
S_1 &=& \hat{I}\hat{I}\hat{I}\hat{I}\hat{Z}\hat{Z}\hat{I}\hat{Z}\hat{Z},\quad 
S_2 = \hat{I}\hat{Z}\hat{Z}\hat{I}\hat{I}\hat{I}\hat{I}\hat{I}\hat{I},\\
S_3 &=& \hat{I}\hat{Z}\hat{Z}\hat{Z}\hat{I}\hat{Z}\hat{Z}\hat{I}\hat{Z},\quad 
S_4 = \hat{I}\hat{I}\hat{I}\hat{I}\hat{Z}\hat{Z}\hat{I}\hat{I}\hat{I},\\   
S_5 &=& \hat{I}\hat{Z}\hat{Z}\hat{I}\hat{I}\hat{I}\hat{Z}\hat{Z}\hat{I},\quad 
S_6 = \hat{Z}\hat{I}\hat{Z}\hat{I}\hat{I}\hat{I}\hat{Z}\hat{Z}\hat{I},\\    
S_7 &=& \hat{X}\hat{X}\hat{X}\hat{X}\hat{X}\hat{X}\hat{I}\hat{I}\hat{I},\quad 
S_8 = \hat{X}\hat{X}\hat{X}\hat{I}\hat{I}\hat{I}\hat{X}\hat{X}\hat{X}
\end{eqnarray*}

The syndrome measurement circuit enables the identification of errors by measuring these stabilizers and determining the error syndromes associated with different error patterns. The full circuit is depicted in Fig.~\ref{fig:circ_9q}.

\subsection{Qutrit codes}\label{sec:qutrit_codes}

\textbf{Three-qutrit code: }To correct either $\hat{X}$ or $\hat{Z}$ errors, a MARL agent is trained directly on the Knill--Laflamme conditions [main text Eq.\ref{eq:K-L_conditions}] to synthesize the encoding stage of the cycle. For the case of qutrits, the codewords for the $X$ Pauli noisy channel are given by
\begin{eqnarray*} 
\ket{0_L} &=& \ket{000}+\ket{121}+\ket{212},\\
\ket{1_L} &=& \ket{000}+\omega\ket{121}+\omega^2\ket{212},\\
\ket{2_L} &=& \ket{000}+\omega^2\ket{121}+\omega\ket{212}. 
\end{eqnarray*}
while for $Z$ error they are given by
\begin{eqnarray*}
\ket{0_L} = \ket{000} + \ket{011} + \ket{022} + \ket{102} + \ket{110} + \ket{121} + \\ + \ket{201} + \ket{212} + \ket{220}, \\
\ket{1_L} = \ket{001} + \ket{012} + \ket{020} + \ket{100} + \ket{111} + \ket{122} + \\ +\ket{202} + \ket{210} + \ket{221}, \\
\ket{2_L} = \ket{002} + \ket{010} + \ket{021} + \ket{101} + \ket{112} + \ket{120} + \\ + \ket{200} + \ket{211} + \ket{222}.
\end{eqnarray*}
Note that $\omega=\exp(2\pi i/d)$ is the primitive of the $d$th root of the unity. This corresponds to the known three-qutrit repetition code as a mathematical object for each Pauli channel, recovered here by the agent as a full QEC cycle from the Knill--Laflamme conditions. The agent then synthesizes the syndrome-measurement stage characterized by the observables.
\begin{eqnarray*}
    S_1 = \hat{Z}^2_1 \hat{Z}^2_2,\quad S_2 = \hat{Z}^2_1 \hat{Z}^2_3. \\ 
    S_1 = \hat{X}^2_1 \hat{X}_3,\quad S_2 =  \hat{X}_2^2 \hat{X}_3.
\end{eqnarray*}
The syndrome measurements associated to these stabilizers and the subsequent recovery operations are summarized in Table \ref{tab:syndrome_combined_horizontal_qutrit}.

\begin{table}[!b]
\centering
\begin{tabular}{|c|c|c||c|c|c|}
\hline
\textbf{Pauli $\hat{X}$ Error} & \textbf{Stab 1} & \textbf{Stab 2} & \textbf{Pauli $\hat{Z}$ Error} & \textbf{Stab 1} & \textbf{Stab 2} \\
\hline
$\hat{I}$       & 0          & 0          & $\hat{I}$       & 0          & 0          \\ \hline
$\hat{X}_0$     & 0          & $\omega^2$ & $\hat{Z}_0$     & $\omega^2$ & 0          \\ \hline
$\hat{X}_0^2$   & 0          & $\omega$   & $\hat{Z}_0^2$   & $\omega$   & 0          \\ \hline
$\hat{X}_1$     & $\omega^2$ & $\omega$   & $\hat{Z}_1$     & $\omega^2$ & $\omega^2$ \\ \hline
$\hat{X}_1^2$   & $\omega$   & $\omega^2$ & $\hat{Z}_1^2$   & $\omega$   & $\omega$   \\ \hline
$\hat{X}_2$     & $\omega^2$ & $\omega^2$ & $\hat{Z}_2$     & $\omega^2$ & $\omega$   \\ \hline
$\hat{X}_2^2$   & $\omega$   & $\omega$   & $\hat{Z}_2^2$   & $\omega$   & $\omega^2$ \\ \hline
\end{tabular}
\caption{Syndrome table for qutrit Pauli $X_j$ and $Z_j$ errors.}
\label{tab:syndrome_combined_horizontal_qutrit}
\end{table}

\textbf{Circuit for erasure channel: }We begin by considering the case of an erasure error in a qutrit, modeled by the following map
\begin{eqnarray*}
\mathcal{N}(\rho) &=& (1 - p) \rho + \sum_{i=1}^2 p_{\hat{X}^i}\, \hat{X}^i \rho\, \hat{X}^{i \dagger} + \sum_{j=1}^2 p_{\hat{Z}^j}\, \hat{Z}^j \rho\, \hat{Z}^{j \dagger} \\
&+& \sum_{i,j=1}^2 p_{\hat{X}^i \hat{Z}^j}\, \hat{X}^i \hat{Z}^j \rho\, (\hat{X}^{i} \hat{Z}^{j})^\dagger.
\end{eqnarray*}
\label{eq: Pauli_qutrits}
where $X^i$, $Z^j$ represent the powers of the  $X$, $Z$ Pauli operators for qutrits, respectively. Our MARL agent synthesizes an encoding circuit whose codewords are
 
\begin{figure}[!t]
    \centering 
    \includegraphics[width=\columnwidth]{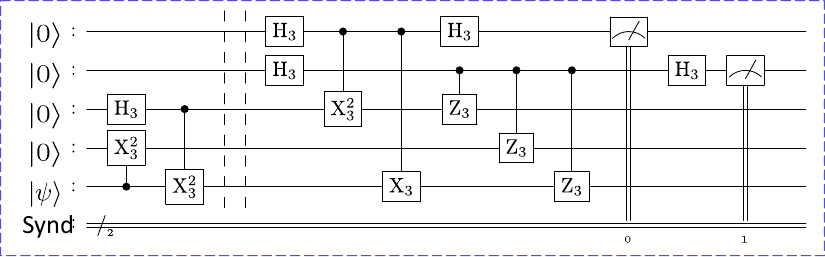} 
    \caption{The encoder circuit for the qutrit erasure code is shown before the two dashed lines, followed by the syndrome measurement circuit.} 
    \label{fig: Qt_Er}
\end{figure}
\begin{eqnarray*} 
\ket{0_L} = \ket{000}+\ket{102}+\ket{201},\\
\ket{1_L} = \ket{021}+\ket{120}+\ket{222},\\
\ket{2_L} = \ket{012}+\ket{111}+\ket{210}. 
\end{eqnarray*}
The stabilizers associated with this code are:
\begin{eqnarray*}
S_1 = \hat{X}^2_1 \hat{X}_3,\quad S_2 = \hat{Z}_1 \hat{Z}_2 \hat{Z}_3.
\end{eqnarray*}
The corresponding syndrome measurements are summarized in Table \ref{tab:syndrome_eras_qutrit}.
\begin{table}[b!]
\centering
\begin{tabular}{|c|c|c|}
\hline
\textbf{Error} & \textbf{Stab 1} & \textbf{Stab 2} \\ \hline
$\hat{I}$    & 0        & 0        \\ \hline
$\hat{X}$    & 0        & $\omega$ \\ \hline
$\hat{X}^2$  & 0        & $\omega^2$ \\ \hline
$\hat{Z}$    & $\omega^2$ & 0        \\ \hline
$\hat{Z}^2$  & $\omega$   & 0        \\ \hline
\end{tabular}
\caption{Syndrome Table for Qutrit erasure code}
\label{tab:syndrome_eras_qutrit}
\end{table}

\begin{figure*}[ht]
\centering
\includegraphics[width=\textwidth]{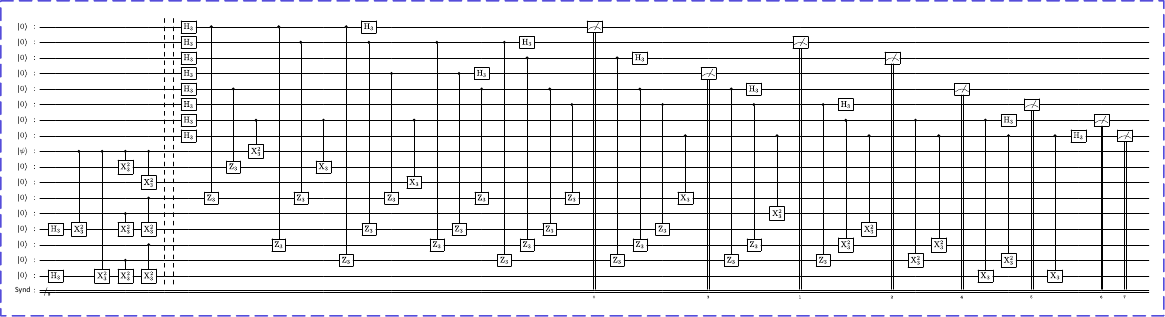}
\caption{The full qutrit QEC cycle synthesized by the MARL framework. The first stage implements a three-qutrit code correcting phase errors; the second stage re-encodes each resulting qutrit with a three-qutrit code correcting shift errors, yielding the nine-qutrit construction shown here together with the corresponding syndrome-measurement and recovery stages.}
\label{fig:circ_9t}
\end{figure*}

\textbf{Nine-qutrit code: }As in the qubit case, this benchmark corresponds to the known nine-qutrit analogue of a Shor-type construction as a mathematical object. From the agent's perspective, however, MARL discovers the full cycle—encoding, syndrome generators, and recovery map—from scratch. Similar to the qubit case, the nine-qutrit code is the most general code that can correct any single Pauli error and combinations of them 
as $\{ \hat{X}, \hat{X}^2, \hat{Z}, \hat{Z}^2, \hat{X}\hat{Z}, \hat{X}\hat{Z}^2, \hat{X}^2\hat{Z}, \hat{X}^2\hat{Z}^2 \}$
 on each qutrit. In this case, the MARL agent synthesizes an encoding circuit whose initial codeword is depicted in the first layer of Fig.~\ref{fig:circ_9t}. The agent then extends the construction to build the syndrome-measurement observables given by the generators.

\begin{eqnarray*}
S_1 &=& \hat{I} \hat{I} \hat{I} \hat{Z}^2 \hat{Z} \hat{I} \hat{Z}^2 \hat{Z}^2 \hat{Z}, \quad 
S_2 = \hat{I} \hat{I} \hat{I} \hat{Z}^2 \hat{I} \hat{Z}^2 \hat{Z}^2 \hat{Z}^2 \hat{Z}, \\
S_3 &=& \hat{I} \hat{I} \hat{I} \hat{Z} \hat{I} \hat{Z} \hat{Z}^2 \hat{Z}^2 \hat{Z}, \quad 
S_4 = \hat{I} \hat{I} \hat{I} \hat{Z}^2 \hat{I} \hat{Z}^2 \hat{I} \hat{Z} \hat{Z}, \\
S_5 &=& \hat{I} \hat{Z}^2 \hat{Z} \hat{Z}^2 \hat{I} \hat{Z}^2 \hat{Z}^2 \hat{Z}^2 \hat{Z}, \quad 
S_6 = \hat{Z} \hat{Z} \hat{I} \hat{Z}^2 \hat{I} \hat{Z}^2 \hat{Z}^2 \hat{Z}^2 \hat{Z}, \\
S_7 &=& \hat{X}^2 \hat{X} \hat{X} \hat{I} \hat{I} \hat{I} \hat{X}^2 \hat{X}^2 \hat{X}, \quad 
S_8 = \hat{I} \hat{I} \hat{I} \hat{X} \hat{X}^2 \hat{X}^2 \hat{X}^2 \hat{X}^2 \hat{X}.
\end{eqnarray*}
Finally, the last part of the circuit depicted in Fig.~\ref{fig:circ_9t} realizes the recovery stage, so that each measured syndrome is mapped to the correction of the corresponding Pauli error.

\section{Stabilizer formalism and notation}
\label{sec:stabilizer_appendix}

\subsection{Quantum error correction and the no-cloning constraint}

Quantum information cannot be protected by simple repetition: the 
no-cloning theorem~\cite{Wootters1982} forbids copying an unknown 
quantum state, and quantum noise encompasses errors beyond classical 
bit-flips~\cite{MacKay2003}. QEC instead encodes logical states 
into a larger Hilbert space $\mathcal{H}_d^{\otimes n}$ such that 
errors act only on recoverable degrees of freedom~\cite{Lidar_Brun_2013}. 
A noise channel acting on the physical system is described by its 
Kraus representation,
\begin{equation}
\mathcal{N}[\rho] = \sum_k \hat{E}_k \rho \hat{E}_k^\dagger, 
\quad \sum_k \hat{E}_k^\dagger \hat{E}_k = \mathbb{I},
\end{equation}
where the Kraus operators $\{\hat{E}_k\}$ capture all possible error 
processes. For qubit systems, any single-qubit error can be 
decomposed in the Pauli basis $\{I, X, Y, Z\}$, so correcting all 
Pauli errors on each qubit is sufficient for universal error 
correction.

\subsection{The Pauli group and stabilizer codes}

The $n$-qubit Pauli group $\mathcal{P}_n$ consists of all $n$-fold 
tensor products of Pauli operators with phases $\{\pm 1, \pm i\}$. 
A stabilizer code~\cite{gottesman1997stabilizer} is defined by an 
abelian subgroup $\mathcal{S} \subset \mathcal{P}_n$, called the 
stabilizer group, that does not contain $-\mathbb{I}$. The code 
space $\mathcal{H}_C$ is the simultaneous $+1$ eigenspace of all 
elements of $\mathcal{S}$:
\begin{equation}
\mathcal{H}_C = \bigl\{ \ket{\psi} \in \mathcal{H}_d^{\otimes n} 
: \hat{S}\ket{\psi} = \ket{\psi},\; \forall\, \hat{S} \in 
\mathcal{S} \bigr\}.
\end{equation}
Since $\mathcal{S}$ is abelian and generated by $n-k$ independent 
Pauli operators $\mathcal{S} = \langle \hat{S}_1, \ldots, 
\hat{S}_{n-k} \rangle$, the code space has dimension $2^k$, 
encoding $k$ logical qubits into $n$ physical qubits. The 
centralizer $\mathcal{C}(\mathcal{S})$ of $\mathcal{S}$ in 
$\mathcal{P}_n$ --- the set of Pauli operators that commute with 
all elements of $\mathcal{S}$ --- contains $\mathcal{S}$ itself 
as a subgroup. The quotient $\mathcal{C}(\mathcal{S})/\mathcal{S}$ 
defines the logical operators: operators in 
$\mathcal{C}(\mathcal{S}) \setminus \mathcal{S}$ act nontrivially 
on the logical degrees of freedom without being detected by 
stabilizer measurements.

\subsection{Knill-Laflamme conditions}

A noise channel $\mathcal{N}$ with Kraus operators $\{\hat{E}_k\}$ 
is correctable on $\mathcal{H}_C$ if and only if the 
Knill-Laflamme (KL) conditions hold~\cite{Knill_2000}:
\begin{equation}
\bra{i}\hat{E}^\dagger_k \hat{E}_l \ket{j} = C_{kl}\delta_{ij},
\end{equation}
where $\{|i\rangle\}$ is any orthonormal basis for $\mathcal{H}_C$ 
and $C_{kl}$ is a Hermitian matrix independent of the logical 
indices $i,j$. The two conditions encoded in this equation have 
distinct operational meanings. The off-diagonal condition 
($i \neq j$) requires that no error leaks information about the 
logical state into the environment: the erroneous states 
$\hat{E}_k\ket{i}$ and $\hat{E}_l\ket{j}$ must be orthogonal for 
$i \neq j$, so that measuring the error does not collapse the 
logical superposition. The diagonal condition ($i = j$) requires 
that any two errors either produce orthogonal states on 
$\mathcal{H}_C$, making them perfectly distinguishable, or act 
identically on it up to the scalar $C_{kl}$, making them 
indistinguishable but harmless.

Stabilizer codes satisfy the KL conditions structurally. For any 
correctable error pair $\hat{E}_a, \hat{E}_b \in \mathcal{P}_n$, 
the operator $\hat{E}_a^\dagger \hat{E}_b$ is itself a Pauli 
operator and falls into one of two cases. Either 
$\hat{E}_a^\dagger \hat{E}_b \in \mathcal{S}$, in which case it 
acts as the identity on $\mathcal{H}_C$ and 
$\bra{i}\hat{E}_a^\dagger \hat{E}_b\ket{j} = \delta_{ij}$; or 
$\hat{E}_a^\dagger \hat{E}_b$ anticommutes with some 
$\hat{S}_\ell \in \mathcal{S}$, in which case it maps 
$\mathcal{H}_C$ to an orthogonal subspace and 
$\bra{i}\hat{E}_a^\dagger \hat{E}_b\ket{j} = 0$. Both cases 
yield $\bra{i}\hat{E}_a^\dagger \hat{E}_b\ket{j} \propto 
\delta_{ij}$, satisfying the KL conditions.

\subsection{Syndrome measurement and the coset structure}

The key insight of syndrome measurement is that errors can be 
identified without learning anything about the logical state. Each 
Pauli error $\hat{E} \in \mathcal{P}_n$ either commutes or 
anticommutes with each stabilizer generator $\hat{S}_\ell$. The 
syndrome of $\hat{E}$ is the binary vector
\begin{equation}
\mathbf{s}(\hat{E}) = (s_1, \ldots, s_{n-k}) \in \{0,1\}^{n-k}, 
\quad s_\ell = \begin{cases} 0 & \text{if } 
[\hat{E}, \hat{S}_\ell] = 0, \\ 
1 & \text{if } \{\hat{E}, \hat{S}_\ell\} = 0. \end{cases}
\end{equation}
Measuring the stabilizer generators on an ancilla register returns 
this syndrome without disturbing the logical state, because the 
measurement commutes with all logical operators. The syndrome 
partitions the Pauli group into cosets of $\mathcal{S}$: all errors 
in the same coset $\hat{E}\mathcal{S}$ produce the same syndrome. 
Errors with trivial syndrome ($\mathbf{s} = \mathbf{0}$) either 
lie in $\mathcal{S}$ itself,  acting trivially on $\mathcal{H}_C$, or lie in $\mathcal{C}(\mathcal{S}) \setminus \mathcal{S}$, 
constituting undetectable logical errors. For each nontrivial 
syndrome, a decoder must select one representative correction 
operator from the corresponding coset. The standard circuit 
implementation measures each stabilizer generator $\hat{S}_\ell$ 
by preparing an ancilla in $\ket{+}$, applying a 
controlled-$\hat{S}_\ell$ gate between the ancilla and the data 
register, and measuring the ancilla in the $X$ basis. Since any 
Pauli stabilizer generator can be decomposed into a product of 
single-qubit Pauli gates and two-qubit controlled-Pauli gates, the 
gate set $\{H\text{-}CX\text{-}H,\, H\text{-}CZ\text{-}H,\, 
H\text{-}CY\text{-}H,\, I\}$ is sufficient to implement any 
stabilizer measurement circuit.

\subsection{Recovery and minimum weight decoding}

Given a syndrome string $\mathbf{s}$, the recovery operation 
selects a correction operator $\hat{R}(\mathbf{s})$ from the 
corresponding coset of $\mathcal{S}$. Since all operators in the 
same coset are equivalent up to stabilizer multiplication, and 
stabilizers act trivially on $\mathcal{H}_C$, any choice of 
representative restores the state to the code space. However, the 
choice matters for logical correctness: if the true error $\hat{E}$ 
and the chosen correction $\hat{R}$ lie in different cosets of 
$\mathcal{C}(\mathcal{S})$, their product $\hat{R}\hat{E}$ is a 
nontrivial logical operator, introducing an uncorrectable logical 
error. The optimal recovery strategy selects the most probable 
error consistent with the observed syndrome, which under independent 
identically distributed depolarizing noise corresponds to minimum 
weight decoding: choosing the coset representative with the fewest 
non-identity Pauli operators. Formally, the recovery map can be 
written as
\begin{equation}
\hat{R}(\mathbf{s}) = \argmax_{\hat{P} \in \hat{E}_\mathbf{s}
\mathcal{S}} \Pr(\hat{P}),
\end{equation}
where $\hat{E}_\mathbf{s}$ is any fixed representative of the 
coset with syndrome $\mathbf{s}$. The resulting corrected state 
satisfies $\mathcal{R}[\mathcal{N}[\rho]] = \rho_c$ for all logical 
states $\rho_c \in \mathcal{H}_C$, provided the error weight does 
not exceed the correction capacity of the code. Crucially, 
$\mathcal{R} \neq \mathcal{N}^{-1}$: the recovery is a conditional 
map that uses classical syndrome information, and its correctness 
depends on the prior probability distribution over errors, not on 
inverting the noise channel directly.

\section{MARL implementation details}\label{sec:Methods}
\label{sec:General_overview}
The main goal of the Reinforcement Learning (RL) framework for QEC is to build an architecture that is able to discover a new encoder and therefore a new stabilizer code, design the syndrome measurement circuit for that code and  discover a recovery procedure to correct possible errors. 
We aim to create this architecture following the steps: defining an agent for the encoder, one for the syndrome measurement and one for the recovery procedure based on the syndromes. From a RL perspective this setup can be seen as a sequential decentralized training task in the MARL \cite{marl-book} picture, where every agent relies on each other, except the first one, even if the various agents are trained independently. In order to build the whole framework we have to let the various parts interact between each other as highlighted in the main text Fig.1.
The key elements of every block are:
\begin{itemize}
    \item \textbf{Agent's Environment:} The agent's environment is defined by the evolution of the quantum circuit that the agent needs to construct. To maintain generality, we describe the environment as a numerical representation of the quantum circuit itself. For certain tasks, additional information may be included in the state. For instance, in the recovery procedure, information about the syndrome of the error to be corrected have been added.

    \item \textbf{Agent's Task:} Each block within the quantum circuit involves a specific task for the agent. For the encoder, the task is to maximize the Knill-Laflamme conditions by creating orthogonal subspaces for error recovery, particularly when dealing with non-degenerate codes. In the syndrome measurement, the agent's task is to construct a circuit that projects the codeword onto one of these subspaces via measurement. Lastly, for the recovery procedure, the task is to identify a sequence of operators which can restore the corrupted codeword to its original state.

    \item \textbf{Reward Function:} The reward function is crucial for optimizing the agent’s policy. For the encoder, the reward function aims to maximize the number of fulfilled KL conditions. In the syndrome measurement, the reward function checks whether each candidate stabilizer is a valid syndrome operator, i.e. whether it distinguishes correctable errors without revealing the logical information. For the recovery procedure, a natural choice for the reward function is the fidelity between the codeword without errors and the corrected codeword after the recovery process.
\end{itemize}

In the next subsections we will delve deeper into the design of every single step that compose our chain of agents.
\subsection{Encoder}\label{sec:encoder_method}
\begin{figure}[!t]
    \centering
    \includegraphics[scale=0.9]{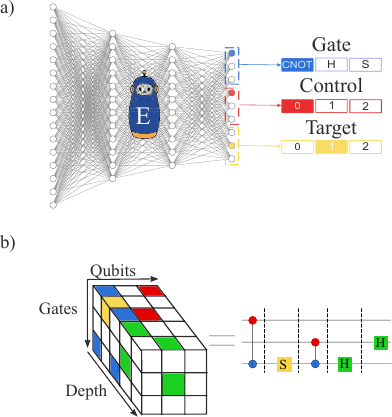}
    \caption{Encoder operation. (a) architecture of the encoder agent in detail. We feed a circuit representation (3D matrix) into a Neural Network. The outcomes are a scalar representing the RL value function and the gate that the agent wants to place. (b) A quantum circuit represented as a 3D matrix.}
    \label{fig:Encoder_agent}
\end{figure}

As previously emphasized, the encoder represents the initial component within this framework. Our philosophy in training this agent is to keep the framework as general as possible. Therefore, the state provided to the RL agent is not a representation of the stabilizer which defines the code, but rather a representation of the encoder circuit that we aim to build. In our approach, we represent the circuit as a three-index tensor $(i,j,k)$, where $i$ refers to the gate type, $j$ to the qubit index and $k$ to the position of the gate within the circuit [see Fig.~ \ref{fig:Encoder_agent}b)]. We start by setting the initial circuit equal to the identity. The interaction between agent and environment takes place in a fixed number of iterations $t_{\rm{steps}}$, which fixes the maximum depth of our circuit. At each iteration $k$, the agent performs an action, i.e., it places one of the possible three gates from the Clifford group $\mathcal{C} = \{CNOT, S, H\}$ on one or two of the available qubits on the circuit at a position $k$. Therefore, we obtain $U_t = U_{t-1}\dots U_{0}$, the unitary representing the encoder circuit. As a result, the logical states are constructed as $\ket{d}_{\rm{log}}=U_t\ket{d}\bigotimes_{j} \ket{0}_{j}$. In a second step, we need to check if the logical states satisfy the Knill-Laflamme conditions, i.e.,
\begin{eqnarray}
KL = \begin{cases}
&KL_1: \bra{\psi_{i}}\ket{\psi_{j}} = 0, \: \forall i\neq j \:,\\
&KL_2:\bra{\psi_{i}}\hat{E}_a\ket{\psi_{j}} = 0, \: \forall i, j, \: \forall \hat{E}_a \:,\\
&KL_3: \bra{\psi_{i}}\hat{E}_a^{\dagger} \hat{E}_b\ket{\psi_j} = 0\:, \: \forall i, j, \forall \hat{E}_{a} \neq \hat{E}_{b} \in \mathcal{E}\quad.
\end{cases}
\label{eq:KL_reward}
\end{eqnarray}
The first KL condition ensures the orthogonality between the codespaces, the second guarantees orthogonal subspaces for errors occurring on one of the possible codewords, and the last one enforces orthogonality between all errors. The encoder agent stops if all the KL conditions are satisfied. Otherwise, we assign a -1 to every KL condition that has not been satisfied The sum of these terms is then normalized by the total number of KL conditions.

To guarantee that the RL agent converges to the optimal encoder circuit, we need to design a reward function that smoothly or monotonically approaches to the condition where all the KL conditions are satisfied. We obtain that by weighting our reward with a discount factor at every iteration $\gamma_t = t/t_{\rm{steps}}$, where $\gamma_t \in (0,1]$. Thus, the reward is defined as 
\begin{equation}
    R_t = -\gamma_t \sum_{k=1}^3 KL_{k} + r_{\text{success}},
    \label{eq: reward_kl}
\end{equation}
where $r_{\text{success}}$ is an additional term that provides a boost when the task is successfully completed. This hyperparameter is zero during unfinished episodes and it takes positive values when the episode concludes successfully. However, the precise value of this parameter needs to be tuned during the training. We highlight that the RL agent is provided with negative rewards at each time step. As a result, the RL agent tends to converge to the optimal encoder circuit which also has minimum depth.
 
To accelerate the learning process of the encoding circuit, we employ \emph{task-specific curriculum learning}. This technique involves gradually increasing the complexity of the errors that the RL agent has to handle throughout the training. For instance, let us consider the total set of error that we wanted to correct $N = \{\hat{E}_a, \dots, \hat{E}_z\}$, the idea is to decimate this in subset as $N_\text{task 1} = \{\hat{E}_a, \dots, \hat{E}_e\}$ and feed to the RL learning agent so that it finds the optimal encoder for them, afterwards, we feed another portion of the error subset $N_\text{task 2} = \{\hat{E}_a, \dots, \hat{E}_g\}$ and again let the agent find the adequate circuit for all these error, we repeat this procedure until we have considered all the errors. In particular, this procedure was applied to the 4 qubits code. In this scenario, by constantly increasing the complexity of the exploration we avoid that the learning agent get stuck in local minima, making the process more efficient.

\subsection{Syndrome measurement circuit}\label{sec:syndrome_method}
After finding the quantum circuit that encodes our sensitive data qubit(s) into logical one(s), we need to determine what are the observables to measure so that we get the information about which error is affecting the code. These syndromes measurement has a two-fold objective; provide information about the type and location of the errors, and \emph{discriminating} them in different subspaces. Thus, for a correcting error encoding $k$ logical systems into $n$ physical ones, the stabilizer formalism state will be $n-k$ stabilizers whose structure corresponds to tensor products of generalized Pauli matrices. W.l.o.g. in this analysis we consider the qubit case where the measurement occurs by defining the projectors $ \hat{P}_i = (I + \hat{S}_i)/2 $, where $\hat{S}_i$ representes the $i$-th stabilizer. However, as this projector is not unitary, the implementation of such operation requires $k$ auxiliary systems. The procedure for measuring the syndromes is the following; we start with all the auxiliary states in the eigenstates of $\hat{X}$ i.e., $(|0\rangle + |1\rangle)/\sqrt{2}$. Afterwards, we perform controlled-$S_i$ gates where the target is the logical state and the controls are the auxiliary ones. Then, we perform a Hadamard gate to obtain the superposition $\ket{0}_a \hat{P}_i \ket{\psi}+\ket{1}_a(I-\hat{P}_i)\ket{\psi}$. Thus, if we measure the state $\ket{0}_a$ ($\ket{1}_a$) we may know if the state lies on the $+1(-1)$ codeword subspace, respectively. 

\begin{figure}[!b]
    \centering
    \includegraphics[scale=0.42]{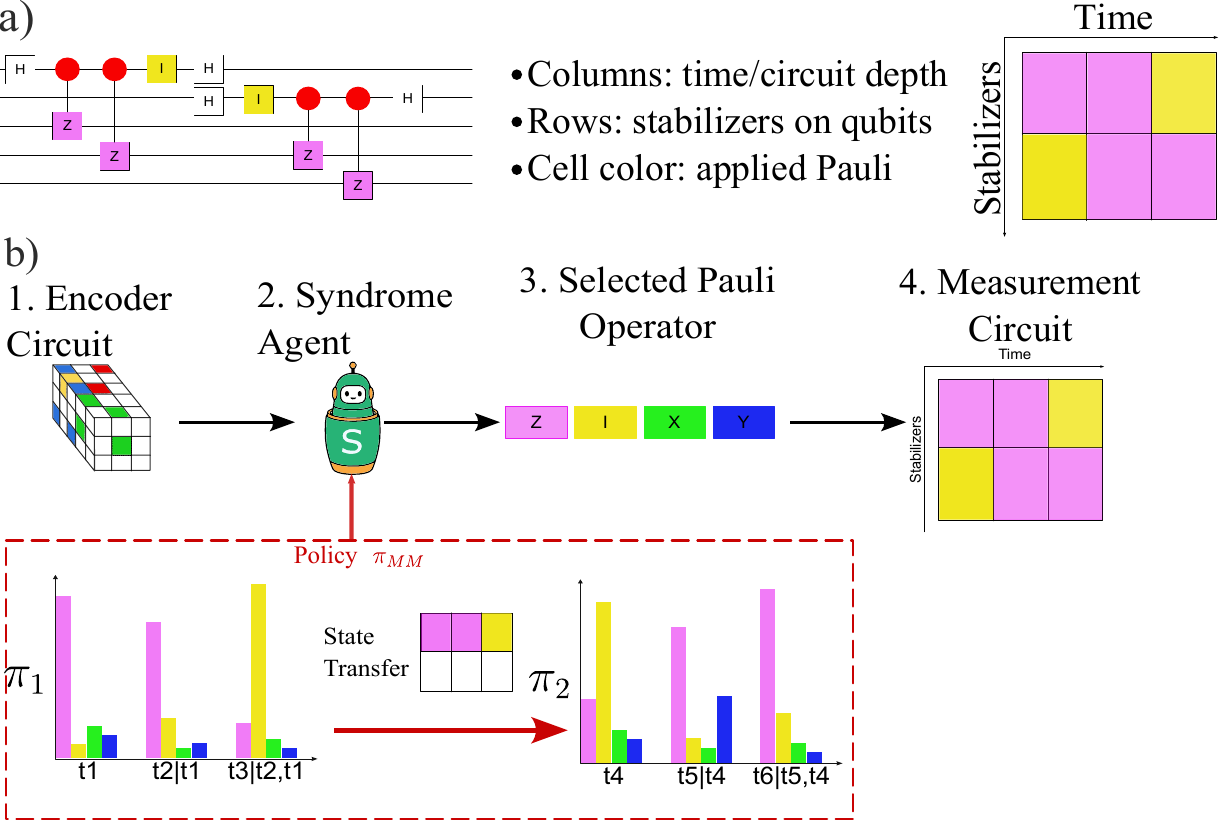}
    \caption{(a): Encoding of a syndrome measurement circuit into a 2D matrix representation. (b): Illustration of the M\&M curriculum learning process performed by the Measurement Agent. The agent receives the quantum circuit representation as input, enabling it to reconstruct the relevant codewords. Its actions consist of selecting Pauli operators that form the stabilizer group. The output is the matrix defining the measurement circuit structure. The policy $\pi$ is highlighted in the red box.}
    \label{fig:Syndrome_meas}
\end{figure}

To build the circuit, we utilize a similar approach to the one used for the encoder.
The state of this agent is defined by a 2D tensor $(j,k)$ -- see Fig.~\ref{fig:Syndrome_meas} (b) -- where at each entry is associated a number linked with the $k$-th Pauli operator which composes the $S_j$ stabilizer, indeed $j$ refers to the ancilla qubit we are currently using and $k$ is again the part of the circuit where the gates are applied. Our RL agent exploits as action a set of quantum gates acting on the Hilbert space generated by the physical and the auxiliary subsystems. The actions are defined in the following set: $ \mathcal{A}_s = \{H\text{-}CX\text{-}H, H\text{-}CZ\text{-}H, H\text{-}CY\text{-}H, I\}$. At each time step $t$ the agent selects an action from the set $\mathcal{A}_s$, the selected set of gates is acting on the $j$-th ancilla qubit (single qubit gate and control of the two qubit gate) and on the $k$-th physical qubit (target of the two qubit gate).

After generating the circuit, we measure an observable formed by a string of tensor products of Pauli matrices. To check if $S_\text{new}$ is a proper syndrome operator we need to check:

\begin{equation}
\begin{split}
&\text{C}_1: \bra{\psi_{\text{log}i}}\hat{S}_\text{new}\ket{\psi_{\text{log}j}} = \delta_{ij} \\
&\text{C}_2:\bra{\psi_{\text{log}i}}E_a^{\dagger} \hat{S}_b\ket{\psi_{\text{log}_i}} = 0 ,\text{ for at least one }E_a \in N ,\\
&\text{C}_3: \hat{S}_\text{new} \neq \Pi_i \hat{S}_i^\alpha \quad \forall \hat{S}_i \in \mathcal{S}, \forall \alpha \in \{0,\dots, d-1\},\\
&\text{C}_4: [\hat{S}_\text{new}, \hat{S}_i] \quad \forall \hat{S}_i \in \mathcal{S}.
\end{split}
\end{equation}
We define the reward function as $\mathcal{R}_t =1$ if all the conditions are satisfied and zero otherwise. This reward function applies universally to all auxiliary systems subsequently.

The policy for this RL agent corresponds to the concatenation of $ n-k $ different ones $ \pi_j $ -- see Fig.~\ref{fig:Syndrome_meas}a -- so that it is trained for every stabilizer for each auxiliary system. By slightly adjusting the final policy formula provided by the Mix\&Match curriculum learning technique, we derive the following formulation for the ultimate policy:

\begin{equation}
\label{eq:MM_syndrome}
    \pi_{\text{mm}}(a|s,t) = \sum_{i=1}^K w_i(t) \pi_i(a|s)
\end{equation}

Here, $ w_i $ represents weight assigned to each policy. We introduce a timestep dependency to these coefficients, ensuring that $ w_i $ equals 1 when the time slice for stabilizer $ i $ is active. This deterministic policy formulation effectively enables us to encode the entire syndrome measurement process. To reduce the agent's state space at each iteration of the training policy $\pi_i$, we place a gate where the target corresponds to the $t$-th data-qubit. An episode ends when for each ancilla the agent reaches the total number of gates that can be placed which is equivalent to $n$.\\

\subsection{Recovery}\label{sec:recovery_method}

\begin{figure}[!t]
    \centering
    \includegraphics[scale=0.8]{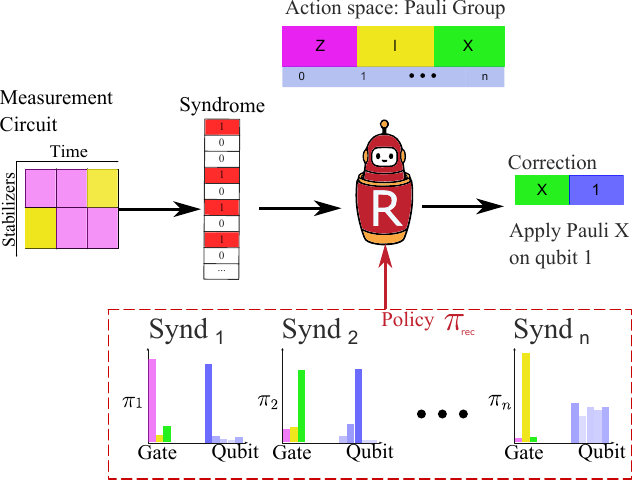}
    \caption{Visualization of the Recovery Agent's functioning. It receives the encoder circuit as input, enabling the computation of codewords associated with the measured syndrome. Available actions include selecting Pauli gates and target qubits for correction. The agent outputs the corresponding correction, guided by the M\&M curriculum learning strategy based on the input syndrome. }
    \label{fig:block_scheme_Rec}
\end{figure}

The final step in the framework involves the recovery procedure. We recall that for a given encoding map $\mathcal{E}$ and a noise channel $\mathcal{N}$, the recovery acts as $\mathcal{R}[\rho(t)]$ such that $\mathcal{R}[\mathcal{N}[\rho(t)]]=\rho_{c}(t)$ so that the corrected error is close to the original logical state. Note that the recovery action is deeply linked to the errors detected through the syndrome measurement proposed in the previous section. To build the recovery circuit, let us define the codespace projector as follows $\hat{P}_C = \prod_{\hat{S}_i \in S} (I + \hat{S}_i)/2$ which is nothing but the product of all syndromes mapping the logical state onto the $+1$ subspace. Next, we consider the action of the noise channel $\mathcal{N}$, described in terms of Pauli errors, which were detected through the stabilizers measurement obtaining a string $\mathbf{s}$ defined as 
\begin{equation}
    \mathbf{s}(\hat{E}_i) = (s_1, s_2, \ldots, s_m), \quad s_j = 
    \begin{cases} 
        0 & \text{if } \hat{E}_i \hat{S}_j = \hat{S}_j \hat{E}_i, \\
        1 & \text{if } \hat{E}_i \hat{S}_j = -\hat{S}_j \hat{E}_i.
    \end{cases}
\end{equation}
Notice that the string $\mathbf{s}$ changes if an error occurs in one of the computational units. We define the projector to the error-space associated with the error $E_i$ as, 
\begin{equation}
    \hat{P}_{\mathbf{s}_{E_{i}}} = \prod_{\hat{S}_j \in S} \frac{[I + (-1)^{s^{i}_{j}} \hat{S}_j]}{2}.
\end{equation}
Here, $s^{i}_{j}$ represents the $j$-th component of the syndrome vector $\mathbf{s}(E_i)$. 
The recovery map $ \mathcal{R} $ is designed based on the value of the such string so that if $s_j=1$ we apply the operation $\mathcal{R}_{\mathbf{s}}(\sigma) = \hat{E}_{\mathbf{s}} \sigma \hat{E}_{\mathbf{s}}^\dagger$. Then, full recovery map $ \mathcal{R} $ is defined by the channel:
\begin{equation} 
\mathcal{R}(\rho) = \sum_{i=0}R_i \rho R_i^\dagger = \sum_{i=0} \hat{E}_{s_{E_i}} \hat{P}_{s_{E_i}} \rho \hat{P}_{s_{E_i}}^\dagger \hat{E}_{s_{E_i}}^\dagger.
\end{equation}
The RL agent must be able to find the recovery operation $\hat{E}_s$ from the circuit representation of the encoder and a given set of syndrome strings $\mathbf{s}$. Furthermore, as we are correcting only bit- and phase-flip errors, the two possible the recovery actions are given by $\mathcal{A}_r = \{ \hat{X}, \hat{Z} \}$. For this stage of the MARL, the policy can be defined as aconditional distribution of finding $\hat{E}_i$ given $\mathbf{R}(E_i)$: $\pi(E_i | s_t = \mathbf{s}_{E_i})$ and the total policy is a weighted sum of this policy for all corrupted systems
\begin{equation}
    \label{eq:MM_recovery}
    \pi_{\text{mm}}(a|s,\mathbf{s}_{E_i}) = \sum_{i=1}^K w_i(\mathbf{s}_{E_i}) \pi_i(a|s).
\end{equation}
In this context, $w(\mathbf{s}_{E_i})$ is set to 1 when the syndrome $\mathbf{s}_{E_i}$ is measured and zero otherwise. We illustrate the training procedure in Fig.~\ref{fig:block_scheme_Rec}. The reward for each policy is again based on the fidelity.

The training of our RL agent follows two different approaches: the elementary approach (Sec. \ref{sec:elementary_approaches}) and the modular approach, where in both cases we represent the circuit either as a 2D or 3D array. The main difference between them is related by how they handle policy training and deployment: In the modular approach, we create a neural network representing the agent, which is trained using the M\&M curriculum learning method, where we optimize for every syndrome its policy $\pi_i$. For a more in-depth discussion, see Section \ref{sec:elementary_approaches}.\\

The algorithm employed for every RL task is the popular implementation in StableBaselines3 \cite{stable-baselines3} of the Proximal Policy Optimization (PPO) algorithm  \cite{schulman2017proximalpolicyoptimizationalgorithms}.

\subsection{Concatenated codes}
In the context of concatenated codes, we propose an extension to our framework that incorporates a more modular approach to error correction by splitting the encoder $\mathcal{E}$ into two distinct sub-encoders, $\mathcal{E}'$ and $\mathcal{E}''$. This structured division allows for a targeted error correction strategy, where each sub-encoder is specialized to handle a specific class of errors. The first encoder, $\mathcal{E}'$, is trained to correct a certain type of errors, such as bit-flip errors, while the second encoder, $\mathcal{E}''$, is responsible for correcting a different type of errors, such as phase-flip errors. By training these two encoders independently, we can efficiently address complex error scenarios in quantum or classical codes, where different types of errors may occur simultaneously.

Once the individual encoders $\mathcal{E}'$ and $\mathcal{E}''$ are trained, they are concatenated or composed together to form the complete encoder, denoted as $\mathcal{E} = \mathcal{E}' \circ \mathcal{E}''$. This composition reflects the hierarchical nature of the encoding process, where the first sub-encoder $\mathcal{E}'$ maps the input to an intermediate encoded state, which is then further processed by the second sub-encoder $\mathcal{E}''$. The overall result is a robust encoding scheme capable of correcting multiple classes of errors through this layered approach.

Following the construction of the complete encoder $\mathcal{E}$, the next step in our framework is to apply the subsequent error correction procedures, specifically the syndrome measurement and the recovery map. After identifying the error syndromes, the recovery map is applied to correct the detected errors.
 
\begin{figure*}[!t]
    \centering
    \includegraphics[width=0.8\textwidth]{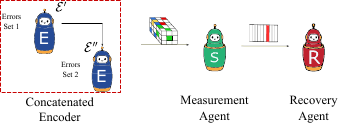}
    \caption{Schematic representation of the concatenated error correction framework. The encoder $\mathcal{E}$ is split into two sub-encoders, $\mathcal{E}'$ and $\mathcal{E}''$, each responsible for correcting different sets of errors. After composing the two encoders $\mathcal{E} = \mathcal{E}' \circ \mathcal{E}''$, the syndrome measurement circuit is applied to detect errors, followed by the recovery map to restore the data to its error-free state.}
    \label{fig:Concat_Codes}
\end{figure*}

\subsection{Elementary learning strategies}\label{sec:elementary_approaches}
\textbf{Encoder: }The initial reward function we developed is designed to encourage exploration. It impose a standard baseline reward of $r_\text{base} \leq 0$ for each step. Moreover it adds small negative rewards for repeated actions of the same type and small positive rewards for using the CNOT gate, which improves information spread within the system. We refer to these small quantities respectively as $r_\text{penalty} \leq 0$ and $r_\text{boost} \geq 0$. Additionally, if the KL criteria are met, a significant positive reward $r_\text{success}$ is given and the episode terminates. Conversely, if the maximum number of gates $\text{gates}_\text{max}$ is reached, a negative reward is applied $r_\text{failure}\leq 0$. This reward function is summarized by the following equation:

\begin{equation}
 R = \begin{cases}
r_\text{base}+r_\text{boost} & \text{for using a CNOT gate} \\
r_\text{base}+r_\text{penalty} & \text{for repeated actions} \\
r_\text{success} & \text{if KL criteria are met} \\
r_\text{failure} & \text{if } \text{gates}_\text{max} \text{ is reached}
\end{cases}   
\end{equation}

\textbf{Syndrome Measurement: }This approach builds upon the previous encoder methodology related to syndrome measurement. In this extended approach, the state inputted into the RL  agent remains a 3D matrix representing the circuit components. This matrix has three dimensions: one for the type of control gate applied, one for the position in the circuit, and one for identifying which ancilla and data qubit the action applies to. 

The length of the circuit is fixed because we need to measure $k$ stabilizers, and each stabilizer comprises $n$ tensor products of Pauli operators. Thus, the total circuit depth, and consequently the maximum number of actions the agent needs to take, is $T = \text{gates}_\text{max} = n \times k$.

In this approach, the agent not only selects the type of gate sequence to insert into the circuit but also determines the location within the circuit. The position in the circuit is determined by the timestep $t$, meaning it is part of the agent's action to choose where to place the control on the ancilla and which data qubit to target.

Since non-degenerate codes correspond to unique syndromes for every error, a different error corresponds to each syndrome. To construct our reward function, when the agent reaches the maximum number of actions, we apply to our codeword a Pauli error to each data qubit once per time. We then execute all these circuits, computing the various syndromes. If the syndromes are unique, the agent receives a positive reward $r_\text{success}$; otherwise, it incurs a negative penalty $r_\text{failure}$. The reward function is summarized by the following equation.

\begin{equation}
 R_t = \begin{cases}
0 & \text{if t $\neq$ T} \\
r_\text{success} & \text{if t = T and unique syndromes} \\
r_\text{failure} & \text{else}
\end{cases}   
\end{equation}

\textbf{Recovery: }In the elementary approach, we train a single agent with a unified policy. The circuit for the correction is described using a 3D matrix. The procedure involves applying every possible error to our encoded codeword, measuring the syndrome, and allowing the agent to choose the appropriate corrective action at each timestep $t$. The agent selects a gate from $\mathcal{A}_r$ and determines the qubit for correction. The reward for the agent's action is given by the fidelity with respect to the original logical codeword, defined as:

\begin{equation}
    \label{eq:fidelity}
    R_t = F_t =  \text{tr}\left[\sqrt{\sqrt{\rho} \rho_{\text{target}} \sqrt{\rho}}\right]^2.
\end{equation}

\textbf{Complexity Analysis: }We compare the computational complexity of the two proposed reinforcement learning strategies for syndrome measurement: the \textit{Modular approach} and the \textit{Elementary approach}. In the Modular approach, each of the $n - k$ stabilizers is handled by an independent agent. At each timestep $t$, the data qubit is fixed to the $t$-th index, and the control ancilla is determined by the stabilizer itself. Therefore, the agent only selects one of the $g$ possible gates from the action set $\mathcal{A}_s$. Since each agent must place a gate on each of the $n$ data qubits, the number of possible action sequences per agent is $g^n$, and the total search space complexity across all stabilizers becomes $\mathcal{O}(g^n \cdot (n - k)) \sim \mathcal{O}(g^n n)$.

In contrast, the Elementary approach uses a single agent to learn a global policy across all stabilizers. At each timestep, the agent selects a gate type ($g$ choices), a control ancilla ($n - k$ choices), and a target data qubit ($n$ choices), leading to $g \cdot n \cdot (n - k)$ possible actions per step. The episode length is $T = n \cdot (n - k)$, as the agent must construct $n - k$ stabilizers, each involving $n$ Pauli terms. Therefore, the total number of possible action sequences over an episode grows as $\mathcal{O}((g \cdot n \cdot (n - k))^{n(n-k)}) \sim O(n^{n^2})$, which means that our approach reduces the scaling from $\text{exp}(n^2 \ln(n))$ to $\text{exp}(n)$. While the Elementary approach provides greater flexibility and potentially more optimal circuits, it introduces a significantly higher computational burden due to its enlarged action and state space.

Comparing the recovery approaches we take advantage that in the context of non-degenerate quantum error correcting codes, each correctable error corresponds uniquely to a syndrome, forming a one-to-one mapping. In a single-agent reinforcement learning approach, the agent receives as input the full state $(s, q, P)$, where $s \in \{0,1\}^{n-k}$ is the syndrome, $q$ is the qubit position, and $P$ is the Pauli correction being applied. This results in a large and diverse input space of size approximately $2^{n-k} \cdot n \cdot 4 \rightarrow O(2^{n-k} n)$. The agent must learn a policy 
\begin{equation}
\pi(s, q, P) \to a,
\end{equation}
which requires generalization across syndromes that may correspond to entirely unrelated errors. In contrast, the Mix and Match approach assigns a dedicated agent $\pi_s$ for each syndrome $s$, reducing the input state to $(q, P)$ and the policy to
\begin{equation}
\pi_s(q, P) \to a.
\end{equation}
This simplifies the learning task by removing the need for syndrome generalization and avoids destructive interference between updates related to different syndromes. Each agent operates in a significantly smaller input space of size $n \cdot 4 \rightarrow n$, allowing for faster convergence and better policy specialization.

\section{BRAVE implementation details}
\label{sec:meth_var}
\subsection{Variational approach}

We describe how we implemented the variational approach to optimize QEC components: encoder, syndrome measurement, and recovery. This approach dynamically changes these circuits based on the modification of the noise profile produced by fluctuations due to hardware drift, environmental changes, or external interference. We illustrate the approach in the main text Fig.1( Stage 2), where all the circuits implementing the different QECC stages now depends on adaptive parameters $\boldsymbol{\theta}$ so that it modifies the encoder $\mathcal{E}'$ as
\begin{equation*}
    \mathcal{E} = \hat{U}_{\boldsymbol{\theta}}[\mathcal{E}'(\rho)]\:.
\end{equation*}
Due to the stabilizer formalism, the change in the structure in the encoder will affect the remaining parts in the code. Thus, the new syndromes and recovery action are given by
\begin{equation*}
    S_i' = \hat{U}_{\boldsymbol{\theta}} S_i \hat{U}^\dagger_{\boldsymbol{\theta}}, \quad E_\mathbf{s}' = \hat{U}_{\boldsymbol{\theta}} E_\mathbf{s} \hat{U}_{\boldsymbol{\theta}}^\dagger \:.
\end{equation*}
In the context-free controller of Algorithm~\ref{alg:brave_cf}, used for the illustrative trajectories of Fig.~\ref{fig:Fig_2_1}, the feedback is the sampled fidelity: after completing the QEC cycle, we compute the overlap with the initial state and provide this value to a RL bandit. This feedback mechanism guides the optimization process. If the encoder fidelity decreases, indicate that the current parameters are no longer effective, force the retraining of the agent. Similarly, low recovery fidelity prompts adjustments to the recovery parameters.Thus the bandit selects actions according to its policy, dynamically adapting the QEC components as demanded. The layer is re-optimized with Powell's method against the Knill–Laflamme conditions, warm-started from the previous $\boldsymbol{\theta}$. Results are averaged over 50 trajectories. The contextual controller used for the results of Sec.~\ref{sec:brave_results} is specified in Appendices~\ref{sec:controller_exact} and~\ref{sec:controllers}.

Motivated by the structure of the experimental hardware, we assume that the unitary $\hat{U}_{\boldsymbol{\theta}}$ is given by a sequence of variational layers where each of them represent an unitary $\hat{U}_{\boldsymbol{\theta}}$. For two- and three-level systems, we parametrize $\hat{U}_{\boldsymbol{\theta}}$ in terms of the $\text{SU}(d)$ generators \cite{Bronzan1988}
\begin{eqnarray}
    \hat{U}_{\boldsymbol{\theta}}=\exp\bigg[i\sum_{k=1}^{d^2-1}\theta_{k}\hat{\lambda}_{k}\bigg],
\end{eqnarray}
where $\hat{\lambda}_{k}$ is the $k$th generator of $\text{SU}(d)$ \cite{Tilma_2002} corresponding to the Pauli matrices for qubits, and the Gell-Mann matrices for qutrits, respectively. This adaptive retraining strategy allows the QEC system to dynamically optimize its components, addressing the unique characteristics of the noise and adjusting to fluctuations over time. By dynamically refining the encoder and recovery circuits, this approach ensures that the QEC system remains resilient and responsive to changing noise conditions, offering a more robust solution for quantum information protection.

Figures~\ref{fig:full_sweep_qubits}, ~\ref{fig:full_sweep_qutrits} quantify the advantage of this adaptive strategy for the contextual controller with exact drift (Appendix~\ref{sec:controller_exact}). Across all regimes, the variationally enhanced protocol achieves lower logical infidelity than standard QEC, with gains that increase with the noise transition time. This trend reflects the ability of the bandit-driven layer to track slow drifts in the noise and realign the code during operation, effectively mitigating the breakdown of static Knill--Laflamme conditions.

\begin{figure}
    \centering
    \includegraphics[width=\linewidth]{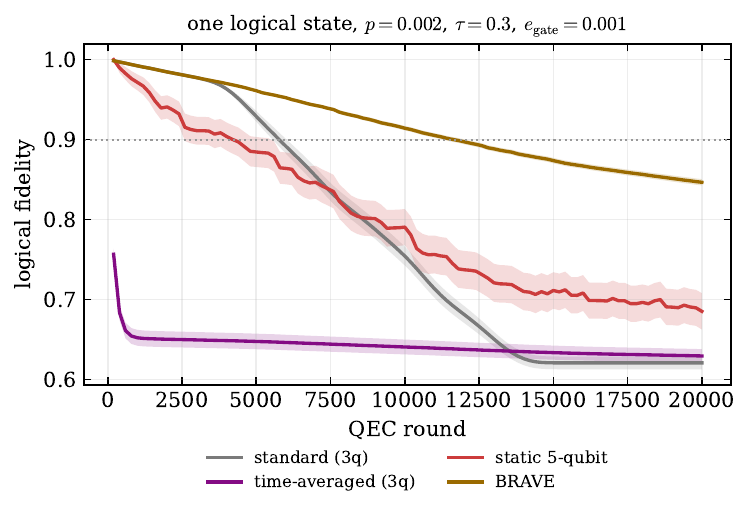}
    \caption{\textbf{Continuous protection of single logical qubit under drift noise}Mean logical fidelity versus QEC round, averaged over 200 trajectories with Haar-random initial states ($p = 0.002, \tau = 0.3, e_{gate} = 10^{-3}$). The state is never decoded. BRAVE decides every M = 200 rounds whether to realign the code frame, and each update applies one transversal single-qubit layer. Shaded bands show the standard error, and the dotted line marks F = 0.9. BRAVE stays above this threshold about twice as long as the standard 3-qubit code ($\approx11 600$ vs $\approx 5 800$ rounds), with about 25 updates.}
    \label{fig:continuous}
\end{figure}

\begin{figure*}
    \centering
    \includegraphics[width=\textwidth]{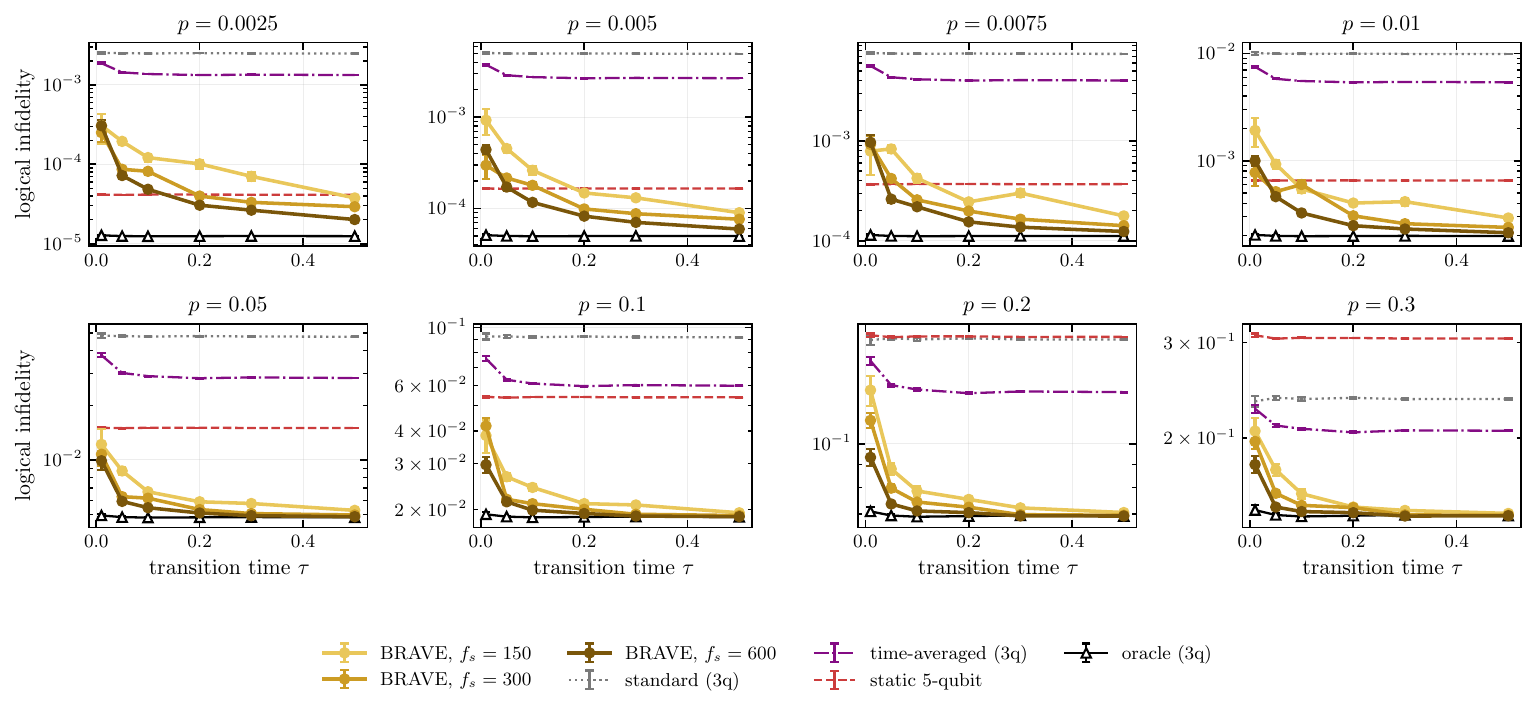}
    \caption{Logical infidelity of the three-qubit code against the transition time
    $\tau$, one panel per error probability, in the exact-drift setting of
    Fig.~\ref{fig:brave_main} ($M = 100$, $e_{\mathrm{gate}} = 10^{-3}$). BRAVE at
    $f_s = 150$, $300$, $600$ (light to dark gold), with the unrotated code (grey,
    dotted), the time-averaged code (purple, dash-dotted), the static five-qubit
    code (red, dashed) and the oracle (black); baselines are pooled over $f_s$.
    BRAVE lies below the unrotated code in every configuration. For short
    transition times the finite update rate leaves a residual misalignment costing
    a logical error of first order in $p$, so that at low error probability the
    five-qubit code is preferable: this occurs in 29 of the 144 configurations, all
    with $p \leq 10^{-2}$ and $\tau \leq 0.3$, and in none at $\tau = 0.5$. Error
    bars: standard error over $50$ trajectories.}
    \label{fig:full_sweep_qubits}
\end{figure*}

\begin{figure*}
    \centering
    \includegraphics[width=\textwidth]{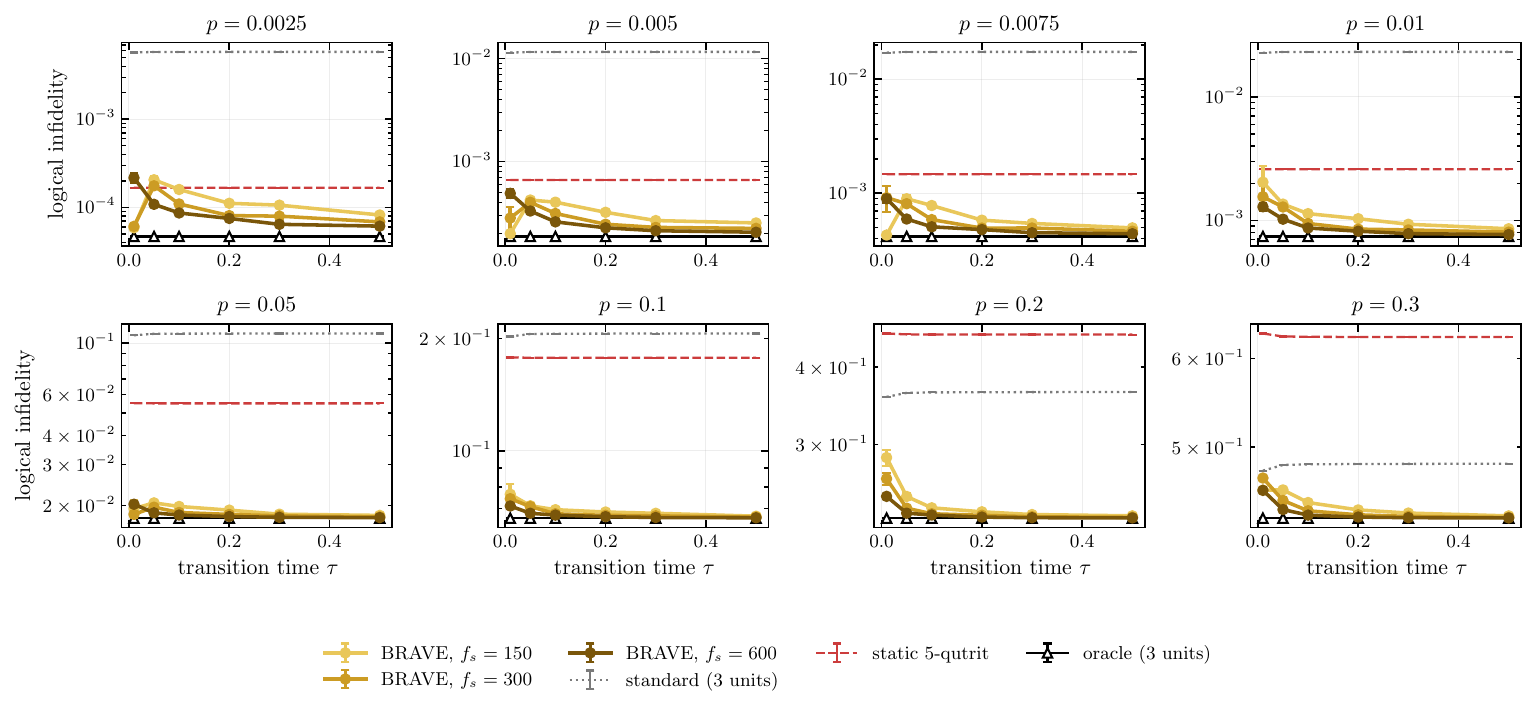}
    \caption{The same sweep for qutrits, with the colours of the qubit figure: BRAVE
    (gold), the unrotated code (grey, dotted), the static five-qutrit code (red,
    dashed) and the oracle (black); the time-averaged code is a qubit-only
    baseline. The behaviour reproduces the qubit case. The five-qutrit code is
    preferable to BRAVE only for the shortest transition times at the lowest error
    probability, $p = 2.5\times10^{-3}$ and $\tau \leq 0.05$ (3 of the 144
    configurations). Error bars: standard error over $50$ trajectories.}
    \label{fig:full_sweep_qutrits}
\end{figure*}

In order to check the improvement in the logical qubit lifetime we run a simulation at $p=0.002$, $\tau = 0.3$, $e_{gate}=0.001$ (Fig.~\ref{fig:continuous}). In this setup BRAVE keeps the mean logical fidelity above $0.9$ for about $11\,600$ QEC rounds, against about $5\,800$ for the standard bit-flip code and about $4\,200$ for the five-qubit perfect code, that is, about $\times 2$ and $\times 2.8$ longer, respectively.

\subsection{Code optimized for the time-averaged channel}
\label{sec:time_avg}
A fixed code cannot follow the drift, but it can be matched to the channel averaged over it, which is a stronger baseline than a code left at its initial calibration. For the qubit channel of Eq.~\eqref{eq:qubit_alpha_channel}, and using $\hat{Z}\hat{X} = i\hat{Y}$, the error operator is
\begin{equation}
    (\hat{Z}\hat{X})^{\alpha}\hat{X} = e^{i\varphi \hat{Y}}\hat{X} = \cos\varphi\,\hat{X} + \sin\varphi\,\hat{Z},
    \qquad \varphi = \frac{\pi\alpha}{2},
\end{equation}
so that the instantaneous channel is $\mathcal{N}_t(\rho) = (1-p)\rho + p\,(\mathbf{n}_t\cdot\boldsymbol{\sigma})\,\rho\,(\mathbf{n}_t\cdot\boldsymbol{\sigma})$ with $\mathbf{n}_t = (\cos\varphi, 0, \sin\varphi)$: the error axis rotates in the $xz$ plane from $\hat{X}$ to $\hat{Z}$. Expanding and averaging over the transition, with $s = t/\tau$ and $\alpha = \sin^2(\pi s/2)$, the diagonal components give $\langle\cos^2\varphi\rangle = \langle\sin^2\varphi\rangle = 1/2$, since $\varphi$ is distributed symmetrically about $\pi/4$, while the cross component is
\begin{equation}
    c \equiv \langle \sin\varphi\cos\varphi\rangle
      = \Big\langle \tfrac{1}{2}\sin(\pi\alpha)\Big\rangle
      = \frac{1}{2}\int_0^1 \cos\Big(\frac{\pi}{2}\cos\pi s\Big)\,ds \simeq 0.236.
\end{equation}
The averaged channel is therefore described, in the $(\hat{X},\hat{Z})$ plane, by the matrix
\begin{equation}
    M = \begin{pmatrix} 1/2 & c \\ c & 1/2 \end{pmatrix},
\end{equation}
whose eigenvalues are $1/2 \pm c$ with eigenvectors $(1,\pm1)/\sqrt{2}$. Since a vector of components $(a,b)$ in this plane represents the operator $a\hat{X} + b\hat{Z}$, the two eigenvectors correspond to $\hat{H} = (\hat{X}+\hat{Z})/\sqrt{2}$ and $\hat{V}_- = (\hat{X}-\hat{Z})/\sqrt{2}$, and using
\begin{align}
    \hat{X}\rho\hat{X} + \hat{Z}\rho\hat{Z} &= \hat{H}\rho\hat{H} + \hat{V}_-\rho\hat{V}_-, \\
    \hat{X}\rho\hat{Z} + \hat{Z}\rho\hat{X} &= \hat{H}\rho\hat{H} - \hat{V}_-\rho\hat{V}_-,
\end{align}
the averaged channel takes the diagonal form
\begin{equation}
    \langle\mathcal{N}\rangle(\rho) = (1-p)\rho
        + p\left(\tfrac{1}{2}+c\right)\hat{H}\rho\hat{H}
        + p\left(\tfrac{1}{2}-c\right)\hat{V}_-\rho\hat{V}_- .
    \label{eq:avg_channel}
\end{equation}
Since $1/2 + c > 1/2 - c$, the dominant averaged error lies along $\hat{H}$, and the best fixed code is the one whose correctable axis is $\hat{H}$. This is obtained from the bit-flip code by the transversal layer $\hat{u} = R_y(-\pi/4)$, for which $\hat{u}\hat{X}\hat{u}^{\dagger} = \hat{H}$ and $\hat{u}\hat{Z}\hat{u}^{\dagger} = -\hat{V}_-$, so that the stabilizer generators become $S_1 = \hat{V}_-^{(1)}\hat{V}_-^{(2)}$ and $S_2 = \hat{V}_-^{(2)}\hat{V}_-^{(3)}$, the sign cancelling in each product, and the recovery operator is $\hat{H}$ on the flagged qubit. This construction is specific to the qubit channel, and the time-averaged code is used as a baseline for qubits only.

\subsection{Controller: exact-drift setting}
\label{sec:controller_exact}

In the sweep of Fig.~\ref{fig:brave_main} the controller knows $\alpha(t)$.
There is no posterior, so the spread used in the syndrome-only setting
(Appendix~\ref{sec:controllers}) is identically zero, and the context is built
from the exact misalignment instead. With $\varphi_{\mathrm{frame}}$ the angle
of the deployed layer and $\delta_k=\varphi(t_k)-\varphi_{\mathrm{frame}}$, the
context is
\begin{equation}
  \mathbf{x}_k=\Big(\,g_k,\ \sqrt{g_k},\ n_{\mathrm{since}}/n_{\max}\Big),
  \qquad
  g_k=\frac{\mathrm{LI}(\delta_k)-\mathrm{LI}(0)}{p},
  \label{eq:context_exact}
\end{equation}
where $\mathrm{LI}(\delta)$ is the exact logical infidelity per round at
misalignment $\delta$, $n_{\mathrm{since}}$ is the number of control steps since
the last update and $n_{\max}=50$. For the three-qubit code
$g_k\simeq 2\sin^2\delta_k$, so that the first component is the gain of
realigning in units of $p$ and the second is proportional to the misalignment
itself, $\sqrt{g_k}\simeq\sqrt{2}\,|\sin\delta_k|$; it takes the place of the
posterior spread. For qutrits the same definitions hold with $\delta_k$ the
misalignment in $\alpha$ and $\mathrm{LI}$ the exact infidelity of the
three-qutrit code.

The policy is Eq.~\eqref{eq:policy}, initialized at $\mathbf{w}=0$, $b=-1$. An
update sets $\varphi_{\mathrm{frame}}=\varphi(t_k)$ through the closed-form
layer. The reward is Eq.~\eqref{eq:reward} with $L_k$ the exact infidelity of
the layer in use, and the parameters follow the REINFORCE rule
\begin{equation}
  (\mathbf{w},b)\leftarrow(\mathbf{w},b)
  +\eta\,(r_k-\bar r)\,\big(a_k-\pi(\textit{update}\mid\mathbf{x}_k)\big)\,(\mathbf{x}_k,1),
  \label{eq:reinforce_exact}
\end{equation}
with $\eta=0.05$ and $\bar r$ the running mean of the rewards. One policy is
trained on $30$ trajectories for each $(p,\tau,f_s)$ and then frozen; the
reported infidelities are averages over $50$ independent trajectories. The
layer starts aligned at $\alpha=0$ and no update is forced. Each control step
contains $M=100$ QEC rounds and $e_{\mathrm{gate}}=10^{-3}$, so that
$\lambda\simeq1.3\times10^{-5}$ for qubits and $1.7\times10^{-5}$ for qutrits,
Eq.~\eqref{eq:lambda}; on an update step the infidelity is the average over the
$M$ rounds, one of which carries the update layer. The oracle realigns at every
step at no cost. The complete procedure, for both settings, is summarized in
Algorithm~\ref{alg:brave}.

For the slowest drift, $\tau=0.5$, the misalignment accumulated between
consecutive control steps is small, the context varies little along a
trajectory, and the trained policy is dominated by its bias: it applies the
update in roughly a quarter to a third of the control steps, almost
independently of the context.

\begin{algorithm}[H]
\caption{BRAVE: contextual update controller}
\label{alg:brave}
\begin{algorithmic}[1]
\Require error probability $p$, update cost per round $\lambda$ [Eq.~\eqref{eq:lambda}],
         rounds per control step $M$, learning rate $\eta$
\State Initialize $\mathbf{w}\gets\mathbf{0}$, $b\gets -1$
\For{each trajectory}
  \State $\varphi_{\rm frame}\gets 0$, $n_{\rm since}\gets 0$ \Comment{layer aligned at $t=0$}
  \State \textsc{Syndrome-only}: initialize the particle filter over $(t_0,\log\tau)$
  \For{each control step $k$}
    \State Estimate the error axis $\hat\varphi_k$
           [exact drift: $\hat\varphi_k=\varphi(t_k)$; syndrome-only: Eq.~\eqref{eq:circmean}]
    \State Form the context $\mathbf{x}_k$
           [Eq.~\eqref{eq:context_exact} or Eq.~\eqref{eq:context}]
    \State $\pi_k\gets\sigma(\mathbf{w}\cdot\mathbf{x}_k+b)$;
           sample $a_k\sim\mathrm{Bernoulli}(\pi_k)$ \Comment{$1=$ update}
    \If{$a_k=1$}
      \State Realign: $\varphi_{\rm frame}\gets\hat\varphi_k$, closed-form layer
             $\hat u'$; apply $\big(\hat u'\hat u^{\dagger}\big)^{\otimes n}$
      \State $n_{\rm since}\gets 0$
    \Else
      \State $n_{\rm since}\gets n_{\rm since}+1$
    \EndIf
    \State Run $M$ QEC rounds with the layer held fixed; record the number
           $m_k$ of detection events
    \State \textsc{Syndrome-only}: update the particle weights with
           $\mathcal{L}(\delta_i\mid m_k)$ [Eq.~\eqref{eq:weight_update}]
    \State $L_k\gets$ infidelity per round of the layer in use
           [exact drift: $\mathrm{LI}(\delta_k)$; syndrome-only: posterior expectation]
    \State $r_k\gets-\big(L_k+\lambda\,a_k\big)/p$ \Comment{reward}
    \If{training}
      \State Update the running average $\bar r$ of the reward
      \State $(\mathbf{w},b)\gets(\mathbf{w},b)
             +\eta\,(r_k-\bar r)\,(a_k-\pi_k)\,(\mathbf{x}_k,1)$
    \EndIf
  \EndFor
\EndFor
\State Freeze $(\mathbf{w},b)$ after the training trajectories; evaluate on
       independent trajectories
\end{algorithmic}
\end{algorithm}

\subsection{Logical infidelity introduced by an update}
\label{app:update_cost}
To model the additional noise introduced by an update, we assume that each of its three single-unit gates is followed independently by a depolarizing channel, with total nonidentity Pauli-error probability \(e_{\rm gate}\). This gate-noise model is separate from the time-dependent physical-noise channels in Eqs. (15)–(16). In the frame of the repetition code, recovery corrects a single-unit shift error \(X^a\), whereas an error \(X^aZ^b\) with \(b\ne0\) leaves a nontrivial logical Pauli. For qubits, two of the three nonidentity Paulis therefore contribute to logical error, giving \(P_{\rm L}\simeq3(2/3)e_{\rm gate}=2e_{\rm gate}\). For qutrits, six of the eight nonidentity generalized Paulis contribute, giving \(P_{\rm L}\simeq3(6/8)e_{\rm gate}=9e_{\rm gate}/4\). A nontrivial logical Pauli occurring with probability \(P_{\rm L}\) contributes \(dP_{\rm L}/(d+1)\) to the fidelity loss averaged uniformly over pure logical states. Hence \(\Delta_{\rm gate}\simeq4e_{\rm gate}/3\) for qubits and \(\Delta_{\rm gate}\simeq27e_{\rm gate}/16\approx1.7e_{\rm gate}\) for qutrits, neglecting terms of order \(e_{\rm gate}^2\) and \(p e_{\rm gate}\). Since an update adds one such layer within a batch of \(M\) QEC rounds, its average cost per round is \(\lambda=\Delta_{\rm gate}/M\).

\subsection{Syndrome-based implementation of BRAVE}
\label{sec:syndrome_impl}

This appendix details the procedure summarized in Sec.~\ref{sec:brave_method}, in which the online stage uses only quantities produced by the hardware during operation. The QEC cycle is run continuously and the syndrome record is processed in batches of $M$ rounds; after each batch the controller updates its estimate of the drift and decides whether to realign the variational layer. The codeword fidelity enters nowhere in this loop and is computed only to report performance.

\subsubsection{Detection statistics}
\label{sec:pdet_app}
Let $\delta$ denote the misalignment between the layer and the instantaneous error axis. In the frame of the code the error on each qubit is $\cos\delta\,\hat{X}' + \sin\delta\,\hat{Z}'$, and only the $\hat{X}'$ component anticommutes with the stabilizers of the three-qubit code: each qubit is therefore flagged with probability $q = p\cos^{2}\delta$, while the $\hat{Z}'$ component is undetectable and acts directly as a logical error. A round of the code shows a non-trivial syndrome unless none or all three qubits are flagged, which gives $P_{\mathrm{det}} = 1 - (1-q)^3 - q^3 = 3q(1-q)$, that is Eq.~\eqref{eq:pdet}. The number $m$ of detection events in a batch of $M$ rounds is then binomial, so that
\begin{equation}
    \mathcal{L}(\delta \mid m) \propto P_{\mathrm{det}}(\delta)^{m}\,\big[1 - P_{\mathrm{det}}(\delta)\big]^{M-m},
    \label{eq:likelihood}
\end{equation}
and $m$ is a sufficient statistic for $\delta$: no information is lost by discarding the individual syndrome strings and keeping only their count. This is what makes the estimation free of experimental overhead, since the counts are a by-product of the error correction itself.

\subsubsection{Bayesian filter}
\label{sec:filter}
The drift is parametrized by its start time $t_0$ and its duration $\tau$, through $\alpha(t) = \sin^{2}[\pi(t-t_0)/2\tau]$ within the transition and $\alpha = 0$ or $1$ outside it, with the physical error probability $p$ assumed known from device characterization. The posterior over $(t_0, \log\tau)$ is represented by $N_p = 2000$ weighted particles, with a uniform prior on $t_0$ over the observation window and a log-uniform prior on $\tau$. After each batch the weights are multiplied by Eq.~\eqref{eq:likelihood} evaluated at the misalignment that each particle predicts, and renormalized. Each particle carries a hypothesis $(t_0^{(i)}, \log\tau^{(i)})$ about the drift and a weight $w_i$, initialized uniformly at $w_i = 1/N_p$. At the $k$th batch, taken at time $t_k$, particle $i$ predicts the drift parameter $\alpha_i = \alpha(t_k; t_0^{(i)}, \tau^{(i)})$, hence the error angle $\varphi_i = \pi\alpha_i/2$ and, given the angle $\varphi_{\mathrm{frame}}$ of the layer currently deployed, the misalignment $\delta_i = \varphi_i - \varphi_{\mathrm{frame}}$ that it implies. Each particle therefore predicts a different detection probability $P_{\mathrm{det}}(\delta_i)$ through Eq.~\eqref{eq:pdet}, and the observed count $m_k$ discriminates between them: the weights are updated multiplicatively by the likelihood of Eq.~\eqref{eq:likelihood} and renormalized,
\begin{equation}
    w_i \;\leftarrow\; \frac{w_i\,\mathcal{L}\big(\delta_i \mid m_k\big)}{\sum_j w_j\,\mathcal{L}\big(\delta_j \mid m_k\big)} .
    \label{eq:weight_update}
\end{equation}
In practice the update is carried out on the logarithm, $\ln\mathcal{L} = m_k\ln P_{\mathrm{det}}(\delta_i) + (M-m_k)\ln[1-P_{\mathrm{det}}(\delta_i)]$, with the maximum over particles subtracted before exponentiating, since for $M = 10^{4}$ the likelihoods span many orders of magnitude and would otherwise underflow. Two features of Eq.~\eqref{eq:weight_update} are worth noting. The weights depend on the deployed layer through $\delta_i$, so the information a batch carries depends on how misaligned the code currently is: a perfectly aligned layer sits at the maximum of $P_{\mathrm{det}}$, where the detection rate is least sensitive to the drift, which is the origin of the quartic scaling derived in Appendix~\ref{sec:info_app}. And the update is exact rather than approximate, since $m_k$ is a sufficient statistic, so the only approximation in the filter is the finite number of particles. When the effective sample size $N_{\mathrm{eff}} = (\sum_i w_i^2)^{-1}$ falls below $N_p/2$ the particles are resampled and moved with a Liu--West kernel, which shrinks them towards their weighted mean and adds a matched jitter, so that the posterior does not collapse onto a single particle as the evidence accumulates.

The layer is then set from the posterior in closed form. Since the error axis is defined modulo $\pi$, the estimate is taken as the circular mean of the predicted angles,
\begin{equation}
    \hat\varphi = \tfrac{1}{2}\arg \sum_i w_i\, e^{2 i \varphi_i},
    \qquad
    \sigma_\varphi = \tfrac{1}{2}\sqrt{-2\ln\Big|\sum_i w_i e^{2 i \varphi_i}\Big|},
    \label{eq:circmean}
\end{equation}
where $\varphi_i = \pi\alpha_i/2$ is the angle predicted by particle $i$ and $\sigma_\varphi$ is the corresponding circular spread, which the controller uses as a measure of its own uncertainty. The layer $\hat{u} = R_y(-\hat\varphi)$ follows immediately, with no numerical optimization in the loop.

\subsection{Update controller: syndrome-only setting}
\label{sec:controllers}
The controller acts once per batch, with the binary action $a \in \{\mathrm{keep}, \mathrm{update}\}$. The state on which it acts is not the syndrome record itself but the summary of it carried by the posterior, together with the layer in use. Concretely, at the $k$th batch the environment state is the triple $\big(\hat\varphi_k,\, \sigma_{\varphi,k},\, \varphi_{\mathrm{frame}}\big)$, comprising the estimated error angle, its spread from Eq.~\eqref{eq:circmean} and the angle of the deployed layer, to which is added the number $n_{\mathrm{since}}$ of batches elapsed since the last update. From these the controller forms the context
\begin{equation}
    \mathbf{x}_k = \Big(\underbrace{\Delta\hat{L}_k/p}_{\text{expected gain}},\;\;
                        \underbrace{\sigma_{\varphi,k}}_{\text{uncertainty}},\;\;
                        \underbrace{n_{\mathrm{since}}/n_{\max}}_{\text{staleness}}\Big),
    \label{eq:context}
\end{equation}
with the expected gain of realigning given by
\begin{equation}
    \Delta\hat{L}_k = \sum_i w_i\,\mathrm{LI}\big(\varphi_i - \varphi_{\mathrm{frame}}\big) - \sum_i w_i\,\mathrm{LI}\big(\varphi_i - \hat\varphi_k\big),
    \label{eq:gain}
\end{equation}
where $\mathrm{LI}(\delta)$ is the logical infidelity of the code at misalignment $\delta$, tabulated once for each $p$, and the two sums are its posterior expectation under the layer currently deployed and under the one that an update would install. The three features therefore answer three different questions: how much is to be gained by realigning, how reliable that estimate is, and how long the layer has been left untouched. The first is normalized by $p$ so that the same policy applies across error probabilities, and $n_{\mathrm{since}}$ is normalized by $n_{\max} = 50$ to keep the three of comparable magnitude.

The policy is logistic in these features,
\begin{equation}
    \pi(\mathrm{update} \mid \mathbf{x}_k) = \sigma\big(\mathbf{w}\cdot\mathbf{x}_k + b\big),
    \qquad \sigma(z) = \big(1 + e^{-z}\big)^{-1},
    \label{eq:policy_syn}
\end{equation}
that is Eq.~\eqref{eq:policy}, and is memoryless given $\mathbf{x}_k$: all the history of the process enters through the posterior, which is itself a sufficient summary of the syndrome counts observed so far, so the decision problem is a contextual bandit rather than a full reinforcement-learning problem. The only quantities the controller carries across batches are the parameters $(\mathbf{w}, b)$, which are frozen after training, the reward baseline $\bar{r}$, used during training only, and the counter $n_{\mathrm{since}}$. This is the difference from the context-free controller of Algorithm~\ref{alg:brave_cf}, whose action probabilities depend on no state at all and which therefore has to rediscover, through its reset mechanism, information that is here supplied explicitly.

After each batch the controller receives
\begin{equation}
    r = -\big(\hat{L} + \lambda\,a\big)/p,
    \label{eq:reward_ctrl}
\end{equation}
with $\hat{L}$ the posterior expected infidelity of the layer actually in use and $\lambda$ the cost of an update per round, $\lambda = \Delta_{\mathrm{gate}}/M \simeq 4e_{\mathrm{gate}}/(3M)$, Eq.~\eqref{eq:lambda}, and its parameters are updated by the REINFORCE rule
\begin{equation}
    (\mathbf{w}, b) \leftarrow (\mathbf{w}, b) + \eta\,(r - \bar{r})\,\nabla \log \pi(a \mid \mathbf{x}),
    \label{eq:reinforce}
\end{equation}
with learning rate $\eta = 0.05$ and $\bar{r}$ a running average of the reward. A policy is trained on $40$ simulated trajectories for each configuration and then frozen, and all reported results are obtained on independent trajectories. The trained weights follow an interpretable pattern: the coefficient of the predicted gain is positive and grows with the number of rounds per batch, from $3.1$ at $M = 10$ to $4.5$ at $M = 10^{4}$ for $p = 5\times10^{-3}$, while the bias rises from $-6.2$ to $-2.5$ over the same range. The controller therefore becomes both more responsive and less reluctant to act as the estimate becomes more reliable, which is behaviour that a fixed rule would have to be given by hand. A policy trained at one $(p, M)$ does not transfer to another, so one is trained per configuration.

\subsubsection{Cost of the estimation}
\label{sec:info_app}
Expanding Eq.~\eqref{eq:pdet} around alignment gives $P_{\mathrm{det}} \simeq 3p\,(1 - \delta^2)$ for small $p$ and $\delta$, so that the Fisher information per round about the misalignment is
\begin{equation}
    I(\delta) = \frac{\big(\partial_\delta P_{\mathrm{det}}\big)^2}{P_{\mathrm{det}}\,(1 - P_{\mathrm{det}})} \simeq 12\,p\,\delta^{2},
\end{equation}
and resolving a misalignment $\delta$ requires $N \gtrsim 1/\big[I(\delta)\,\delta^2\big] \sim 1/(12\,p\,\delta^{4})$ rounds; for $p = 10^{-2}$ and $\delta = 0.1$ this is of order $10^{5}$ rounds. The quartic dependence on $\delta$ and the linear dependence on $p$ explain the thresholds in $M$ reported in Sec.~\ref{sec:brave_results}, and identify the syndrome statistics, rather than the classical computation, as what limits how fast a drift can be followed.

\subsection{Marginal Efficiency and Cost--Performance Structure}
\label{sec:bandit_eval}

This appendix concerns the context-free controller of Algorithm~\ref{alg:brave_cf}, whose reward is the sampled fidelity. We now analyze the performance of the bandit alone to assert the efficiency of the algorithm. We perform the analysis by jointly examining (i) mean logical fidelity, (ii) retraining cost, and (iii) the \emph{marginal efficiency} of retraining. This provides a more operational view of how additional retraining translates into performance gains.

\subsubsection{Performance and Cost Revisited}

\begin{figure}[t]
\centering
\includegraphics[width=0.45\textwidth]{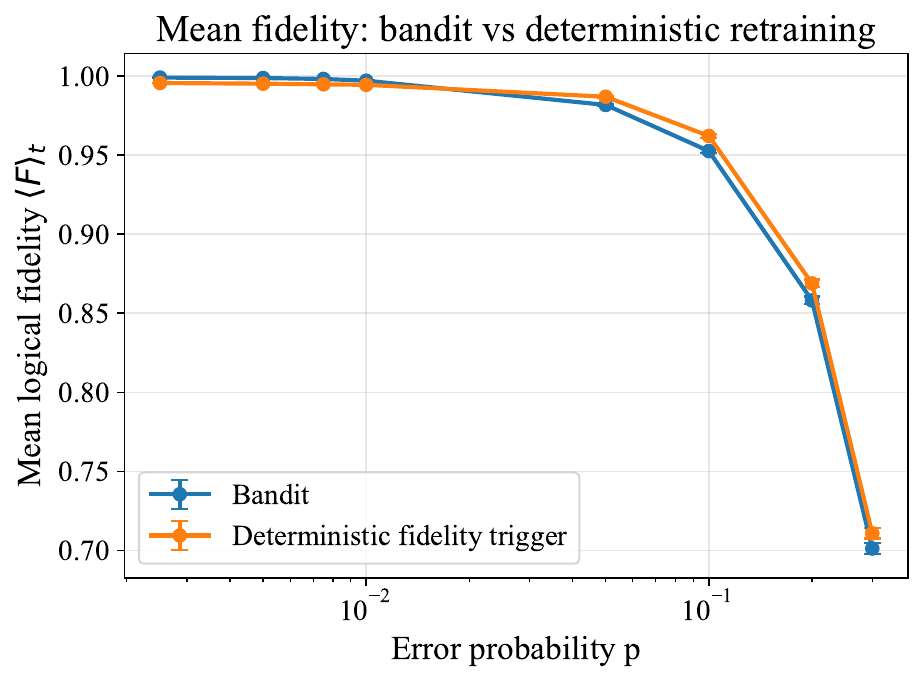}
\includegraphics[width=0.45\textwidth]{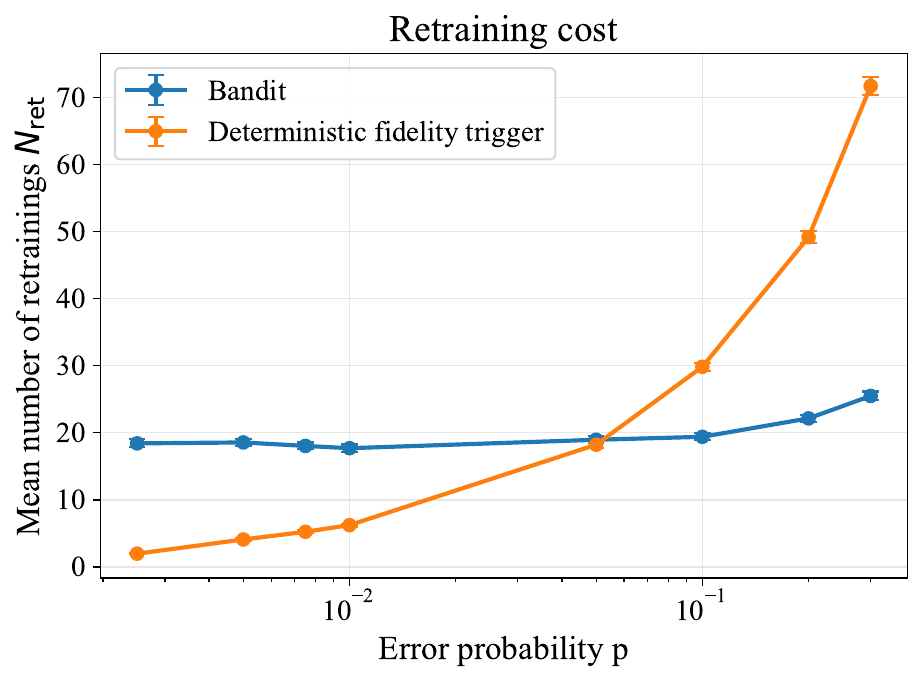}
\caption{
\textbf{Performance and retraining cost.}
Above: Mean logical fidelity $\langle F \rangle_t$.
Below: Mean number of retraining events $N_{\mathrm{ret}}$.
Error bars denote the standard error of the mean over $50$ trajectories.
}
\label{fig:bandit_perf_cost_updated}
\end{figure}

Figure~\ref{fig:bandit_perf_cost_updated} reveals two distinct operating regimes in both fidelity and retraining cost. At low error rates, both policies achieve near-perfect fidelity, with the bandit showing a slight advantage, albeit at the expense of a higher number of retraining events. As $p$ increases, the deterministic policy attains consistently higher fidelity; however, this comes with a rapidly escalating retraining rate that grows superlinearly with the noise level.

In contrast, the bandit policy exhibits a much weaker dependence on $p$, effectively regularizing the retraining frequency. This indicates that the bandit implicitly adapts to the noise level by allocating a nearly constant retraining budget.

\subsubsection{Marginal Efficiency of Retraining}

To quantify how useful additional retraining events are, we introduce the marginal efficiency
\begin{equation}
\eta(p) = \frac{\Delta \langle F \rangle_t}{\Delta N_{\mathrm{ret}}},
\end{equation}
where $\Delta \langle F \rangle_t = \langle F \rangle_t^{\mathrm{det}} - \langle F \rangle_t^{\mathrm{bandit}}$ and $\Delta N_{\mathrm{ret}} = N_{\mathrm{det}} - N_{\mathrm{bandit}}$.

To avoid numerical instabilities when $\Delta N_{\mathrm{ret}}$ is small, we discard points with $|\Delta N_{\mathrm{ret}}| < 2$.

\begin{figure}[t]
\centering
\includegraphics[width=0.5\textwidth]{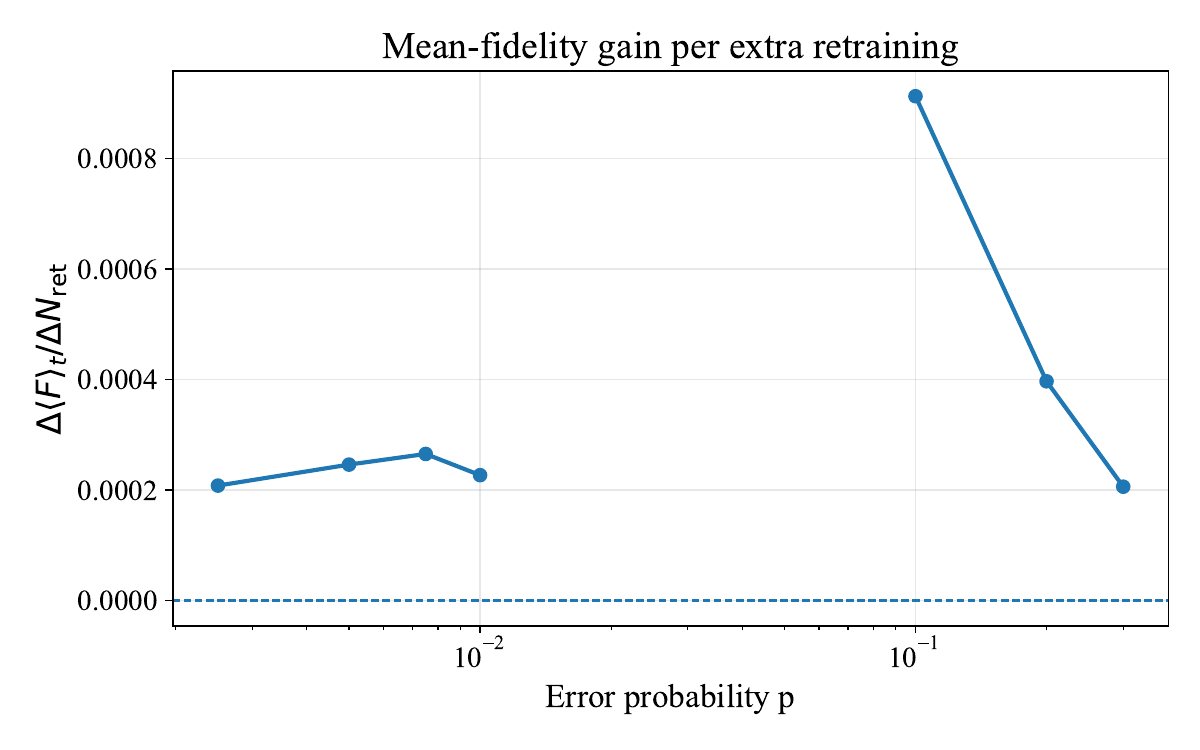}
\caption{
\textbf{Marginal efficiency of deterministic retraining.}
The quantity $\eta(p)$ measures the gain in mean logical fidelity per additional retraining event relative to the bandit policy. Points with small retraining differences are excluded to ensure stability.
}
\label{fig:marginal_efficiency}
\end{figure}

\subsubsection{Regime Analysis}

Combining Figs.~\ref{fig:bandit_perf_cost_updated} and \ref{fig:marginal_efficiency}, we identify three regimes:

\paragraph{Low-noise regime ($p \lesssim 10^{-2}$).}
Both policies achieve $\langle F \rangle_t \approx 1$, with a slight advantage
for the bandit, but here the deterministic trigger is the more economical
policy: it retrains only a few times per trajectory, since the fidelity rarely
falls below the threshold, while the bandit retrains at its nearly constant
rate. The marginal efficiency is small, $\eta \sim 2\text{--}3 \times 10^{-4}$,
so each additional retraining of the bandit improves the mean fidelity
negligibly. In this regime the bandit is therefore over-provisioned, and a
threshold rule is sufficient; the advantage of the bandit lies at high noise,
where the deterministic trigger is not economical.

\paragraph{Intermediate regime ($p \sim 5\times10^{-2}$).}
Here, the retraining cost gap closes ($\Delta N_{\mathrm{ret}} \approx 0$), and the marginal efficiency becomes ill-defined. This corresponds to a crossover region where both policies behave similarly in terms of resource usage, and neither provides a clear advantage.

\paragraph{High-noise regime ($p \gtrsim 10^{-1}$).}
In this regime, the deterministic policy significantly increases retraining frequency (Fig.~\ref{fig:bandit_perf_cost_updated}), while achieving only moderate fidelity gains. The marginal efficiency peaks around $p \approx 0.1$, reaching $\eta \sim 10^{-3}$, indicating that retraining is most valuable in this region.

However, for larger $p$, $\eta(p)$ decreases again, revealing diminishing returns: despite a large increase in retraining events, each additional retraining contributes progressively less to improving fidelity. This highlights an intrinsic inefficiency of deterministic strategies in strongly noisy regimes.

\subsubsection{Interpretation}

This joint analysis reveals that the key advantage of the bandit policy is not only reducing the total number of retraining events, but also \emph{avoiding inefficient retraining}. In particular:

\begin{itemize}
    \item At low noise, the bandit maintains a steady retraining rate and reaches a
    slightly higher fidelity than the deterministic trigger, which retrains only
    when the fidelity has already dropped below threshold.
    \item At high noise, it prevents entering a regime of diminishing returns where retraining becomes increasingly inefficient.
\end{itemize}

Overall, the bandit policy effectively concentrates retraining efforts in regimes where they are most impactful, resulting in a more balanced and scalable performance--cost tradeoff.

\subsection{Regret bounds for gradient bandits}\label{sec:regret_bandits}
\begin{figure}[ht!]
  \centering
  \includegraphics[width=\linewidth]{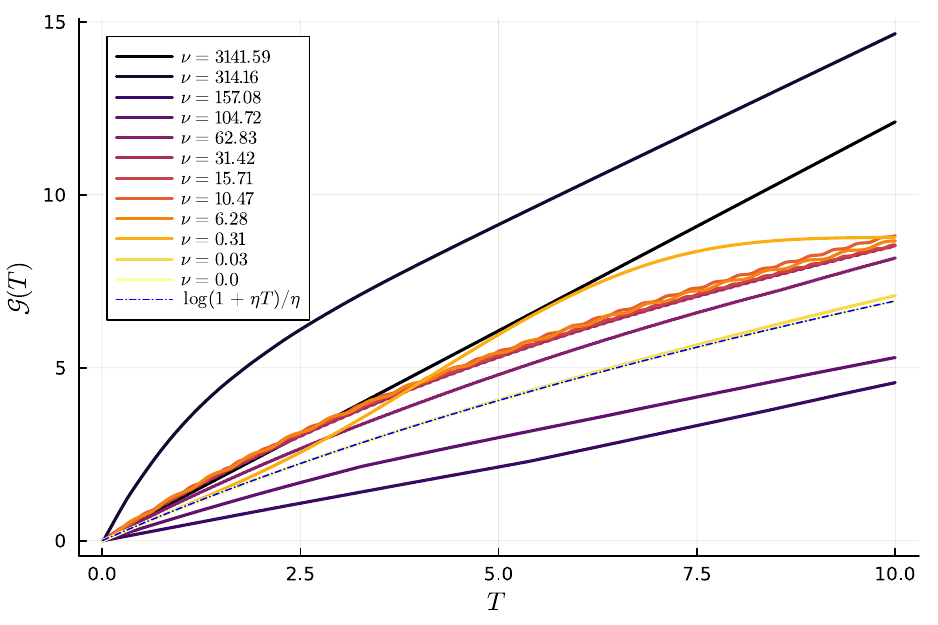}
  \caption{Representation of the cumulative regret for a gradient bandit algorithm simulated with the methodology given in Ref.~\cite{walton2020shortnotesoftmaxpolicy} and a time-dependent reward. Here we plot the cumulative regret for different values of $\nu = \pi/\tau$ and a value of the learning rate of the gradient bandit $\eta = 0.1$. We also plot (dashed blue line) the analytical solution for the case $\nu = 0$. We see here that different values of $\nu$ can reduce or increase the scaling of the regret, but the corresponding curves all lie above the curve that corresponds to $\nu = 0$, where there is no time-dependent noise. It appears that, e.g., for $\tau = 0.01$ the scaling grows asymptotically faster than logarithmic. }
  \label{fig:regret_bounds}
\end{figure}

The analysis of this appendix applies to the context-free gradient bandit of Algorithm~\ref{alg:brave_cf}, for which the policy dynamics can be written as a differential equation.
Bandit algorithms are usually analyzed in terms of the so-called regret. The regret plays a similar role for bandit algorithms as the value function in RL algorithms. More specifically, for a bandit policy $\pi_a(t)$ with $a=1,...,N$ with corresponding reward $r_a(t)$ and optimal reward $r^*$, we can define the total cumulative regret as:
\begin{align}
    \mathcal{G}(T) = \int^T_0 \sum_{a=1}^N \pi_a(t) (r^* - r_a(t)) dt
\end{align}
and the instantaneous regret as:
\begin{align}
    g(t) = \sum_{a=1}^N \pi_a(t) (r^* - r_a(t)).
\end{align}
In several multi-armed bandit algorithms with stationary rewards, specific upper bounds as a function of the total runtime $T$ can be derived, e.g. for the UCB algorithm \cite{Sutton1998}. The optimal bound has a logarithmic scaling $ \mathcal{G}(T) \sim \log(T)$, which guarantees optimal convergence. In the case of (policy) gradient bandit algorithms, an upper bound can be derived -- see Ref.~\cite{walton2020shortnotesoftmaxpolicy} -- by considering a differential equation for the regret and assuming stationary rewards:

\begin{align}
    \pdv{g(t)}{t} = \sum_{a=1}^N  (r^* - r_a)  \pdv{\pi_a(t)}{t},
\end{align}
with the policy
\begin{align}
    \pi_a(t) = \frac{e^{\beta H_a(t)}}{\sum_{a=1}^N e^{\beta H_a(t)}},
\end{align}
which leads to the differential inequality
\begin{align}
    \pdv{g(t)}{t} \leq - \eta \pi_{a^*}^2 g(t)^2,
\end{align}
where $\eta$ represents the learning rate. This latter inequality can be solved and integrated analytically to give the bound $g(t) \sim \frac{1}{\eta^2}\log(T)$. For the bandit problem considered in this work, however, modifications to the treatment described above are necessary: First, the reward is non-stationary, as it oscillates in time due to the time-dependent noise parameter. Here the drift is periodic, $\alpha(t) = \sin^2(\nu t)$ with $\nu = \pi/\tau$, so that $\tau$ is the full period of the oscillation $\hat X \to \hat Z \to \hat X$, rather than the duration of a single transition as in Sec.~\ref{sec:brave_results}. Secondly, the gradient bandit that we implement resets itself periodically, so the update due to the gradient bandit is only valid for the periods in which no resets take place. In the case of non-stationary rewards, the reward itself becomes a time-dependent stochastic process with mean $r_a(t)$ and optimum $r^*(t)$, and we have
\begin{align}
    \pdv{g(t)}{t} = \sum_{a=1}^N  (r^* - r_a)  \pdv{\pi_a(t)}{t} + \sum_{a=1}^N \pi_a(t) \pdv{\Delta r_a(t)}{t}.
\end{align}
The second term of the equation on the right hand side corresponds to the difference between the optimal and the given fidelity of the quantum circuit, that is
\begin{align}\label{eq:infidelity_as_reward}
    \Delta r_a(t) = 1 - \Tr{\rho_{\text{QEC}} \Pi},
\end{align}
for an output state $\rho_{\text{QEC}}$ that is acted upon by the QEC circuit and the error channel and a target pure state $\Pi = \ket{\phi_{\text{target}}} \bra{\phi_{\text{target}}}$. If we consider w.l.o.g. one qubit evolving under a unitary $U$ ($\rho_U = U \rho_{0} U^{\dagger}$) acted upon by the error channel given in Kraus representation in the main text Eq.\ref{eq:qubit_alpha_channel} we have
\begin{align}
    \rho_{\text{out}} = \hat{E}_1 \rho_U \hat{E}_1^{\dagger} + \hat{E}_2 \rho_U \hat{E}_2^{\dagger}
\end{align}
and its derivative with respect to time that uses the chain rule on $\alpha(t)$:
\begin{align}
    \pdv{}{\alpha(t)}\rho_{\text{QEC}} = - i \frac{\pi}{2} p \left[Y, e^{i\frac{\pi}{2} \alpha(t) Y}X \rho_U X e^{-i\frac{\pi}{2} \alpha(t) Y} \right] \pdv{\alpha(t)}{t}
\end{align}
A bound for the derivative of Eq.~\ref{eq:infidelity_as_reward} is given by:
\begin{align}
    \Big\vert  \pdv{\Delta r_a(t)}{t} \Big\vert  = \pi \Im{ \tr{Y e^{i\frac{\pi}{2} \alpha(t) Y}X \rho_U X e^{-i\frac{\pi}{2} \alpha(t) Y} \Pi}} ,
\end{align}
which can be bounded from above as
\begin{align}
    \Big\vert \pdv{\Delta r_a(t)}{t} \Big\vert \leq  2 \pi p \cdot s(t)^2 \nu \sin(2\nu t)
\end{align}
where $s(t) = \vert \cos(\sin(\frac{\pi}{2}  \nu t)) \vert + \vert \sin(\sin(\frac{\pi}{2}  \nu t)) \vert$ and $p$ is the error probability. The instantaneous regret results in the following differential equation for the upper bound:
\begin{eqnarray}
      \pdv{g(t)}{t}  = -\eta \pi_{\text{max}} g(t)^2 + 2 p \cdot s(t)^2 \pi \nu \sin (2\nu t),
\end{eqnarray}

where $\pi_{\text{max}} = \text{max}_t \left[ \pi_*(t) \right]$ = 1. We can see that the second term in the equation modifies the bound, but the function is bounded. The learning rate $\eta$ of the gradient bandit, the noise frequency $\nu$ and the structure of the spectrum of the Hamiltonian defining the gradient of the quantum cost function. In Fig.~\ref{fig:regret_bounds} we show some simulations of the cumulative regret for different values of $\tau$ ($\nu$) and $\eta$. We observe that the behaviour for small frequencies resembles the typical expected logarithmic scaling of the regret as a function of the total time $T$. For larger frequencies more pronounced oscillations start to appear and the logarithmic scaling seems to be less pronounced. This reflects the two classes of non-stationary bandit problems studied in
Ref.~\cite{wei2018abruptlychangingslowlyvaryingmultiarmedbandit}.: in a slowly-varying environment the mean rewards change by a small
amount between consecutive steps, while in an abruptly-changing one they switch at isolated breakpoints. Our drift is smooth, so a slow drift (small $\nu$) belongs to the first class, while a fast drift, which moves the reward appreciably between consecutive control steps, approaches the second. These simulations were realized in the limit of de-facto infinite sampling rate: this is somewhat a requirement to treat the gradient bandit algorithm using differential equations.

\begin{algorithm}[H]
\caption{Context-free BRAVE controller (gradient bandit with reset)}\label{alg:brave_cf}
\begin{algorithmic}[1]
\State Initialize preferences $\mathcal{H} \gets [h_0, h_1]$ \Comment{e.g., keep, train}
\State Set baseline reward $\bar{F}$ \Comment{e.g., target fidelity}
\State Compute action probabilities $\pi \gets \operatorname{softmax}(\mathcal{H})$
\For{each time step $t$}
    \State Compute or update the quantum noise model
    \State Sample action $a_t \sim \pi$ \Comment{$0 =$ keep, $1 =$ retrain}
    \If{$\,t = 0 \lor a_t = 1\,$}
        \State Reset bandit: $\mathcal{H} \gets [h_0, h_1]$
        \State Retrain $\mathbf{\theta}$ params using current noise
    \EndIf
    \State Run encoding, apply noise, perform recovery
    \State Measure fidelity $F_t$ with respect to target state
    \State Update preferences:
    \Statex $\mathcal{H}_0 \gets \mathcal{H}_0 + (F_t - \bar{F})(1 - \pi_0)$
    \Statex $\mathcal{H}_1 \gets \mathcal{H}_1 - (F_t - \bar{F})\pi_1$
    \State Update action probabilities: $\pi \gets \operatorname{softmax}(\mathcal{H})$
\EndFor
\end{algorithmic}
\end{algorithm}

 For retraining purposes we assume to resample the cost function using a sample rate $f_s$. The full modified gradient bandit method is described as pseudo-code in Algorithm \ref{alg:brave_cf}. Empirically, we still observe that the gradient bandit alone is not sufficient to achieve a fast convergence of the policy, so we introduce reset steps that are activated when $a_t = 1$ (\textit{retrain}). These steps reset the bandit weights allowing for faster convergence in the presence of strong noise oscillations and thus help increase convergence times.

\section{Reinforcement-learning background}\label{sec:RL}
Reinforcement Learning (RL) is a subset of machine learning that concentrates on training intelligent agents to make sequential decisions by interacting with their environment. Unlike supervised learning, where models rely on labeled data, or unsupervised learning, which uncovers hidden patterns, RL focuses on learning optimal \emph{policies} by agents interacting with environments that offer feedback in the form of rewards \cite{Sutton1998}. The aim is to maximize a weighted sum of rewards over successive steps of a trajectory, aligning with long-term objectives.

In the field of RL, the state space, denoted as $S$, encompasses all potential states an agent can access. This space can be discrete or continuous and encapsulates pertinent information regarding the environment's current status. To ascertain the action transitioning the agent from the current state $s_t$ at time $t$ to the subsequent state $s_{t+1}$, the agent relies on a policy. This policy can be formally expressed as a probability distribution of selecting an action $a$ given a particular state $s$, denoted as $\pi(a|s)$.

In order to optimize the agent's policy, we establish a figure of merit also called reward function, labeled as $R$, which quantifies the immediate consequence of the agent's actions. In essence, $R(s, a)$ maps a state-action pair $(s, a)$ to a real number, denoting the reward obtained when executing action $a$ from state $s_{t+1}$. The agent's primary objective is to maximize the expected cumulative reward, often called expected return, formally expressed as the so-called discounted reward:
\begin{align}
    G = \mathbb{E}_{\pi} \left[\sum_{t=0}^{\infty} \gamma^t R(s_t, a_t, s_{t+1})\right],
\end{align}
where $\gamma \in [0,1)$ serves as a discount factor. 

To facilitate policy optimization, two central concepts are often used: the \emph{state-value function} $V(s)$ and the \emph{action-value function} $Q(s, a)$. The state-value function $V(s)$ estimates the expected return starting from state $s$ and thereafter following policy $\pi$, given by
\begin{equation}
    V(s) = \mathbb{E}_{\pi} \left[\sum_{t=0}^{\infty} \gamma^t R(s_t, a_t, s_{t+1}) \mid s_0 = s \right].
\end{equation}
Similarly, the action-value function $Q(s, a)$ evaluates the expected return starting from state $s$, taking action $a$, and then following policy $\pi$, defined as
\begin{align}
    Q(s, a) = \mathbb{E}_{\pi} \left[\sum_{t=0}^{\infty} \gamma^t R(s_t, a_t, s_{t+1}) \mid s_0 = s, a_0 = a \right].
\end{align}
These functions provide the foundation for many RL algorithms by guiding the agent toward actions that yield higher long-term rewards. Both functions $Q(s,a)$ and $V(s)$ can be combined to form the advantage function $A(s,a) = Q(s,a) - V(s)$.

\subsection{Multi-armed bandits}
The multi-armed bandit problem is a foundational challenge in RL where an agent must choose from a set of actions (or "arms"), each with an unknown reward distribution. The goal is, as for the other RL frameworks, to maximize cumulative rewards over time by balancing exploration (trying new actions) and exploitation (choosing the best-known actions). Bandit problems help model decision-making in uncertain environments, offering insights into more complex reinforcement learning tasks. 

In particular, in this subsection we will focus on the role of a particular type of Bandits algorithm called Gradient Bandits.
Gradient Bandits use optimization techniques to balance exploration and exploitation by adjusting the probabilities of selecting actions through policy-based methods. Instead of estimating the reward value for each action, the algorithm maintains a set of preferences $ H(a) $ for each action $ a $. These preferences are converted into probabilities using the \emph{softmax function}:

\begin{equation}
\label{eq: bandit_policy}
\pi(a_t = a) = \pi_a(t) = \frac{e^{H(a_t)}}{\sum_{b=1}^k e^{H(b)}}
\end{equation}

This ensures that actions with higher preferences are chosen more often, but all actions retain some probability of being selected, facilitating continued exploration. There are several approaches for multi-armed bandit problems with stationary rewards, e.g., $\epsilon$-greedy approaches and Thompson sampling, and gradient bandits -- see Ref.~\cite{slivkins2024introductionmultiarmedbandits} for a comprehensive introduction. 
In the latter approach the preferences $H(a)$ are updated using gradient ascent based on the received reward $ r_t $ and a baseline reward $ \bar{R_t} $, which helps normalize updates. The update rule is:
\begin{eqnarray}
H(a_t) \leftarrow H(a_t) + \eta \cdot (r_t - \bar{R_t}) \cdot \nabla_H \log \pi(a_t)
\end{eqnarray}

Here, $\eta$ is the learning rate, and $ \nabla_H \log \pi(a_t) $ represents the gradient of the log-probability of the chosen action. This update maximizes the expected reward by increasing the preference for actions that yield higher rewards over time. The softmax ensures a balance between exploration and exploitation throughout the learning process. 
The convergence ratio of a multi-armed bandit algorithm can be determined by studying the scaling with time $T$ (i.e., the horizon) of the so-called regret, i.e.,
\begin{align}
    \mathcal{G}(T) = \int^T_0 \sum_a (r_a - r_*)\pi_a(t) dt,
\end{align}
where $\pi_a(t)$ is defined in Eq.~\ref{eq: bandit_policy}, $r_a$ is the reward obtained under the current policy and $r_*$ the reward obtained when using the optimal policy. For the algorithms discussed above, under the stationarity condition of the reward, the regret scales as $\mathcal{G}(t) \sim \log(T)$, which ensures particularly fast convergence in time $T$. A linear growth of the regret with time, $O(T)$, would correspond to a random strategy, which is why we expect the regret to grow at least sub-linearly. However, the convergence is still affected by parameters such as the exporation parameter $\epsilon$ in the $\epsilon$-greedy case and the learning rate $\eta$ for the gradient bandit algorithm. For a more detailed treatment of the regret scaling in gradient bandit algorithms, see Section \ref{sec:regret_bandits} or Ref.~\cite{walton2020shortnotesoftmaxpolicy}. In cases in which the multi-armed bandit problem is more complex, worse scaling of the regret bounds are common. In the case of contextual bandits, one often has $O(\sqrt{T})$ regret \cite{badanidiyuru2015resourcefulcontextualbandits}. In the case of non-stationary rewards, it is often hard to obtain regret bounds, particularly for gradient bandits. For the gradient bandit problem considered here, we try to study the behaviour of the regret bound in Section \ref{sec:regret_bandits}.

\subsection{Policy-gradient methods}

In the realm of RL, Q-learning and Policy Gradient (PG) methods represent two distinct approaches.
Q-learning estimates the value of state-action pairs and works well in discrete action spaces with fully observable environments. In contrast, PG directly optimizes policies, making it more effective for continuous action spaces and non-fully observable environments, where discretization and complete state information can be challenging for Q-learning. In this work we concentrate on PG methods: this section provides a brief introduction about the foundational principles of PG approaches.

We recall that a policy $\pi(a|s)$ determines the probability of selecting action $a$ given the current state $s$ of the RL environment. In the PG framework we assume that the each policy is associated to a set of parameters $\boldsymbol{w}$, e.g., neural network weights. 

As previously highlighted, the objective of RL lies in finding an optimal policy that maximize the expected return; in the PG framework this goal can be achieved by iteratively updating the parameters $\boldsymbol{w}$ via the RL policy gradient update rule:

\begin{equation}
\delta w_j = \eta \frac{\partial\mathbb{E}[R]}{\partial w_j} = \eta \bigg[\mathbb  R \sum_t \pdv{}{w_j} \ln \pi_{\boldsymbol{w}}(a_t | s_t) \bigg].  
\label{eq: policy_gradient}
\end{equation}

Here, $\eta$ denotes the learning rate parameter which is a real value parameter on which the converging properties of the RL algorithm depend and $\mathbb{E}[\cdot]$ represents the expectation value computed over all possible rewards.
These fundamental elements form the core policy gradient approach. 

Furthermore, Eq.~\ref{eq: policy_gradient} serves as the standard recipe for policy based RL in fully observed environments. This approach can be extended to accommodate partially observed environments, where the policy relies solely on observations rather than the complete state information. These observations offer only partial insights into the actual state of the environment, introducing additional complexities in the RL framework.

In our work, we choose to adopt the Stable Baselines3 implementation of the PPO algorithm \cite{stable-baselines3} in which a neural network is used to compute the $\pi_{\boldsymbol{w}}$ policy, with the multidimensional parameter $\boldsymbol{w}$ encompassing all the network's weights and biases. The neural network takes the current state ($s_t$) as an input vector and produces the action probabilities ($\pi_{\boldsymbol{w}}$) as its output. 

\subsection{Curriculum learning}
Curriculum learning \cite{narvekar2020curriculum} in RL involves gradually exposing an agent to increasingly complex tasks or environments during training. Instead of presenting the agent with all tasks at once, curriculum learning introduces a structured learning schedule, starting with simpler tasks and progressing to more challenging ones. This approach aims to facilitate the learning process by allowing the agent to first master simpler concepts before tackling more complex ones. Curriculum learning can enhance the agent's learning efficiency and improve its performance on difficult tasks by leveraging a carefully designed learning trajectory. In this work, we used two different curriculum learning strategies: (i) Task specific curriculum~\cite{curriculum2009}, (ii)  Mix\&Match(MM)~\cite{Mix_Match_Curr}.\\
\begin{itemize}
    \item[(i)] \textbf{Task specific:} In this curriculum framework, we outline a sequence of $n$ tasks denoted as $\mathcal{G} = \{g_1,\dots,g_n\}$, where $g_1$ denotes the most basic task, progressing towards the ultimate objective represented by $g_n$. Initially, the agent focuses on mastering task $g_1$. Once it reaches a certain proficiency level or convergence, the task shifts, requiring the agent to adjust its policy accordingly. This iterative process continues until the agent successfully solves the final task, $g_n$. 
    \item[(ii)] \textbf{Mix\&Match:} This curriculum framework trains an RL agent by leveraging knowledge transfer among $m$ agents. These agents are arranged in a sequence from simple to complex, with each agent parameterized with some shared weights, typically through common lower layers. Once all the agents have been trained successfully the final policy is defined as a weighted sum with coefficients $\alpha_i$ between all the $m$ policies: 
\begin{equation}
    \pi_{mm}(a|s)=\sum_{i=1}^K\alpha_i\pi_i(a|s)\:.
\end{equation}
\end{itemize}
\subsection{Multi-agent reinforcement learning}
Multi-Agent Reinforcement Learning (MARL) is an area of reinforcement 
learning where multiple agents learn to interact with an environment, 
each pursuing its own goals while influencing each other's learning 
process~\cite{marl-book}. Unlike traditional single-agent reinforcement 
learning, where a single agent interacts with the environment, MARL 
involves multiple agents that may have different objectives, policies, 
and learning dynamics. Agents can be cooperative, competitive, or a mix 
of both, and their interactions can lead to complex emergent behaviors, 
making MARL a challenging but rewarding area of research.

Some common approaches in MARL include centralized training with 
decentralized execution (CTDE)~\cite{CTDE}, where agents share 
information during training but act independently during execution; 
decentralized training with decentralized execution 
(DTDE)~\cite{wen2021dtde}, in which agents learn independently without 
sharing information during training or execution; and centralized 
training with centralized execution (CTCE)~\cite{CTCE}, where agents 
share information both during training and execution.

In this work, we adopt a sequential decentralized training paradigm. 
We formalize the overall problem as a multi-agent Markov game with 
(i)~$\mathcal{N} = \{1,2, \dots, N\}$ agents, (ii)~state space 
$\mathcal{S}$, (iii)~joint action space $\mathcal{A} = \mathcal{A}_1 
\times \mathcal{A}_2\times \dots \times \mathcal{A}_N$, 
(iv)~transition function $T(s'|s, a_1, a_2, \dots, a_N)$, and 
(v)~a reward function for each agent $r_i : \mathcal{S}\times\mathcal{A}$. 
Within this framework, each agent $i$ independently learns its own 
policy $\pi_i$, conditioned on its local trajectory of observations, 
actions, and rewards, as well as on the fixed policy of agent $i-1$ 
(the first agent is trained independently). We define the 
\emph{best response} policy $\pi_i^* \in \Pi_i$ of agent $i$ to the 
fixed policies $\boldsymbol{\pi}_{-i}$ of all preceding agents as
\begin{equation}
    V^i_{\pi_i^*, \boldsymbol{\pi}_{-i}} \geq 
    V^i_{\pi_i, \boldsymbol{\pi}_{-i}}
\end{equation}
for all policies $\pi_i \in \Pi_i$.

The learning process for each agent is conducted in isolation, without 
direct access to the internal states or gradients of other agents, as 
in standard independent learners~\cite{Tan1997MultiAgentRL, MARL-dyn}. 
Agents are trained sequentially: once an agent satisfies its training 
objective, its policy is frozen and used to simulate the environment 
for the next agent. If an agent fails to satisfy its objective, 
training is restarted for that agent until the condition is met. 
Each agent therefore incorporates the behavior of all preceding agents 
into its learning process. This sequential procedure yields a composed 
joint policy
\begin{equation}
    \pi = \pi_{N}(\pi_{N-1}(\cdots(\pi_1))),
\end{equation}
where each $\pi_i$ is a best response to the frozen policies of all 
preceding agents.

\section{Qutrit stabilizer algebra}
\label{app:qutrit_algebra}
We present a specific analysis of the qutrit algebra related to the stabilizer formalism. We show how to compute commutators in the Lie-Algebra picture which reduces the computational complexity by $d^{3n}$.
\subsection{The Exchange Property}

The matrices $\hat{X}$ and $\hat{Z}$ satisfy the following exchange property:
\[
\hat{X} \hat{Z} = \omega \hat{Z} \hat{X} \quad \text{and} \quad \hat{Z} \hat{X} = \omega^2 \hat{X} \hat{Z}.
\]

From this we can generalize by iteratively moving each $\hat{Z}$ to the left, one $\hat{X}$ at a time, with each such operation providing a $\omega$ factor, that for $i$ and $j$ terms:
\[
\hat{X}^i \hat{Z}^j = \omega^{ij} \hat{Z}^j \hat{X}^i.
\]
And equivalently for $\hat{Z}^j \hat{X}^i$:
\[
\hat{Z}^j \hat{X}^i = \omega^{ij} \hat{X}^i \hat{Z}^j.
\]

\subsection{The Stabilizer Basis}

Since $\hat{X}^3 = \hat{Z}^3 = \mathbb{I}$, any combination of $\hat{X}$ and $\hat{Z}$ matrices can be rewritten in terms of $\omega^k \hat{X}^i \hat{Z}^j$ with $i, j, k \in \{0, 1, 2\}$, leading to a total of 9 possible matrices with each three possible phases. We therefore store stabilizer matrices via the two integers $i$ and $j$ and possibly the phase via $k$. We keep in mind, that the exponents $i, j, k$ can always be replaced by their modulo 3 value, since $\hat{X}^3 = \hat{Z}^3 = \mathbb{I}$ and $\omega^3 = 1$.

For a system of $n$ qutrits we store the vectors $\{i_1, i_2, \ldots, i_n\}$ and $\{j_1, j_2, \ldots, j_n\}$ as well as a single $k$ for the phase, to represent the Kronecker product of the stabilizer matrices:
\[
\omega^k \bigotimes_{m=1}^n \hat{X}^{i_m} \hat{Z}^{j_m}.
\]

\subsection{Products and Commutators of Stabilizer Matrices}

The product of two stabilizer matrices is another stabilizer matrix. Specifically, we find for the product of the stabilizers $ij$ and $kl$:
\[
\left( \hat{X}^i \hat{Z}^j \right) \left( \hat{X}^k \hat{Z}^l \right) = \omega^{jk} \hat{X}^{i+k} \hat{Z}^{j+l}.
\]

Equivalently we have:
\[
\left( \hat{X}^k \hat{Z}^l \right) \left( \hat{X}^i \hat{Z}^j \right) = \omega^{li} \hat{X}^{i+k} \hat{Z}^{j+l}.
\]

From this we conclude, that the commutator is given by:
\[
\left[ \hat{X}^i \hat{Z}^j, \hat{X}^k \hat{Z}^l \right] = (\omega^{jk} - \omega^{li}) \hat{X}^{i+k} \hat{Z}^{j+l},
\]
and so two operators commute exactly when $\text{mod}_3(jk) = \text{mod}_3(il)$.

In the case of $n$ qutrits, the commutator of two stabilizer matrices is similarly given by:
\[
\left[ \bigotimes_{m=1}^n \hat{X}^{i_m} \hat{Z}^{j_m}, \bigotimes_{m=1}^n \hat{X}^{k_m} \hat{Z}^{l_m} \right] = \left( \omega^{\sum_{m=1}^n j_m k_m} - \omega^{\sum_{m=1}^n i_m l_m} \right) \bigotimes_{m=1}^n \hat{X}^{i_m + k_m} \hat{Z}^{j_m + l_m}.
\]

In order to check if two $n$ qutrit stabilizer matrices commute, we need to check if:
\[
\text{mod}_3 \left( \sum_{m=1}^n j_m k_m \right) = \text{mod}_3 \left( \sum_{m=1}^n i_m l_m \right).
\]

\subsection{Table}

\begin{table}[h]
\centering
\begin{tabular}{|c|c|c|c|}
\hline
Gate            & $S_1$       & $S_2$       & Output                           \\ \hline
$\text{CNOT}_3$ & $I_i$       & $\hat{Z}_j$ & $\hat{Z}_i \otimes \hat{Z}_j$    \\ \hline
$\text{CNOT}_3$ & $\hat{Z}_i$ & $I_j$       & $\hat{Z}_i \otimes I_j$          \\ \hline
$\text{CNOT}_3$ & $I_i$       & $\hat{X}_j$ & $I_i \otimes \hat{X}_j$          \\ \hline
$\text{CNOT}_3$ & $\hat{X}_i$ & $I_j$       & $\hat{X}_i \otimes \hat{X}_j^{2}$ \\ \hline
H               & $\hat{Z}$   & /           & $\hat{X}$                        \\ \hline
H               & $\hat{X}$   & /           & $\hat{Z}$                        \\ \hline
$\text{S}_q \bmod 3$ & $\hat{Z}$   & /           & $\hat{Z}^q$                     \\ \hline
$\text{S}_q \bmod 3$ & $\hat{X}$   & /           & $\hat{X}^q$                     \\ \hline
\end{tabular}
\caption{Stabilizers gate map.}
\label{tab:gates_qutrits_stab}
\end{table}

\end{document}